\documentclass[english,12pt]{article}
\pdfoutput=1
\usepackage[a4paper,bottom=4.2cm,top=1cm,head=3cm,width=18cm,dvipdfm]{geometry}
\evensidemargin=\oddsidemargin

\usepackage[T1]{fontenc}
\usepackage[latin9]{inputenc}
\usepackage{amstext}
\usepackage{babel}
\usepackage{verbatim}
\usepackage{amsfonts}
\usepackage{mathrsfs}
\usepackage{amsmath}
\usepackage{amssymb}
\usepackage{bm}
\usepackage{extarrows}
\usepackage{slashed}
\usepackage{isodateo}
\usepackage{xcolor}
\usepackage{graphicx}
\usepackage[low-sup]{subdepth}
\usepackage{isodateo}

\usepackage[linktocpage=true]{hyperref}
\hypersetup{
	colorlinks=true,
	citecolor=blue,
	linkcolor=blue,
	urlcolor=blue,
	pdfauthor={},
	pdftitle={},
	pdfsubject={}
}

\numberwithin{equation}{section}

\makeatletter
\@ifundefined{textcolor}{}
{%
 \definecolor{BLACK}{gray}{0}
 \definecolor{WHITE}{gray}{1}
 \definecolor{RED}{rgb}{1,0,0}
 \definecolor{GREEN}{rgb}{0,1,0}
 \definecolor{BLUE}{rgb}{0,0,1}
 \definecolor{CYAN}{cmyk}{1,0,0,0}
 \definecolor{MAGENTA}{cmyk}{0,1,0,0}
 \definecolor{YELLOW}{cmyk}{0,0,1,0}
}

\newcommand{\red}{\color{red}}

\newcommand{\blu}{\color{blue}}

\newcommand{\fr}[2]{\mbox{$\frac{\,{#1}\,}{#2}$}}

\def\bge{\begin{equation}}
\def\ede{\end{equation}}
\def\bga{\begin{aligned}}
\def\eda{\end{aligned}}
\newcommand{\beq}{\begin{equation}}
\newcommand{\eeq}{\end{equation}}
\newcommand{\bq}{\begin{equation}}
\newcommand{\eq}{\end{equation}}
\newcommand{\ba}{\begin{array}}
\newcommand{\ea}{\end{array}}
\newcommand{\beqa}{\begin{eqnarray}}
\newcommand{\eeqa}{\end{eqnarray}}
\newcommand{\beqs}{\begin{subequations}}
\newcommand{\eeqs}{\end{subequations}}

\def\nn{\nonumber}

\def\dis{\displaystyle}

\def\({\left(}
\def\){\right)}

\def\leqq{\leqslant}

\def\LG{\langle}
\def\RG{\rangle}
\def\End{\end{document}}

\def\ii{{\tt i}}

\def\al{\alpha}
\def\be{\beta}
\def\ga{\gamma}

\def\ep{\epsilon}

\def\bge{\begin{equation}}
\def\ede{\end{equation}}
\def\bga{\begin{aligned}}
	\def\eda{\end{aligned}}

\def\nn{\nonumber}

\def\dis{\displaystyle}

\def\({\left(}
\def\){\right)}
\def\[{\left[}
\def\]{\right]}

\def\End{\end{document}}

\def\leqq{\leqslant}

\def\al{\alpha}
\def\be{\beta}

\def\lam{\lambda}

\def\ga{\gamma}

\def\ep{\epsilon}
\def\lam{\lambda}

\def\di{\mathrm{d}}

\def\to{\rightarrow}
\def\To{\Rightarrow}
\def\ii{\mathrm{i}}

\def\sm{\text{sm}}
\def\BT{\widetilde{B}}

\def\cut{\Lambda}

\newcommand{\mL}{\mathcal{L}}
\newcommand{\mO}{\mathcal{O}}

\def\LL{\mathcal{L}}

\def\SZZ{\mathcal{Z}}

\def\OBW{\mathcal{O}_{\!\widetilde{B}W}^{}}
\def\OGP{\mathcal{O}_{\!G+}^{}}
\def\OGM{\mathcal{O}_{\!G-}^{}}
\def\OCP{\mathcal{O}_{\!C+}^{}}
\def\OCM{\mathcal{O}_{\!C-}^{}}

\newlength{\halfpagewidth}
\divide\halfpagewidth by 2

\def\End{\end{document}}
\renewcommand{\thefootnote}{\fnsymbol{footnote}}
\makeatother

\begin{document}

\thispagestyle{empty}

\vspace*{-15mm}

\begin{center}
{\Large\bf\boldmath Probing New Physics in Dimension-8 \\[2mm]
Neutral Gauge Couplings at $e^+e^-$ Colliders}

\vspace*{6mm}
{\sc John Ellis}\,$^{a,b,}$\footnote{john.ellis@cern.ch},
~{\sc Hong-Jian He}\,$^{b,c,}$\footnote{hjhe@sjtu.edu.cn},
~{\sc Rui-Qing Xiao}\,$^{b,}$\footnote{xiaoruiqing@sjtu.edu.cn}

\vspace*{3mm}
$^a$\,Department of Physics, Kings College London, Strand, London WC2R 2LS, UK;
\\
Theoretical Physics Department, CERN, CH-1211 Geneva 23, Switzerland;
\\
NICPB, R\"{a}vala 10, Tallinn 10143, Estonia
\\[2mm]
$^b$\,Tsung-Dao~Lee Institute $\&$ School of Physics and Astronomy, \\
Shanghai Key Laboratory for Particle Physics and Cosmology,\\
Shanghai Jiao Tong University, Shanghai 200240, China
\\[2mm]
$^c$\,Institute of Modern Physics and Department of Physics,\\
Tsinghua University, Beijing 100084, China;
\\
Center for High Energy Physics, Peking University, Beijing 100871, China

\end{center}

\vspace*{2mm}
\begin{abstract}
\baselineskip 17pt
\noindent
Neutral triple gauge couplings (nTGCs) are absent in the standard model effective
theory up to dimension-6 operators, but could arise from dimension-8 effective
operators. In this work, we study the {\it pure gauge operators} of dimension-8
that contribute to nTGCs and are independent of the dimension-8 operator involving
Higgs doublets. We show that the pure gauge operators generate both $Z\gamma Z^*$
and $Z\gamma\gamma^*$ vertices with rapid energy dependence $\propto\!E^5$,
which can be probed sensitively via the reaction $e^+e^-\!\!\to\!Z\gamma$\,.
We demonstrate that measuring the nTGCs via the reaction
$e^+e^-\!\!\to\!Z\gamma$ followed by $Z\!\!\to\! q\bar{q}$ decays
can probe the new physics scales of dimension-8 pure gauge operators up to
the range $(1\!-\!5)$\,TeV at the CEPC, FCC-ee and ILC colliders with $\sqrt{s}=(0.25\!-\!1)$\,TeV, and up to the range $(10\!-\!16)$TeV at CLIC
with $\sqrt{s}=(3\!-\!5)$\,TeV, assuming in each case an integrated luminosity
of $5$\,ab$^{-1}$. We compare these sensitivities with the corresponding probes of
the dimension-8 nTGC operators involving Higgs doublets and the dimension-8 fermionic
contact operators that contribute to the $e^+e^-Z\ga$ vertex.
\\[5mm]
%
KCL-PH-TH/2020-28, CERN-TH-2020-076
\\[2mm]
{\sc Science China} (Phys.\,Mech.\,Astro.)
64 (2021) 221062, no.2 $[$\,arXiv:2008.04298\,$].$
\\
DOI:\,10.1007/s11433-020-1617-3
\\[3mm]
Selected as ``Editor's Focus'' and ``Cover Article''.
\\[1mm]
Journal's Research Highlights on this article: \\
https://doi.org/10.1007/s11433-020-1633-x
~~
https://doi.org/10.1007/s11433-020-1630-x
\\
https://doi.org/10.1007/s11433-020-1631-0
~~
https://doi.org/10.1007/s11433-020-1632-7
%
\end{abstract}

\newpage
\baselineskip 18pt
\tableofcontents

\vspace*{5mm}

\setcounter{footnote}{0}
\renewcommand{\thefootnote}{\arabic{footnote}}
\section{\large\hspace{-2mm}Introduction}
\label{sec:1}

The standard model effective field theory (SMEFT), which contains the
standard model (SM) Lagrangian of dimension\,4 and higher-dimensional effective
operators constructed out of the known SM fields\,\cite{dim6,dim6x},
provides a powerful model-independent approach
for probing possible new physics beyond the SM\,\cite{John}.
Operators of dimension $d \!>\! 4$ in the SMEFT have coefficients
that are scaled by inverse powers of the ultraviolet (UV) cutoff $\,\Lambda$\,,\,
larger than the electroweak Higgs vacuum expectation value (VEV),
which is expected to arise from and be comparable to the mass scale of the
underlying new physics beyond the SM (BSM).
The leading terms in an expansion in the dimensions of SMEFT operators are those
of dimension\,5, which may play roles
in generating neutrino masses\,\cite{weinberg-d5}\cite{BEG},
and those of dimension\,6 \cite{reviews},
which have been studied extensively in the context of collider physics\,\cite{SMEFT}.

\vspace*{1mm}

Operators of dimension\,8 or higher have been
classified\,\cite{above6,nTGC1,Degrande:2013kka,dim8class},
but there have been fewer studies of their
phenomenology and possible experimental probes and constraints. This is because their
contributions to scattering amplitudes are typically suppressed by higher powers in
the energy/cutoff expansion, $(E/\cut)^{d-4}$,\,
rendering them generally less relevant and relatively inaccessible to low-energy experiments.
However, there are instances where such higher-dimensional operators may become more accessible,
especially if they make the leading BSM contributions to processes.
Examples include dimension-8 contributions
to light-by-light scattering\,\cite{EMY} and
gluon-gluon$\,\to\!\!\gamma \gamma$\,\cite{EG},
to the reaction $e^+e^-\!\!\to\!Z\ga$
via neutral triple-gauge couplings (nTGCs)\,\cite{Ellis:2019zex},
and to processes at future $pp$ colliders\,\cite{pp-nTGC},
that do not receive contributions from dimension-6 operators.
Other recent studies of dimension\,8 operators also appeared in \cite{d8-others}.

\vspace*{1mm}

In this work, we introduce a new type of dimension-8 {\it pure gauge
operators} for the nTGCs and study their nTGC contributions to the process
$e^+e^-\!\!\to\!Z\ga$\,.\,
In a recent paper\,\cite{Ellis:2019zex}, we studied the contribution
of the {\it Higgs-related} dimension-8 operator
$\,\mathcal{O}_{\widetilde{B}W} \!=
\ii\, H^\dagger  \widetilde{B}_{\mu\nu}W^{\mu\rho} \!\left\{D_\rho,D^\nu\right\}\! H$\,
to $e^+e^-\!\!\to\!Z\ga\,$ with leptonic $Z$ decays,
and estimated the sensitivities of the projected future $e^+ e^-$
colliders for probing the associated new physics cutoff scale $\Lambda$\,.\,
However, in addition to the Higgs-related
dimension-8 operator $\mathcal{O}_{\widetilde{B}W}^{}$,
we find two {pure gauge operators}
of dimension\,8,  $\mO_{G+}^{}$ and $\mO_{G-}^{}$,
that contribute to nTGCs and are independent of the dimension-8 operator
involving Higgs doublets (cf.\ Section\,\ref{sec:2}).
For the present work,
we will study how these operators contribute to the reaction
$e^+e^-\!\!\to\!Z\ga$\,,
and estimate the prospective sensitivities to their corresponding
new physics cutoff scales $\,\Lambda\,$ at the $e^+ e^-$ colliders of
CEPC\,\cite{CEPC}, FCC-ee\,\cite{FCCee}, ILC\,\cite{ILC} and
CLIC\,\cite{CLIC} that are currently under planning.
We find that the accelerators of CEPC, FCC-ee and ILC
with $\sqrt{s}=(0.25\!-\!1)$\,TeV
should be able to probe $\,\Lambda\lesssim\!(1\!-\!5)$TeV,\,
whereas the higher design energy of CLIC with $\sqrt{s}=(3\!-\!5)$\,TeV
should enable it to probe $\Lambda\lesssim\!(10\!-\!16)$TeV,\,
for a sample integrated luminosity $\mathcal{L}\!=\!5\,$ab$^{-1}$.
These sensitivities are substantially stronger than
what we found before\,\cite{Ellis:2019zex}
for the Higgs-related nTGC operator $\,\mathcal{O}_{\widetilde{B}W}^{}$
by using the leptonic channels of the final state $Z$ decays.
For comparison, we further study the related dimension-8 fermionic operators which
contribute to the $e^-e^+Z\ga$ contact vertex and thus the reaction
$e^+e^-\!\!\to\!Z\ga$\,.

\vspace*{1mm}

The layout of this paper is as follows. In Section\,\ref{sec:2}, we
formulate the new type of pure gauge operators of dimension-8 and derive their
contributions to nTGCs. We discuss the distinction between these
pure gauge operators and the Higgs-related dimension-8 operators
studied previously\,\cite{Ellis:2019zex}.
Then, in Section\,\ref{sec:3}, we analyze
${Z\ga}$ production followed by hadronic decays
$\,Z\!\to\! q\bar{q}$\,,\, incorporating contributions of
the dimension-8 pure gauge operator and other operators to nTGCs.
Specifically, in Section\,\ref{sec:3.1}, we analyze cross sections
and angular distributions, and in Section\,\ref{sec:3.2}
we study effects of the dijet angular
resolution. In Section\,\ref{sec:4}, we explore the sensitivities
for probing the nTGCs via hadronic $Z$ decays,
analyzing separately individual probes of $\mO_{G\pm}^{}$,
$\mathcal{O}_{\widetilde{B}W}^{}$,
and the fermionic contact operators.
We also make fits to each pair of the dimension-8 operators
and analyze the correlations between them.
Finally, Section\,\ref{sec:5} presents our conclusions.
The helicity amplitudes of the $Z\gamma$ production
for the SM contributions and for the dimension-8 contributions
are summarized in Appendix\,\ref{app:A}.

\section{\large\hspace{-2mm}Contributions of Pure Gauge Operators to nTGCs}
\label{sec:2}

For studying the neutral triple gauge couplings (nTGCs),
it is customary in the literature to consider dimension-8 operators
involving the SM Higgs doublets\,\cite{Degrande:2013kka}.
Among such Higgs-related dimension-8 operators,
one may choose the following
independent CP-even dimension-8 operator after using the
equations of motion (EOM)\,\cite{Degrande:2013kka,Ellis:2019zex}: 
%
\begin{eqnarray}
\label{eq:nTGC-H}
\mathcal{O}_{\widetilde{B}W} \!&=&
\ii\, H^\dagger  \widetilde{B}_{\mu\nu}W^{\mu\rho}
\!\left\{D_\rho,D^\nu\right\}\! H+\text{h.c.},
\label{eq:obtw}
\end{eqnarray}
%
where $H$ denotes the SM Higgs doublet. The dual U(1) field strength is defined as
$\,\widetilde{B}_{\mu\nu}^{} \!\equiv\! \ep_{\mu\nu\al\be}^{}B^{\al\be}$\,, and analogously
$\,\widetilde{W}_{\mu\nu}^{} \!\equiv\! \ep_{\mu\nu\al\be}^{}W^{\al\be}$,\,
where we denote $\,W_{\mu\nu}^{}\!=\! W_{\mu\nu}^a\tau^a/2\,$,\,
with $\tau^a$ being Pauli matrices.

\vspace*{1mm}

The dimension-8 pure gauge operators classified previously\,\cite{Degrande:2013kka}
have the form ${B}_{\mu\nu}^{}X^{\mu\nu}$, where $X^{\mu\nu}$ contains two
covariant derivatives and two field strengths of $W^{a\mu}$ fields
[cf.\ Eqs.(2.7)-(2.14) of Ref.\,\cite{Degrande:2013kka}].
However, ${B}_{\mu\nu}^{}X^{\mu\nu}$ violates CP conservation.

\vspace*{1mm}

In contrast to Ref.\,\cite{Degrande:2013kka},
we construct the following new set of CP-conserving
pure gauge operators of dimension\,8:
\beqs
\label{eq:OG}
\begin{eqnarray}
\mathcal{O}_{G1}^{} \!&=&\!
\widetilde{B}_{\mu\nu}^{} \!\left<D^\mu D_\alpha W^{\alpha\beta} W_{\!\beta}^{\,\,\nu}\right> ,
\label{OG1}
\\
\mathcal{O}_{G2}^{} \!&=&\!
\widetilde{B}_{\mu\nu}^{} \!\left<D_\beta^{} D_\alpha^{} W^{\alpha\mu}  W^{\beta\nu}\right> ,
\label{OG2}
\\
\mathcal{O}_{G3}^{} \!&=&\!
\widetilde{B}_{\mu\nu}^{}\!\left<D_\alpha^{} W^{\alpha\mu}D_\beta W^{\beta\nu}\right> ,
\label{OG3}
\\
\mathcal{O}_{G4}^{} \!&=&\!
\widetilde{B}_{\mu\nu}^{}\!\left<D_\alpha^{} W^{\alpha\beta}D_\beta W^{\mu\nu}\right> ,
\label{OG4}
\\
\mathcal{O}_{G5}^{} \!&=&\!
\widetilde{B}_{\mu\nu}^{}\!\left<D_\alpha^{} W^{\alpha\beta}
D^\mu W_{\!\beta}^{\,\,\nu}\right> ,
\label{OG5}
\end{eqnarray}
\eeqs
where $\LG...\RG$ denotes the trace over Pauli matrices of the
weak gauge group $SU(2)_W^{}$.
Each of these pure gauge operators has at least one covariant derivative contracted
with the gauge field strength on which this covariant derivative is acting.
Hence they can be converted into a sum of dimension-8 operators with Higgs doublets
(which contribute to nTGCs) and dimension-8 operators with fermions,
by using the EOM of the gauge fields\,\cite{Degrande:2013kka,Ellis:2019zex}.
The additional fermion operator in each sum can be eliminated
using the EOM of gauge fields, if we choose the above pure gauge operators
and the operator with Higgs doublets as two independent sets of operators.

\vspace*{1mm}

The pure gauge operators \eqref{eq:OG} were not considered
in Ref.\,\cite{Degrande:2013kka},
which chose a basis of dimension-8 operators
with Higgs doublets (which contribute to nTGCs)
and dimension-8 operators with fermions
(which do not contribute to nTGCs) as two independent sets.
This choice is actually not ideal for studying nTGCs,
because we find that measurements of nTGCs at high-energy colliders
are extremely sensitive to the pure gauge operators \eqref{eq:OG}.
As we show in the present study, the pure gauge operators
generate both nTGC vertices $Z\ga Z^*$ and $Z\ga\ga^*$, which have enhanced
energy power dependences $\propto E^5$,
and can thus be probed with high precision
via the reaction $e^-e^+\!\!\to\!Z\ga$\,.\,

\vspace*{1mm}

Inspecting further the pure gauge operators \eqref{eq:OG}, we find
that the operator $\mathcal{O}_{G3}^{}$ does not contribute to the nTGCs, and hence is
irrelevant to the current study. Moreover, we find that the operators
$\mathcal{O}_{G4}^{}$ and $\mathcal{O}_{G5}^{}$ are not independent
due to the relation
\beqa
\mathcal{O}_{G4}^{} = 2\mathcal{O}_{G5}^{}\,,
\label{eq:G4-G5}
\eeqa
This follows from the Jacobi identity
$\,D_{\!\beta}^{}W^{\mu\nu}\!+\!D^\mu W^\nu_{\,\,\,\beta}
   \!+\!D^\nu W_{\!\beta}^{\,\,\mu}\!=0\,$,  which leads to
\beqa
\widetilde{B}_{\mu\nu}^{}\!\left<D_\alpha^{}W^{\alpha\beta}D_\beta^{} W^{\mu\nu}\right>
= -\widetilde{B}_{\mu\nu}^{}\!
\langle D_\alpha^{}W^{\alpha\beta}(D^\mu W^{\nu}_{\,\,\,\beta}
 +D^\nu W_{\!\beta}^{\,\,\mu})\rangle
= 2 \widetilde{B}_{\mu\nu}^{}\!
\left<D_\alpha^{} W^{\alpha\beta}D^\mu W_{\!\beta}^{\,\,\nu}\right> .~~~
\eeqa
Furthermore, we can derive the relation,
\beqa
\mathcal{O}_{G5}^{} = -\mathcal{O}_{G1}^{}\,,
\eeqa
because integration by parts gives
\beqa
&&\hspace*{-10mm}
\widetilde{B}_{\mu\nu}\!\LG D_\alpha W^{\alpha\beta}D^\mu W^{\,\,\nu}_{\!\beta}\RG
~\To~
\nn
\\[1.5mm]
&&\hspace*{-10mm}
-\widetilde{B}_{\mu\nu}^{}\!
\LG D^\mu D_\alpha W^{\alpha\beta} W_{\!\beta}^{\,\, \nu}\RG
\!-\! \LG D^\mu\widetilde{B}_{\mu\nu}  D_\alpha W^{\alpha\beta} W_{\!\beta}^{\,\,\nu}\RG
= -\widetilde{B}_{\mu\nu}^{}\!
\LG D^\mu D_\alpha W^{\alpha\beta} W_{\!\beta}^{\,\, \nu}\RG ,
\eeqa
where we have made use of the Bianchi identity
$\,D^\mu\widetilde{B}_{\mu\nu}^{}\!\!=\!0$\, \cite{Degrande:2013kka}.\,
Hence, we only need to study the pure gauge operators
$\{\mathcal{O}_{G1}^{},\,\mathcal{O}_{G2}^{}\}$ for the present analysis.

\vspace*{1mm}

For convenience, we define the following two independent combinations of
the pure gauge operators $\{\mathcal{O}_{G1}^{},\,\mathcal{O}_{G2}^{}\}$:
\beqs
\label{eq:OG+-}
\begin{eqnarray}
\label{eq:OG+}
g \mathcal{O}_{G+}^{} \!\!\!&=&\!\!	
\widetilde{B}_{\!\mu\nu}^{}	 W^{a\mu\rho}
( D_\rho^{} D_\lambda^{} W^{a\nu\lambda} \!+\! D^\nu D^\lambda W^{a}_{\lambda\rho}) ,	
\\[1.5mm]
\label{eq:OG-}
g \mathcal{O}_{G-}^{} \!\!\!&=&\!\! 		
\widetilde{B}_{\!\mu\nu}^{} W^{a\mu\rho}
( D_\rho^{} D_\lambda^{} W^{a\nu\lambda} \!-\! D^\nu D^\lambda W^{a}_{\lambda\rho}) .
\end{eqnarray}
\eeqs
Using the EOM, both of these operators can be related to a sum of operators
with additional Higgs doublets and additional fermion bilinears:
\beqs
\label{eq:OG-EOM}
\begin{eqnarray}
\label{eq:OG+QBW-F}
\mathcal{O}_{G+}^{} \!\!\!&=&\!\!
\{\,\ii H^\dagger\widetilde B_{\mu\nu}{W}^{\mu\rho}\!\left[D_\rho,D^\nu\right]\! H
\!+\ii\,2(D_\rho H)^{\!\dagger}\widetilde B_{\mu\nu}{W}^{\mu\rho}\! D^\nu H+\text{h.c.}\}
+\mO_{\!C-}^{}\,,\hspace*{10mm}
\\[1.5mm]
\mathcal{O}_{G-}^{} \!\!\!&=&\!\!
\mathcal{O}_{\!\widetilde{B}W}^{}\!+\mO_{\!C+}^{}\,,
\label{eq:OG-QBW-F}
\end{eqnarray}
\eeqs
where $\mO_{\!C+}^{}$ and $\mO_{\!C-}^{}$
denote the following dimension-8 fermionic contact operators:
\beqs
\label{eq:C+-}
\beqa
\mO_{\!C+}^{} \!\!\!&=&\!\!
\widetilde{B}_{\!\mu\nu}^{}W^{a\mu\rho}\!
\left[D_{\!\rho}^{}(\overline{\psi_{\!L}^{}}T^a\!\gamma^\nu\!\psi_{\!L}^{})
+D^\nu(\overline{\psi_{\!L}^{}}T^a\!\gamma_\rho^{}\psi_{\!L}^{})\right]\!,
\label{C+}
\\[1.5mm]
\mO_{\!C-}^{} \!\!\!&=&\!\!
\widetilde{B}_{\!\mu\nu}^{}W^{a\mu\rho}\!
\left[D_{\!\rho}^{}(\overline{\psi_{\!L}^{}}T^a\!\gamma^\nu\!\psi_{\!L}^{})
-D^\nu(\overline{\psi_{\!L}^{}}T^a\!\gamma_\rho^{}\psi_{\!L}^{})\right]\!.
\label{C-}
\eeqa
\eeqs
Inspecting the terms inside the $\{\,\cdots\,\}$ on the
right-hand-side (RHS) of Eq.\eqref{eq:OG+QBW-F},
we note that the first Higgs operator contains the commutator
$\left[D_\rho,\,D^\nu\right]$,\,
which can be replaced by a gauge field strength tensor.
This part therefore has the same structure
as the conventional dimension-6 pure gauge operators
once the Higgs doublet is set to its VEV
$\left<H\right>\!=(0,\,v/\!\sqrt{2})^T$,
and thus makes no contribution to the nTGCs.
The second Higgs operator inside the
$\{\,\cdots\,\}$ of Eq.\eqref{eq:OG+QBW-F}
has at least 4 gauge fields in each vertex after setting
the Higgs doublet to its VEV $\LG H\RG$,\,
so it is also irrelevant to the nTGCs.
Since the EOMs apply to on-shell fields in any physical process,
we conclude that the contribution of the operator $\mathcal{O}_{G+}^{}$
to the reaction $\,e^-e^+\!\!\to\! Z\ga$\,
is equivalent to that of the fermionic operator
$\mO_{C-}^{}$ on the RHS of Eq.\eqref{eq:OG+QBW-F},
which contributes to the $eeZ\ga$ contact vertex only.

\vspace*{1mm}

By power counting on both sides of Eq.\eqref{eq:OG+QBW-F},
we find that the amplitude of this reaction has the leading high-energy behavior
$\,\mathcal{T}[e^-e^+\!\!\to\! Z\ga]\!\propto\! E^4$.
We note from the exact helicity amplitudes \eqref{eq:T8-T}-\eqref{eq:T8-L}
presented in Appendix\,\ref{app:A} that
the contributions of $\mathcal{O}_{G+}^{}$
exhibit the following high-energy behaviors:
\beqs
\beqa
\mathcal{T}^{ss'\!,T}_{(8)}\!(\pm\pm)
&\!\!=\!\!& O\!\(\!\frac{E^4}{\cut^4}\!\)\!,
~~~~~~
\mathcal{T}^{ss',T}_{(8)}\!(\pm\mp) \,=\, 0\,,
\hspace*{14mm}
\\
\mathcal{T}^{ss'\!,L}_{(8)}\!(0\pm)
&\!\!=\!\!& O\!\(\!\frac{\,E^3M_Z^{}}{\cut^4}\!\)\!,
\eeqa
\eeqs
where the amplitudes $\mathcal{T}^{ss'\!,T}_{(8)}\!(\lambda\lambda')$
denote the final state $Z_T^{}(\lambda)\gamma_T^{}(\lambda')$
with helicity combinations
$\,\lambda\lambda'=\pm\pm,\pm\mp\,$,
and the amplitudes
$\mathcal{T}^{ss'\!,L}_{(8)}\!(\lambda\lambda')$
denote the final state $Z_L^{}(\lambda)\gamma_T^{}(\lambda')$
with helicity combinations
$\,\lambda\lambda'\!=\! 0\pm\,$.
Hence, this reaction can provide an extremely sensitive probe
of the new physics scale $\Lambda$ associated with the pure gauge operator
$\mathcal{O}_{G+}^{}$ and thus the nTGCs when $\,E^2\!\gg\! M_Z^2$\,.
It also provides the same probe for the fermionic operator
$\mO_{C-}^{}$ since it is not independent according to Eq.\eqref{eq:OG+QBW-F}.

\vspace*{1mm}

On the other hand, inspecting Eq.\eqref{eq:OG-QBW-F},
we first recall that the Higgs operator $\mathcal{O}_{\BT W}^{}$ makes a
contribution to the amplitude
$\,\mathcal{T}[e^-e^+\!\!\to\! Z\ga]\!\propto\! E^3v\,$
as we showed before~\cite{Ellis:2019zex}.\,
Regarding the fermion-bilinear operator $\mO_{C+}^{}$
on the RHS of Eq.\eqref{eq:OG-QBW-F},
it makes a contribution
$\,\mathcal{T}[e^-e^+\!\!\!\to\! Z\ga]\propto E^3M_Z^{}$\,
for the on-shell $Z\ga$ final state.
Hence, the pure gauge operator $\mathcal{O}_{G-}^{}$
also makes a contribution
$\,\mathcal{T}[e^-e^+\!\!\to Z\ga]\propto E^3v\,$
which is similar to that of the Higgs operator $\mathcal{O}_{\BT W}^{}$,\, and
they both vanish when $\,v\!\to\! 0\,$,\,
as we noted in Ref.\,\cite{Ellis:2019zex}.
Furthermore, from the exact helicity amplitudes
\eqref{eq:T8-TT-OG-OBW-OC+}-\eqref{eq:T8-LT-OG-OBW-OC+}
given in Appendix\,\ref{app:A}, we find that
the contributions of the operators
$(\OGM,\,\OBW,\,\OCP)$  display the following high-energy behaviors:
\beqs
\beqa
\mathcal{T}^{ss'\!,T}_{(8)}\!(\pm\pm)
&\!\!=\!\!& O\!\(\!\frac{E^2M_Z^2}{\cut^4}\!\)\!,
~~~~~~
\mathcal{T}^{ss',T}_{(8)}\!(\pm\mp) \,=\, 0\,,
\hspace*{14mm}
\\
\mathcal{T}^{ss'\!,L}_{(8)}\!(0\pm)
&\!\!=\!\!& O\!\(\!\frac{\,E^3M_Z^{}}{\cut^4}\!\)\!,
\eeqa
\eeqs
These observations suggest that the operators
$(\OGM,\,\OBW,\,\OCP)$
are less sensitive to the nTGCs than the operator
$\mathcal{O}_{G+}^{}$ at high-energy $e^+e^-$ colliders,
which we highlight in the following analysis.

\vspace*{1mm}

In summary, Eq.\eqref{eq:OG+QBW-F} shows that $\mO_{C-}^{}$
is equivalent to $\mO_{G+}^{}$ for the reaction $e^-e^+\!\!\to Z\ga$,\,
and thus can be dropped in the current analysis;
while Eq.\eqref{eq:OG-QBW-F} proves that $\mO_{C+}^{}$ is not independent
from $\mO_{G+}^{}$ and $\mO_{\widetilde{B}W}^{}$.
Hence, we are justified to choose $\mO_{G+}^{}$, $\mO_{G-}^{}$,
and $\mO_{\widetilde{B}W}^{}$
as the 3 remaining independent dimension-8 operators
contributing to the $e^-e^+\!\!\to Z\ga$ process.

\vspace*{1.5mm}
\section{\large\hspace{-2mm}Analyzing \boldmath{$Z\ga$}
         Production with Hadronic \boldmath{$Z$} Decays}
\label{sec:3}

In this Section, we analyze systematically the $Z\ga$ production
with hadronic $Z$ decays.
We first present the relevant cross sections and
angular distribution in Section\,\ref{sec:3.1}.
Then, we discuss the effects of including the dijet angular resolution
at $e^+e^-$ colliders in Section\,\ref{sec:3.2}.

\vspace*{1mm}
\subsection{\hspace{-2mm}Cross Sections and Angular Distributions}
\vspace*{1.5mm}
\label{sec:3.1}

The dimension-8 effective Lagrangian can be written as
\vspace*{-2mm}
\beqa
\Delta\mathcal{L}(\text{dim-8})
\,=\, \sum_{j}^{}\frac{\tilde{c}_j}{\,\tilde{\cut}^4\,}\mathcal{O}_j^{}
\,=\, \sum_{j}^{}
\frac{\,\text{sign}(\tilde{c}_j^{})\,}{\,\cut_j^4\,}\mathcal{O}_j^{} \,,
\label{cj}
\eeqa
where each dimensionless coefficient $\,\tilde{c}_j^{}$ may be ${\cal O}(1)$
and has possible signs $\,\text{sign}(\tilde{c}_j^{})=\pm$\,.\,
For each dimension-8 operator $\mathcal{O}_j^{}$\,,
we define the corresponding effective UV cutoff scale
$\,\cut_j^{} \equiv \tilde{\cut}/|\tilde{c}_j^{}|^{1/4}\,$.\,

\vspace*{1mm}

We first derive the Feynman rules for the nTGC vertices
as generated by the pure gauge operator $\mathcal{O}_{G+}^{}$.\,
We find that $\mathcal{O}_{G+}^{}$ contributes to both
the $Z\gamma Z^*$ and $Z\gamma \gamma^*$ vertices in the following forms:
\beqs
\label{eq:Vertex-G+}
\begin{eqnarray}
\label{eq:VertexZAZ-G+}
\Gamma_{Z\gamma Z^*+}^{\alpha\beta\mu}(q_1^{},q_2^{},q_3^{})
\!\!&=&\!\!
-\text{sign}(\tilde{c}_{G+}^{})\frac{\,v(q_3^2\!-\!M_Z^2)\,}{\,M_Z^{}\Lambda^4\,}
\(q_3^2\,q_{2\nu}^{}\epsilon^{\alpha\beta\mu\nu}\!
+2q_2^{\alpha} q_{3\nu}^{}q_{2\sigma}^{}\epsilon^{\beta\mu\nu\sigma}\)\!,
\hspace*{16mm}
\\[2mm]
\label{eq:VertexZAA-G+}
\Gamma_{Z\gamma \gamma^*+}^{\alpha\beta\mu}(q_1^{},q_2^{},q_3^{})
\!\!&=&\!\!
-\text{sign}(\tilde{c}_{G+}^{})\frac{\,s_W^{}v\, q_3^2\,}{\,c_W^{}M_Z^{}\Lambda^4\,}
\(q_3^2\,q_{2\nu}^{}\epsilon^{\alpha\beta\mu\nu}\!
+ 2q_2^{\alpha}q_{3\nu}^{} q_{2\sigma}^{}\epsilon^{\beta\mu\nu\sigma}\)\!,
\hspace*{16mm}
\end{eqnarray}
\eeqs
where $\,\text{sign}(\tilde{c}_{G+}^{})=\pm$\,
denotes the sign of the coefficient
of the operator $\mathcal{O}_{G+}^{}$.\,
From the above,
we observe that the $Z\gamma Z^*$ and $Z\gamma\gamma^*$ vertices
both have strong energy dependences $\varpropto\! E^5$,\,
and thus can be probed sensitively at high energies
via the reaction $\,e^-e^+\!\!\to\!Z\gamma$\,.

\vspace*{1mm}

We note that the contributions
from the initial-state right-handed fermions to
the sum of the amplitudes
$\,\mathcal{T}[f\bar f\!\!\to\! Z^*\!\!\to\!\! Z\gamma]$\,
and $\,\mathcal{T}[f\bar f\!\!\to\! \gamma^*\!\!\to\!\! Z\gamma]\,$ vanish.
Denoting these two amplitudes by
$\,\mathcal T_{Z^*}^R\,$ and $\,\mathcal T_{\gamma^*}^R$\, respectively,
we observe this cancellation by computing their ratio:
\begin{equation}
\frac{\,\mathcal T_{Z^*}^R\,}{\mathcal T_{\gamma^*}^R}
=\frac{\,\Gamma_{f\bar f Z}^R\,}{\,\Gamma_{f\bar f \gamma}^R\,}
\!\times\! \frac{q_3^2}{\,q_3^2\!-\!M_Z^2\,} \!\times\!
\frac{\Gamma_{Z\gamma Z^*+}^{\alpha\beta\mu}}
{\Gamma_{Z\gamma \gamma^*+}^{\alpha\beta\mu}}
=-\frac{s_W^{}}{c_W^{}} \!\times\!\frac{q_3^2}{q_3^2\!-\!M_Z^2}
\!\times\!\frac{\,c_W^{}(q_3^2\!-\!M_Z^2)\,}{s_W^{}q_3^2}=-1\,,
\end{equation}
which leads to
$\,\mathcal T_{Z^*}^R + \mathcal T_{\gamma^*}^R = 0$\,.
This cancellation can be understood from the observation that
the contributions of $\mathcal{O}_{G+}^{}$ and $\mathcal{O}_{C-}^{}$
to the reaction $\,e^-e^+\!\!\to\! Z\ga$\, are equivalent,
as explained below Eq.\eqref{eq:C+-}, and from the fact that
the operator $\mathcal{O}_{C-}^{}$ contains only
the left-handed fermions as shown in Eq.(\ref{C-}).

\begin{figure}[t]
\begin{center}
\includegraphics[width=11cm]{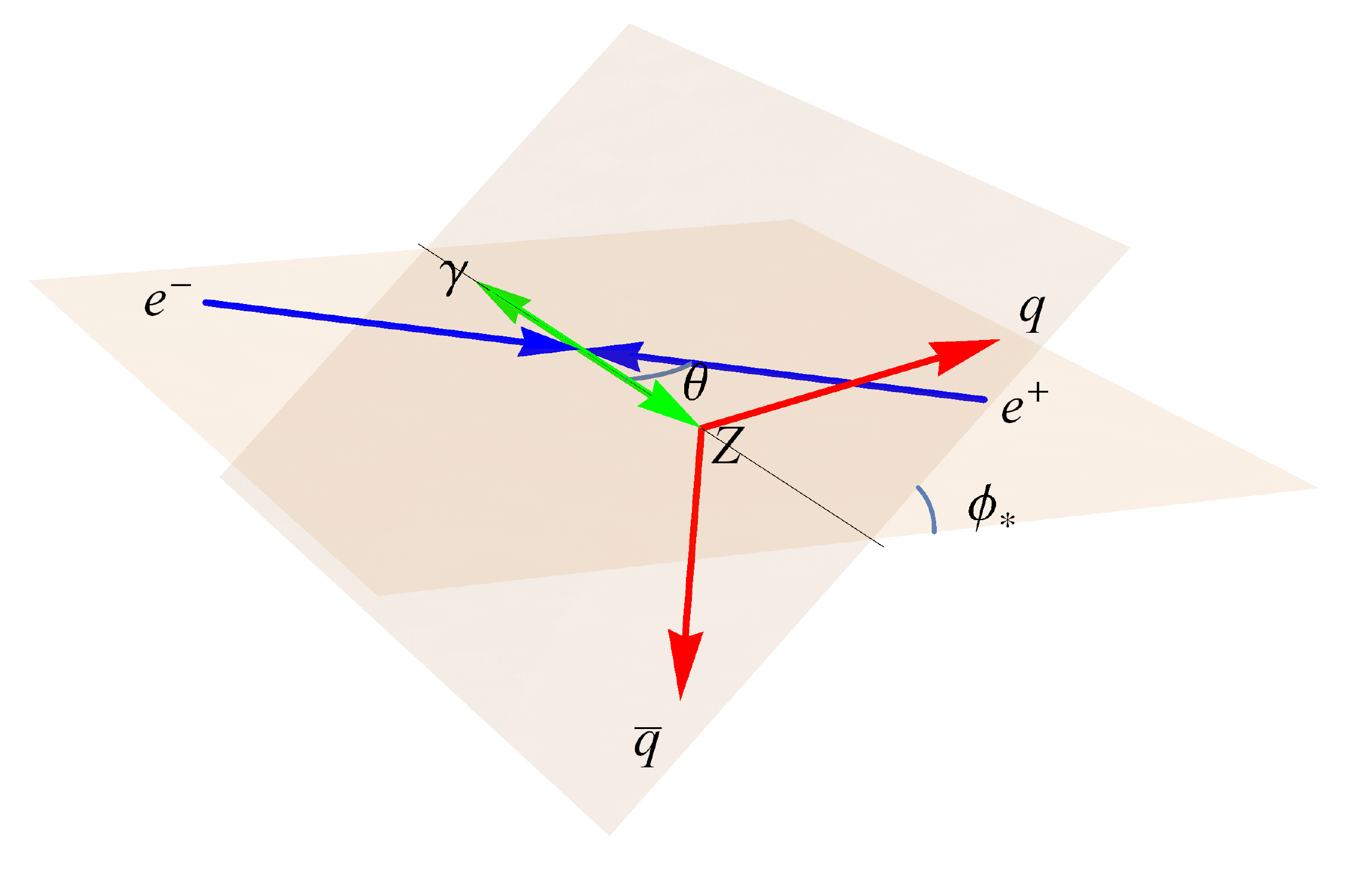}
\vspace*{-6mm}
\caption{\small Kinematical structure of the reaction
		$\,e^+e^-\!\!\to Z\ga\,$ followed by the hadronic decays $\,Z\!\to q\bar{q}\,$,\,
        in the $\,e^+e^-$ collision frame.
		}
\label{figtheta}
\label{fig:1}
\vspace*{-1mm}
\end{center}
\end{figure}

\vspace*{1mm}

\begin{figure*}[t]
	\centering
	\vspace*{-3mm}
	\hspace*{1mm}
	\includegraphics[width=4cm]{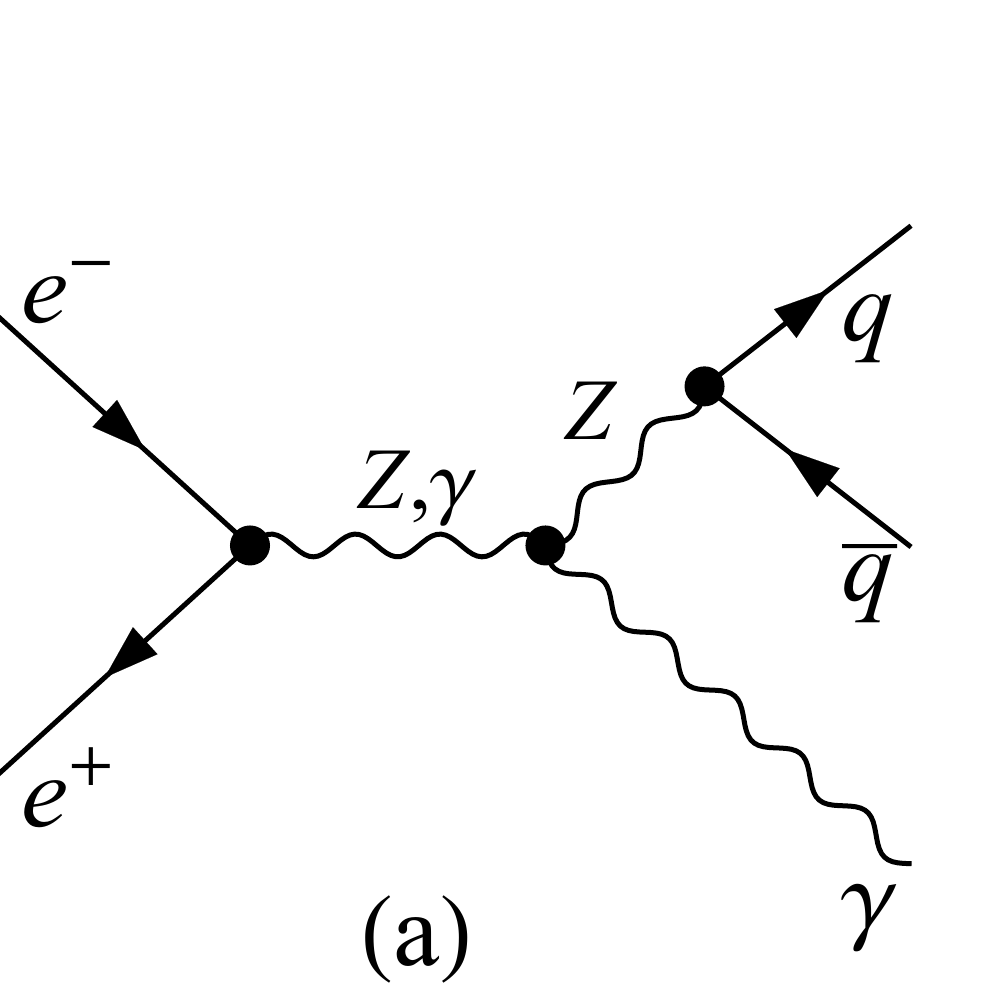}
	\includegraphics[width=4cm]{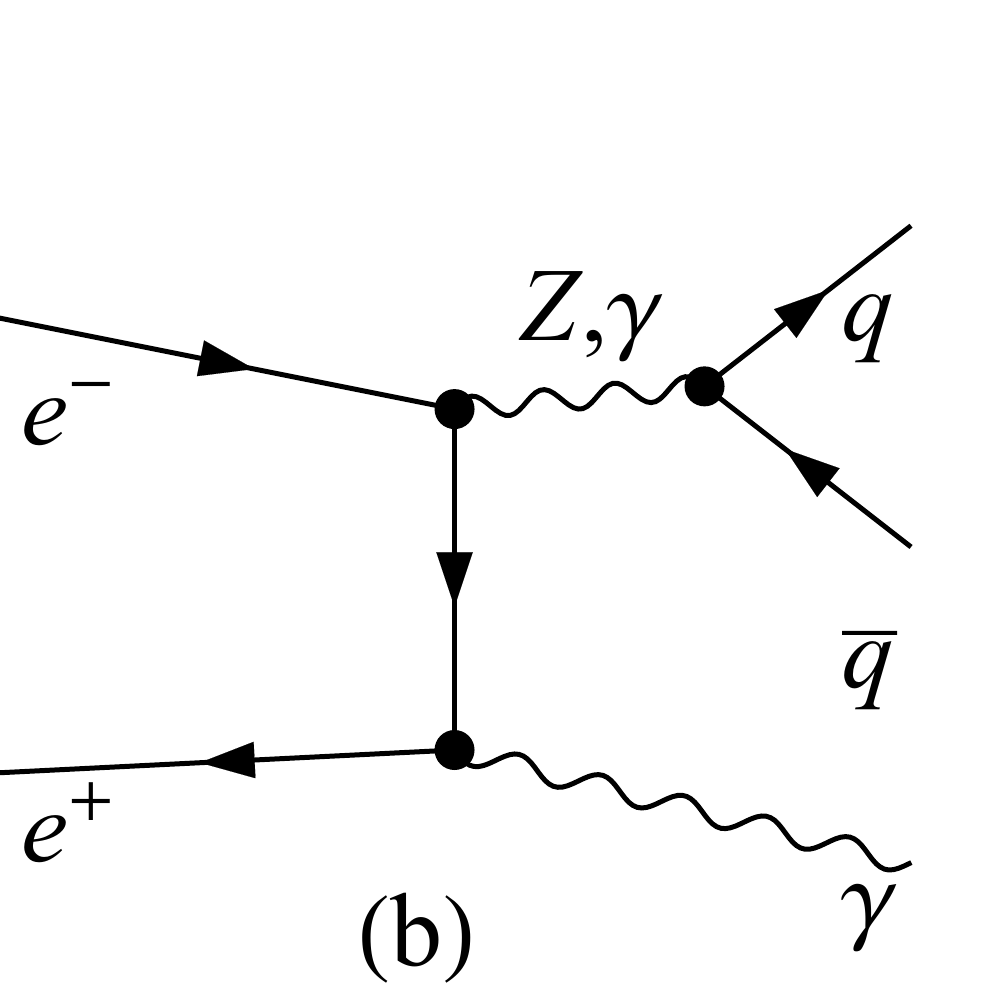}
	\includegraphics[width=4cm]{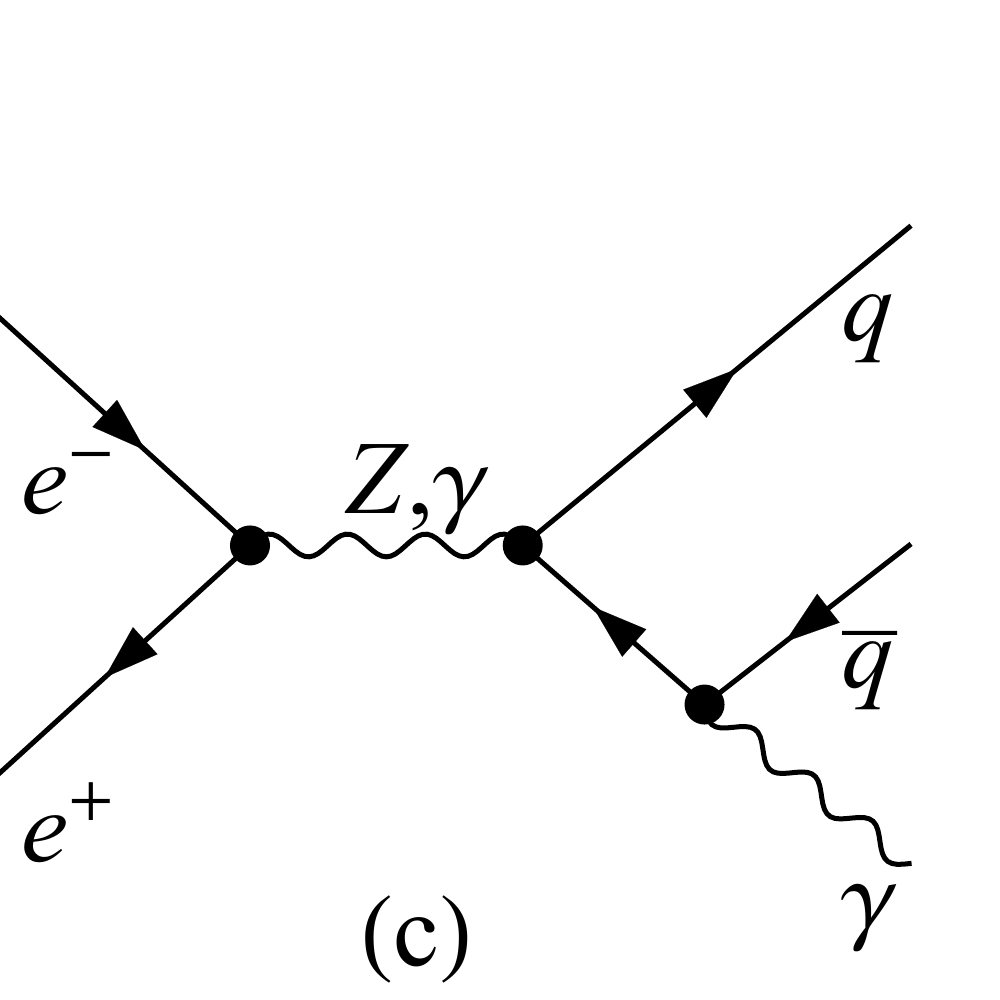}
    \includegraphics[width=4cm]{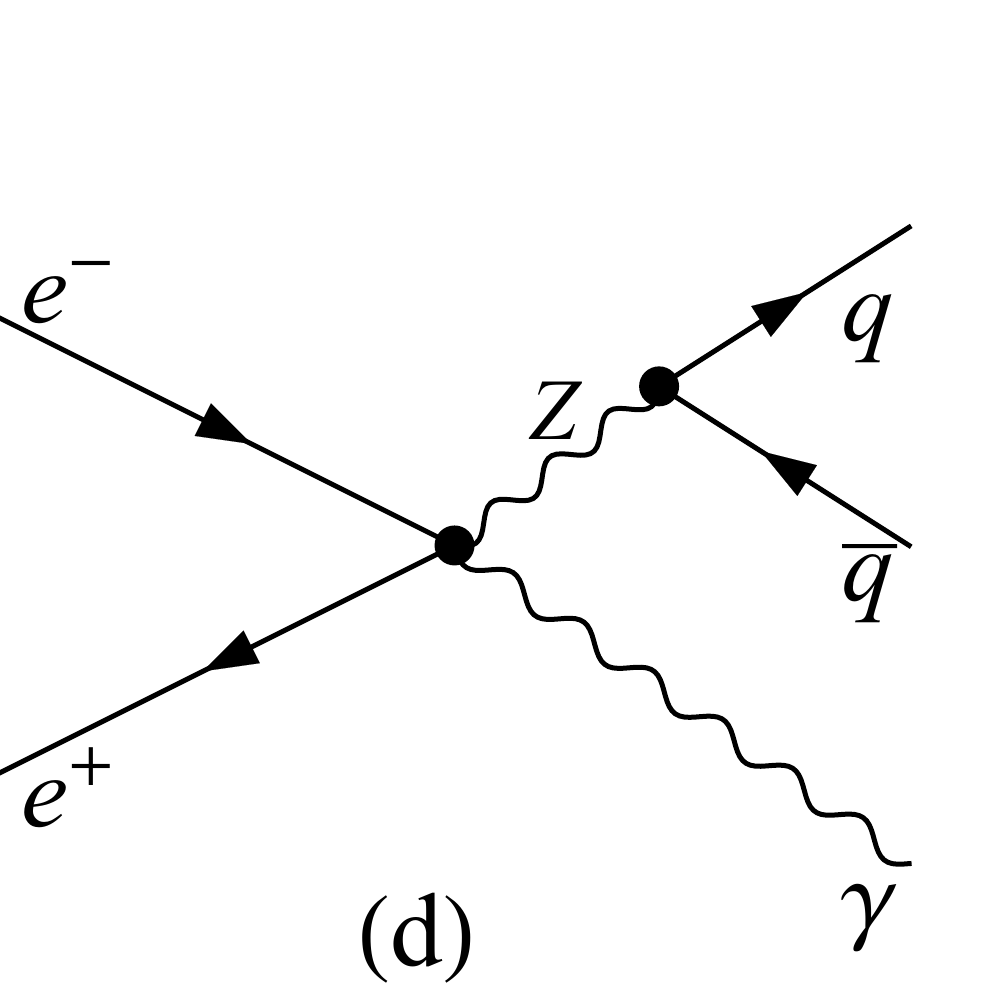}
\vspace*{-2mm}
\caption{\small Feynman diagrams that contribute to the reaction
$e^-e^+\!\!\to\! \gamma\,q\bar{q}$\,.
Type\,(a) provides the signals via the nTGC vertex $Z^*Z\gamma$
or $\gamma^*Z\gamma$, arising from
the relevant dimension-8 operator, while types\,(b) and (c) give the SM backgrounds.
Diagram\,(b) together with a similar $u$-channel diagram for
$\,e^-e^+\!\!\to\! Z\gamma\!\to\!\gamma q \bar q$\, 
presents an irreducible background.
Diagram\,(c) has the $s$-channel gauge-boson exchange and final-state
$\gamma$ radiation, providing a reducible background.
Diagram\,(d) arises from the contact vertex $eeZ\gamma$ which is
generated by the relevant dimension-8 fermion-bilinear operator.
}
\vspace*{2mm}
\label{feyndiag}
\label{SB}
\label{fig:Fdiag}
\label{fig:2}
\end{figure*}

Fig.\,\ref{fig:Fdiag} shows four types of Feynman diagrams that
contribute to the reaction  $\,e^-e^+\!\!\to\! q\bar{q}\,\gamma$\,.
Diagram\,(a) arises from the nTGC contributions of the dimension-8 operators
such as $\mathcal{O}_{G+}^{}$ or $\mathcal{O}_{G-}^{}$\,.
Diagrams (b) and (c) give the SM background contributions, where
diagram\,(c) is a reducible background that can be suppressed effectively by the
invariant-mass cut for the on-shell $Z$ boson.
Finally, diagram\,(d) denotes the possible $eeZA$ contact contribution
from dimension-8 fermion-bilinear operators
such as $\mathcal{O}_{C+}^{}$.
In view of the reducibility of the background from the diagram (c),
for our analytical analysis
we first calculate the on-shell $Z\gamma$ production
given by diagrams (a) and (b),
as we did before when analyzing the nTGC
operator $\mathcal{O}_{\widetilde{B}W}^{}$ \cite{Ellis:2019zex}.
Also, we take the conventional approach of treating each operator individually,
and consider later the contribution of the dimension-8 fermion-bilinear
operators via diagram\,(d).

\vspace*{1mm}

Applying the lower angular cut $\,\sin\theta>\sin\delta\,$ (with $\delta\ll 1$),
we find the following total cross section for $e^+ e^- \!\to Z\gamma$\,,\,
including the contributions of $\mathcal{O}_{G+}^{}$ and
summing over the final state $Z$ polarizations:
\vspace*{-2mm}
\begin{eqnarray}
\label{Zgamma-}
\sigma_{+}^{}(Z\gamma)
\!\!&=&\!\!
\frac{\,e^4(c_L^2\!+\!c_R^2)\!\!\left[-(s\!-\!M^2_Z)^2\!-\!2(s^2\!+\!M_Z^4)
	\ln\sin\!\frac{\delta }{2}\,\right]\,}
{\,8\pi s_W^2c_W^2(s\!-\!M^2_Z)s^2\,}
\hspace*{15mm}
\nn\\[1mm]
\!\!&&\!\!
+~\text{sign}(\tilde{c}_{G+}^{})\frac{\,e^2 c_Lx_{L}M_Z^2\!\(s\!-\!M_Z^2\)\!\,}
{\,4\pi s_W^{}c_W^{}\, s\,}\frac{1}{\,\cut^4\,}
\label{eq:ZA-G+}
\\[1mm]
\!\!&&\!\!
+\,\frac{\,x_L^2(s+M_Z^2)\!\(s\!-\!M^2_Z\)^{\!3}\,}
{\,48\pi\,s\,}\frac{1}{\,\cut^8\,} + O(\delta)\,,
\nn
\end{eqnarray}
%
where $\,\tilde{c}_+^{}$
is the coefficient of the dimension-8 operator\footnote{%
The same formula holds for the operator
$\mathcal{O}_{C-}^{}$ because its contribution to the reaction
$\,e^+ e^- \!\!\to\! Z\gamma$\,
is equivalent to that of $\,\mathcal{O}_{G+}^{}$,
as we explained below Eq.\eqref{eq:C+-}.}
$\mathcal{O}_{G+}^{}$ and the factor $\,x_L^{}\!=\fr{1}{2}\,$
is from the weak coupling of $\,W_3^{}$\,.
In Eq.\eqref{eq:ZA-G+}, the coefficients
$\,(c_L^{},\,c_R^{}) = (-\fr{1}{2}\!+\!s_W^2,\, s_W^2)$
arise from the (left, right)-handed gauge couplings of electrons to
$Z$ boson.
The differential cross section depends on the three kinematical angles
$(\theta,\,\theta_*^{},\,\phi_*^{})$,
where $\,\theta\,$ is the polar scattering angle describing the direction of
the outgoing $Z$ relative to the initial state $e^-$ (cf.\ Fig.\ref{fig:1}),
$\,\theta_*^{}$ denotes the angle between the
direction opposite to the final-state $\gamma$
and the final-state $q$ direction in the $Z$ rest frame,
and $\,\phi_*^{}\,$ is the angle between the scattering plane
and the decay plane of $Z$ in the $e^+e^-$ center-of-mass frame
(cf.\ Fig.\ref{fig:1}).

\vspace*{1mm}

We define the normalized angular distribution functions as follows:
\begin{eqnarray}
f_\xi^j \, = \, \frac{1}{\sigma_{\!j}^{}}\frac{\di\sigma_{\!j}^{}}{\,\di\xi\,}\,,
\end{eqnarray}
where the angles $\,\xi \in (\theta,\,\theta_*^{},\,\phi_*^{})$,\,
and the cross sections
$\,\sigma_j^{}$ ($j=0,1,2$) represent the SM contribution ($\sigma_0^{}$),
the ${\cal O}(\cut^{-4})$ contribution ($\sigma_1^{}$), and the
${\cal O}(\cut^{-8})$ contribution ($\sigma_2^{}$),\, respectively.
For the normalized azimuthal angular distribution functions
$\,f_{\phi_*}^j\,$,\, we derive the following:
{\small
\beqs
\label{eq:f-phi*}
\begin{eqnarray}
\label{eq:f0-phi*}
	\hspace*{-12mm}
f_{\phi_*^{}}^{0} \!\!&=&\!\!
	\frac{1}{2\pi} +\frac{3\pi^2(c_L^2\!-\!c_R^2)^2M_Z^{}\sqrt{s}\,(s\!+\!M_Z^2)\cos\!\phi_*^{}\!
		-8(c_L^2\!+\!c_R^2)^2M_Z^2\,s \cos\!2\phi_*^{}\,}
	{\,16 \pi(c_L^2\!+\!c_R^2)^2\!\left[(s\!-\!M_Z^2)^2\!+2(s^2\!+\!M_Z^4)
    \ln\sin\!\frac{\delta}{2}
		\right]\,}+O(\delta) ,\hspace*{5mm}
	\hspace*{4mm}
\\[2mm]
\label{eq:f1-phi*}
	\hspace*{-8mm}
f_{\phi_*^{}+}^{1} \!\!&=&\!\!
\frac{1}{2\pi} -\frac{\,3\pi (q_L^2\!-\!q_R^2)(M_Z^2+5 s)
\cos\phi_*\,}{256(q_L^2\!+\!q_R^2)M_Z\sqrt s}+
\frac{\,s\cos2\phi_*\,}{8\pi M_Z^2} ,
\label{f1}
\\[2mm]
\label{eq:f2-phi*}
\hspace*{-8mm}
f_{\phi_*^{}+}^{2} \!\!&=&\!\!
	\frac{1}{2\pi} - \frac{\,9\pi (q_L^2\!-\!q_R^2)M_Z^{}\sqrt{s}\cos\!\phi_*^{}\,}
	{\,128(q_L^2\!+\!q_R^2)(s\!+\!M_Z^2)\,}  ,
	\label{eq:phi-}
	\end{eqnarray}
\eeqs
}
\hspace*{-3mm}
where the subscripts ``$+$'' in the notations
$\,f_{\phi_*^{}+}^{1,2}\,$
denote the contributions by the dimension-8 operator $\OGP$.
The coefficients $\,(c_L^{},\,c_R^{})$\,
in the above formulae have already been
defined below Eq.\eqref{eq:ZA-G+},
and the coefficients
\,$(q_L^{},\, q_R^{})\!=(T_3^{}\!-\!Qs_W^2,\, -Qs_W^2)$\, correspond to
the gauge couplings of the (left, right)-handed quarks to $Z$ boson,
with $Q$ being the electric charge of the quark
and $T_3^{}\!=\pm\fr{1}{2}$\,.

\vspace*{1mm}

\begin{figure}[t]
\includegraphics[width=7.7cm,height=5.5cm]{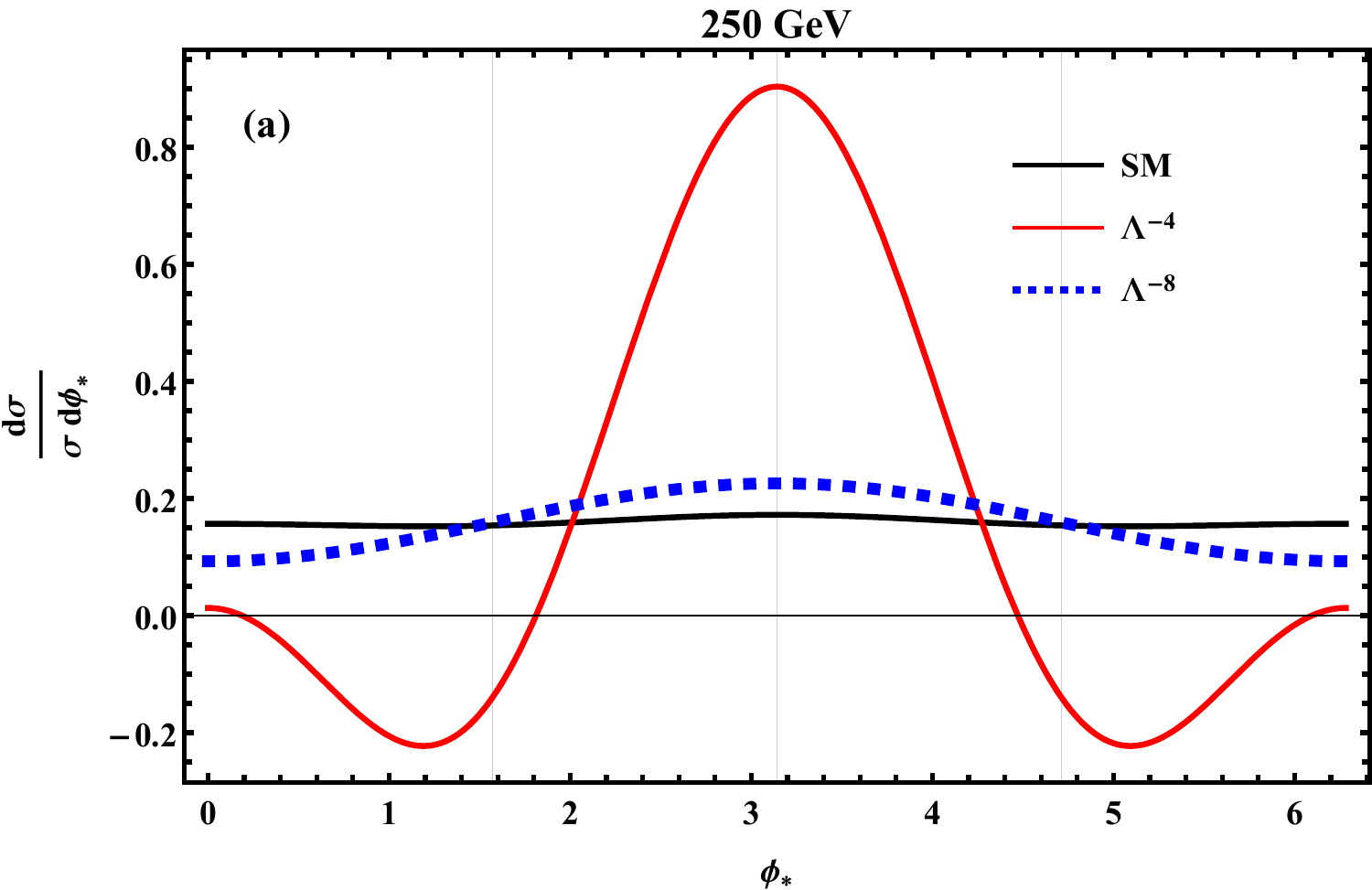}
\includegraphics[width=7.7cm,height=5.5cm]{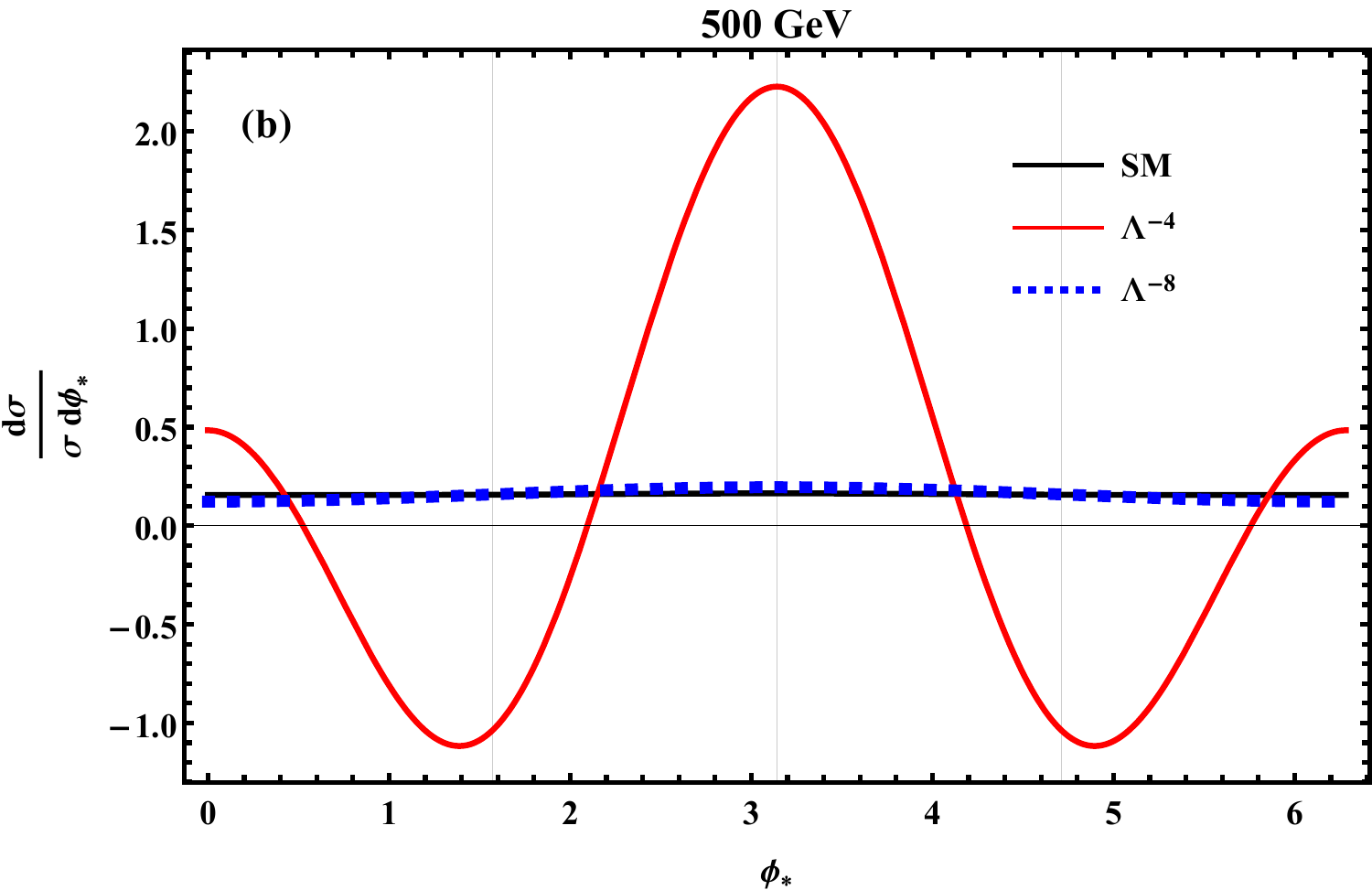}
\\[3mm]
\includegraphics[width=7.7cm,height=5.5cm]{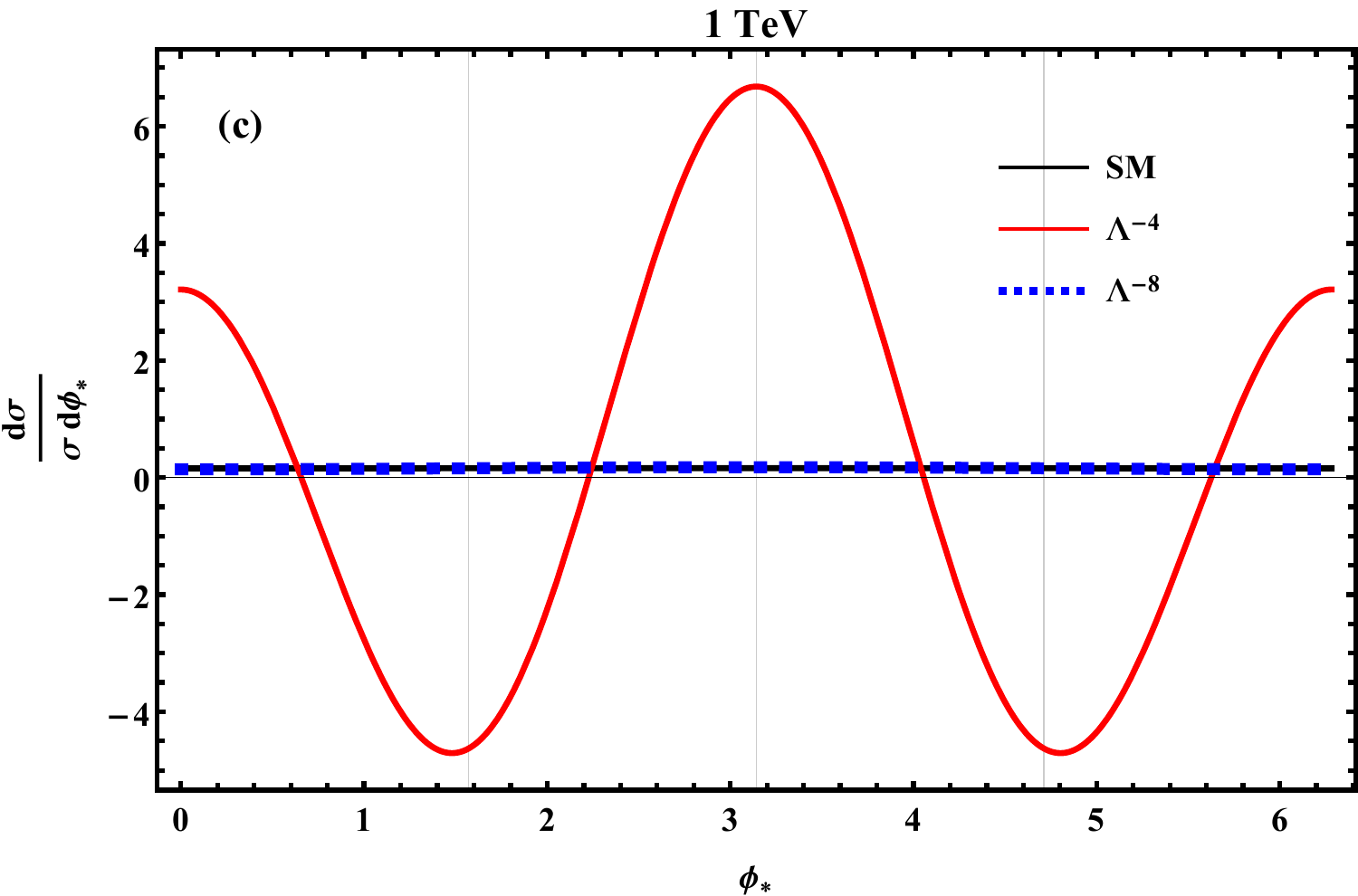}
\includegraphics[width=7.7cm,height=5.5cm]{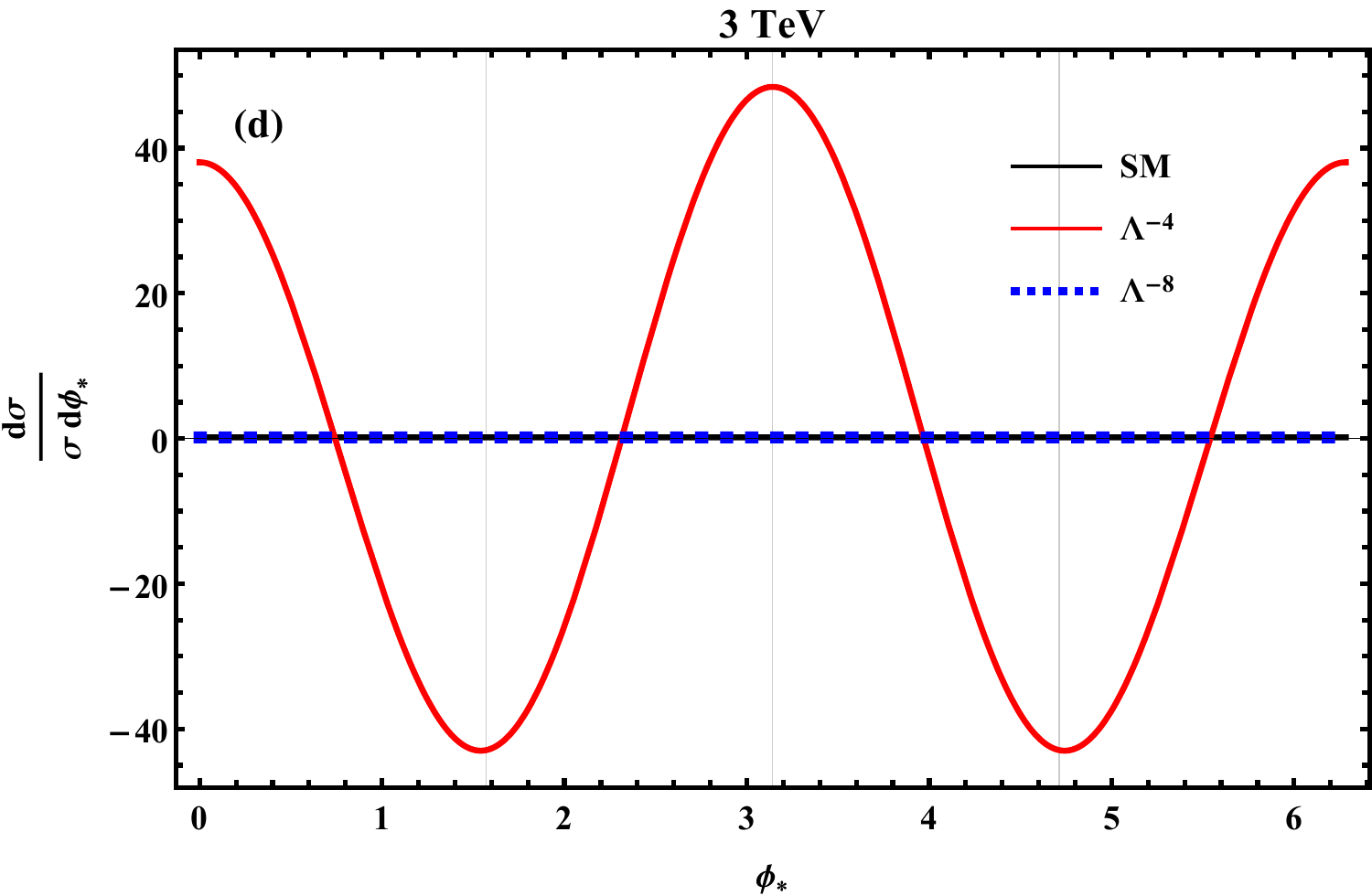}
\vspace*{-2mm}
\caption{\small{%
Normalized angular distributions in the azimuthal angle
$\phi_*^{}$ for $e^-e^+\!\!\to Z\ga$ followed by $Z\!\to d\bar d$ decays,
as generated by $\,\mO_{G+}^{}$\!
at the collision energies}
$\sqrt{s}=(0.25,\, 0.5,\, 1,\, 3)$\,TeV,
{respectively.
In each plot, the black, red, and blue curves denote the contributions from
the SM, the interference term of $\,{\cal O}(\cut^{-4})$, and the quadratic term
of $\,{\cal O}(\cut^{-8})$, respectively,
where we note that the blue and black curves almost coincide.
We have imposed a basic cut on the polar scattering angle, $\sin\theta>\sin\delta$,
with $\delta=0.2$ for illustration.}}
\label{figphi}
\label{fig:44}
\label{fig:3F}
\vspace*{3mm}
\end{figure}

We present these $\phi_*^{}$ distributions in Fig.\,\ref{fig:3F}.
We see that the interference contributions of $O(\cut^{-4})$ (red curves)
are mainly $\,\propto\! \cos 2\phi_*^{}$,\, except in the case of
the relatively lower collider energy $\sqrt{s\,}\!=\!250$\,GeV as shown in plot\,(a),
which has some visible deviations from $\cos 2\phi_*^{}$.\,
This is because the $f^1_{\phi_*^{}+}$ distribution \eqref{eq:f1-phi*}
is dominated by the $\cos 2\phi_*^{}$ term, which has a significant
energy enhancement factor $\propto\! s/\!M_Z^2\,$,
whereas the SM contributions (black curves) are nearly flat.
The same feature also holds for the squared dimension-8 contributions
of $O(\cut^{-8})$, depicted as blue dashed curves, which are quite flat
except for the case of $\sqrt{s\,}\!=\!250$\,GeV,
as also seen in plot\,(a).\footnote{For the following analysis of sensitivities
to probing the new physics scale $\cut$ in Sec.\,\ref{sec:4}, we retain the
$\cut$-dependent contributions only up to ${\cal O}(\cut^{-4})$ and drop
systematically  the ${\cal O}(\cut^{-8})$ contributions, since the latter
are generally negligible, in view of the severe suppression of ${\cal O}(\cut^{-8})$
for the large values of $\cut$ that are probed via the hadronic decay channels
$\,Z\!\to\!q\bar{q}$\, in the present study.}
This is because both the $f_{\phi_*^{}}^{0}$ and $f_{\phi_*^{}+}^{2}$
distributions [Eqs.\eqref{eq:f0-phi*} and \eqref{eq:f2-phi*}]
are dominated by the constant term $\fr{1}{\,2\pi\,}$
and the $\phi_*^{}$-dependent terms are suppressed by $M_Z^{}/\!\sqrt{s\,}$
when $s\gg M_Z^2$\,. So it is expected that only the case of the
lower collider energy $\sqrt{s\,}\!=\!250$\,GeV in plot\,(a) shows a visible
$\,\cos\phi_*^{}\,$ dependence for the $O(\cut^{-8})$ contributions, while in all
the higher-energy cases with $\sqrt{s\,}\!\gtrsim\!500$\,GeV
the SM distributions (black curves) and the squared contributions (blue dashed curves)
are essentially flat.

\vspace*{1mm}

Next, we find that the operator $\mathcal{O}_{G-}^{}$ does not contribute to the
$Z\gamma Z^*$ coupling for on-shell gauge bosons $Z$ and $\gamma$\,.
Computing the $\mathcal{O}_{G-}^{}$ contribution to
the nTGC coupling $Z\gamma\gamma^*$,
we derive the following Feynman vertex:
\begin{eqnarray}
\label{eq:vertex-OG-}
\ii\, \Gamma^{\al\beta\mu}_{Z\gamma \gamma^*-}
({q}_1^{}, {q}_2^{}, {q}_3^{})
\,=\,
-\text{sign}(\tilde{c}_-^{})
\frac{\, s_W^{}v M_Z^{}\,}{\,c_W^{}\Lambda^4\,}
\epsilon^{\alpha\beta\mu\nu} q_{2\nu}^{}{q}_3^2 \, ,
\end{eqnarray}
for on-shell gauge bosons $Z$ and $\gamma$ plus a virtual photon $\gamma^*$.
In the above formula, $\,\text{sign}(c_-^{})=\pm$\,
denotes the sign of the coefficient of the operator $\mathcal{O}_{G-}^{}$.\,

\vspace*{1mm}

For comparison, the Higgs-related dimension-8 operator
$\mathcal{O}_{\widetilde{B}W}^{}$ yields
the following effective $Z\gamma Z^*$ coupling
in momentum space\,\cite{Ellis:2019zex}:
\begin{eqnarray}
\label{eq:vertex-OBW}
\ii\, \Gamma^{\al\beta\mu}_{Z\gamma Z^*\!(\widetilde{B}W)}
({q}_1^{}, {q}_2^{}, {q}_3^{})
\,=\,
\text{sign}(\tilde{c}_{\widetilde{B}W}^{})
\frac{\,v M_Z^{} ({q}_3^2\!-\!M_Z^2)\,}{\,\Lambda^4\,}
\epsilon^{\alpha\beta\mu\nu} q_{2\nu}^{}  \,.
\end{eqnarray}
This operator also makes no contribution to the $Z\gamma\gamma^*$ coupling
for on-shell gauge bosons $Z$ and $\gamma$\,,\,
as we noted before in Ref.\,\cite{Ellis:2019zex}.

\vspace*{1mm}

\vspace*{1mm}

The fermion-bilinear operator $\mathcal{O}_{C+}^{}$
contributes the following effective contact vertex $f\bar{f}Z\gamma$
when the four external fields are on-shell:
\begin{eqnarray}
\label{eq:vertex-OC+}
\ii\, \Gamma^{\al\beta}_{Z\gamma f\bar f}
({q}_1^{}, {q}_2^{})
\,=\,
-\text{sign}(\tilde{c}_{C+}^{})
\frac{\,2M_Z^{2}T_3^{}\,}{\,\Lambda^4\,}
\epsilon^{\al\be\mu\nu} q_{2\nu^{}}^{}\gamma_\mu^{}P_L^{} \,,
\end{eqnarray}
where $\,P_L^{}\!=\!\fr{1}{2}(1\!-\!\gamma_5^{})\,$
and $\,T_3^{}\!=\pm\fr{1}{2}$\,.
For comparison, we consider another fermion-bilinear operator $\mathcal{O}_{C-}^{}$,
and derive its contribution to the effective contact vertex
$f\bar{f}Z\gamma$ with the four external fields being on-shell:
\begin{eqnarray}
\label{eq:Vertex-C-}
\ii\, \Gamma^{\al\beta}_{Z\gamma f\bar f}
({q}_1^{}, {q}_2^{})
\,=\,
-\text{sign}(\tilde{c}_{C-}^{})
\frac{\,2T_3^{}\,}{\,\Lambda^4\,}\!
\(q_3^2\,q_{2\nu}^{}\epsilon^{\alpha\beta\mu\nu}\!
+ 2q_2^{\alpha}q_{3\nu}^{}q_{2\sigma}^{}\epsilon^{\beta\mu\nu\sigma}\)\!
\gamma_\mu^{}P_L^{} \,.
\end{eqnarray}
We can compare this with the nTGC vertices \eqref{eq:Vertex-G+} by
the contribution of $\OGP$. We observe that they share the same kinematic
structure. Namely, Eq.\eqref{eq:Vertex-G+} contributes to the reaction
$\,f\bar{f}\!\to\!Z\ga\,$ via the $s$-channel $Z^*$ and $\ga^*$ exchanges,
and we find that the on-shell amplitude $\,\mathcal{T}[f\bar{f}\!\!\to\!\!Z\ga]\,$
induced by $\OGP$ is exactly the same as that contributed by the above
$f\bar{f}Z\gamma$ contact vertex \eqref{eq:Vertex-C-} of $\OCM$,
as we have expected.

\vspace*{1mm}

Applying a basic angular cut $\,\sin\theta>\sin\delta\,$
(with $\,\delta\ll 1$),\,
we derive the following total cross section for the on-shell $Z\gamma$ production,
including any given nTGC operator $\mathcal{O}_j^{}$ and
summing over the final-state $Z$ polarizations:
%

\begin{eqnarray}
\sigma(Z\gamma)
&\!\!=\!\!&
\frac{~e^4(c_L^2\!+\!c_R^2)\!\!\left[-(s\!-\!M^2_Z)^2\!-\!2(s^2\!+\!M_Z^4)
	\ln\sin\!\frac{\delta }{2}\,\right]\,}
{\,8\pi s_W^2c_W^2(s\!-\!M^2_Z)s^2\,}
\hspace*{15mm}
\nn\\[1mm]
&&
+\,\text{sign}(\tilde{c}_j^{})
\frac{\,e^2 (c_L^{}x_{L}^{}\!\!-\!c_R^{}x_{R}^{})M_Z^2\!
\(s\!-\!M_Z^2\)\!\(s\!+\!M^2_Z\)\,}
{\,8\pi s_W^{}c_W^{}\, s^2\,}\frac{1}{\,\cut^4\,}
\label{Zgamma1}
\\[1mm]
&& +\frac{\,(x_L^2\!+\!x_R^2)M_Z^2\!\(s\!+\!M^2_Z\)\!\(s\!-\!M^2_Z\)^{\!3}\,}
{\,48\pi\,s^2\,}\frac{1}{\,\cut^8\,} + O(\delta)\,,
\nn
\end{eqnarray}
where the relevant coupling coefficients
$(x_L^{},\,x_R^{})$ are defined as
\beqs
\label{eq:xLR-OBWC+G-}
\begin{eqnarray}
(x_L^{},\,x_R^{}) \!\!&=&\!\! (s_W^2,\,s_W^2),
\hspace*{25.5mm}
(\text{for}~\,\mathcal{O}_j^{}\!=\mathcal{O}_{G-}^{}) ,
\label{eq:xLR-OG-}
\\
(x_L^{},\,x_R^{}) \!\!&=&\!\! \(-\fr{1}{2}\!+\!s_W^2,\,s_W^2\)\!,
\hspace*{14.5mm}
(\text{for}~\,\mathcal{O}_j^{}\!=\mathcal{O}_{\widetilde{B}W}^{}) ,
\hspace*{15mm}
\label{eq:xLR-OBW}
\\
(x_L^{},\,x_R^{}) \!\!&=&\!\! (\fr{1}{2},\,0) ,
\hspace*{30.7mm}
(\text{for}~\,\mathcal{O}_j^{}\!=\mathcal{O}_{C+}^{}) .
\label{eq:xLR-OC+}
\end{eqnarray}
\eeqs
Then, we derive the normalized angular distribution functions
$\,f_{\phi_*}^j\,$ for quarks,
{
\beqs
\label{eq:f-phi*q}
\begin{eqnarray}
\hspace*{-8mm}
f_{\phi_*}^{0} \!\!\!&=&\!\!
\frac{1}{2\pi} +
\frac{\,3\pi^2c_{-}^2q_-^2M_Z^{}\sqrt{s}\,(s\!+\!M_Z^2)\cos\!\phi_*^{}\!
-8c_{+}^2q_+^2M_Z^2\,s \cos\!2\phi_*^{}\,}
{\,16 \pi c_{LR+}^2q_+^2\!\left[(s\!-\!M_Z^2)^2\!+2(s^2\!+\!M_Z^4)
\ln\sin\frac{\delta}{2}
\right]\,}+O(\delta),
\hspace*{4mm}
\\[2mm]
\hspace*{-8mm}
f_{\phi_*}^{1} \!\!\!&=&\!\! \frac{1}{2\pi} -
\frac{\,9\pi (c_L^{}x_L^{}\!+\!c_R^{}x_R^{})
(q_L^2\!-\!q_R^2) \sqrt{s}\,\cos\!\phi_*^{}\,}
{\,128(c_L^{}x_L^{}\!-\!c_R^{}x_R^{})(q_L^2\!+\!q_R^2) M_Z^{}\,}+
\frac{s \cos\!2\phi_*^{}\,}
{\,4\pi(s\!+\!M_Z^2)\,} ,
\label{f1q}
\\[2mm]
\hspace*{-8mm}
f_{\phi_*}^{2} \!\!\!&=&\!\!
\frac{1}{2\pi} - \frac{\,9\pi (x_L^2\!-\!x_R^2)(q_L^2\!-\!q_R^2)M_Z^{}\sqrt{s}\,\cos\!\phi_*^{}\,}
{\,128(x_L^2\!+\!x_R^2)(q_L^2\!+\!q_R^2)(s\!+\!M_Z^2)\,}\,,
\end{eqnarray}
\eeqs
}
\hspace*{-3mm}
with the coefficients
\,$(c_{\pm}^2,\,q_\pm^2)\!=\!(c_L^2\pm c_R^2,\,q_L^2\pm q_R^2)$.\,
Here the left-handed and right-handed lepton and quark couplings
to the gauge boson $Z$ are given by
%
\beqa
\(c_L^{},\,c_R^{}\) = \(-\fr{1}{2}\!+\!s_W^2,\, s_W^2\!\)\!,
~~~~~
(q_L^{},\,q_R^{}) = \(T^{}_3\!-\!Qs_W^2,\, -Qs_W^2\!\)\!,
\eeqa
%
where $\,T^{}_3=(\fr{1}{2},\,-\fr{1}{2})$\,
and $\,Q=(\fr{2}{3},\,-\fr{1}{3})\,$
correspond to the (up,\,down)-type quarks, respectively.
In Eq.\eqref{eq:f-phi*q}
we have also imposed a lower cutoff $\,\delta\,(\ll\! 1)$\,
on the polar scattering angle, $\sin\theta>\sin\delta\,$,
which corresponds to a lower cut
on the transverse momentum of the final-state photon,
$\,P_T^\gamma> q\sin\delta$\,.\,

\vspace*{1mm}

\begin{figure}[t]
\includegraphics[width=7.7cm,height=5.5cm]{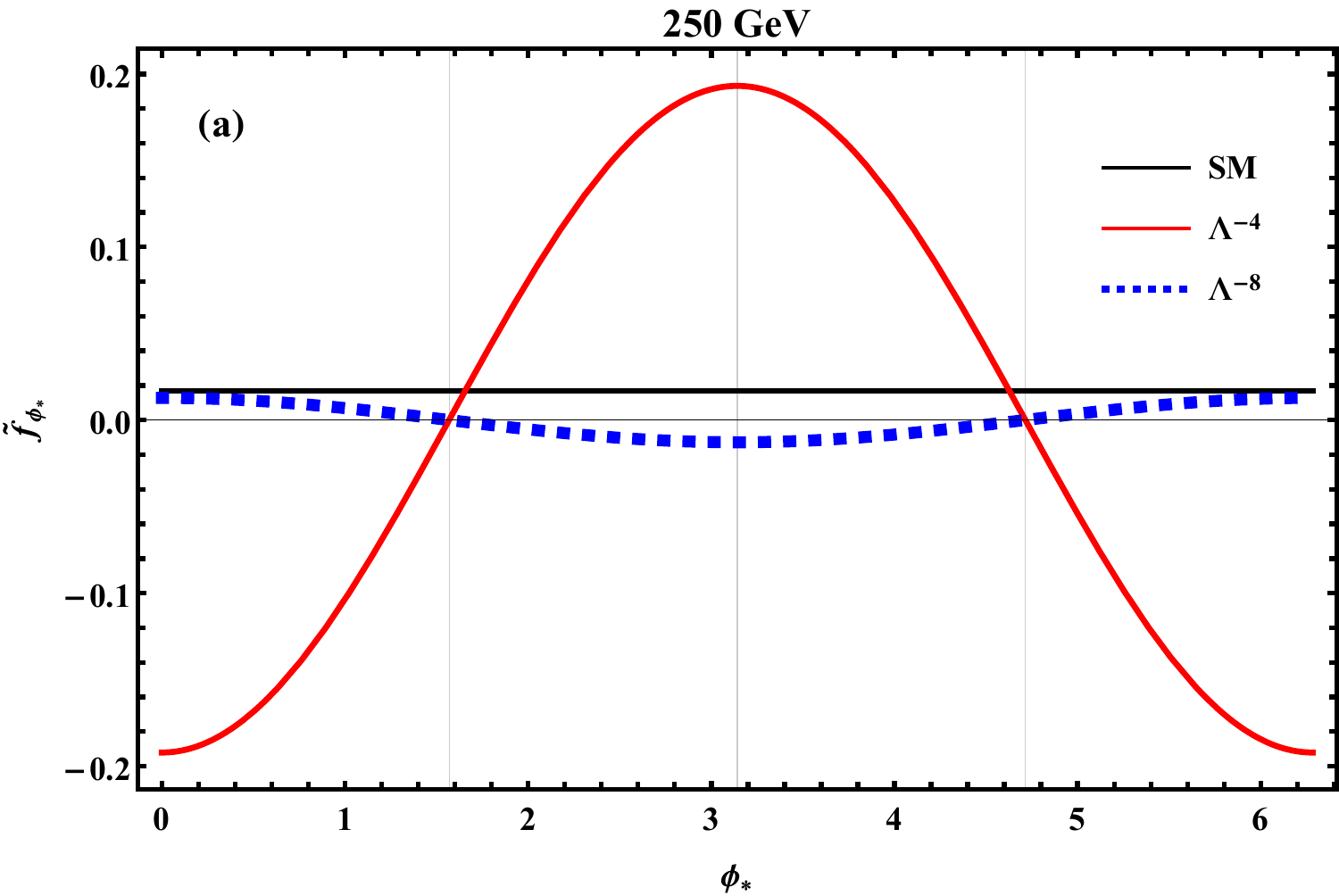}
\includegraphics[width=7.7cm,height=5.5cm]{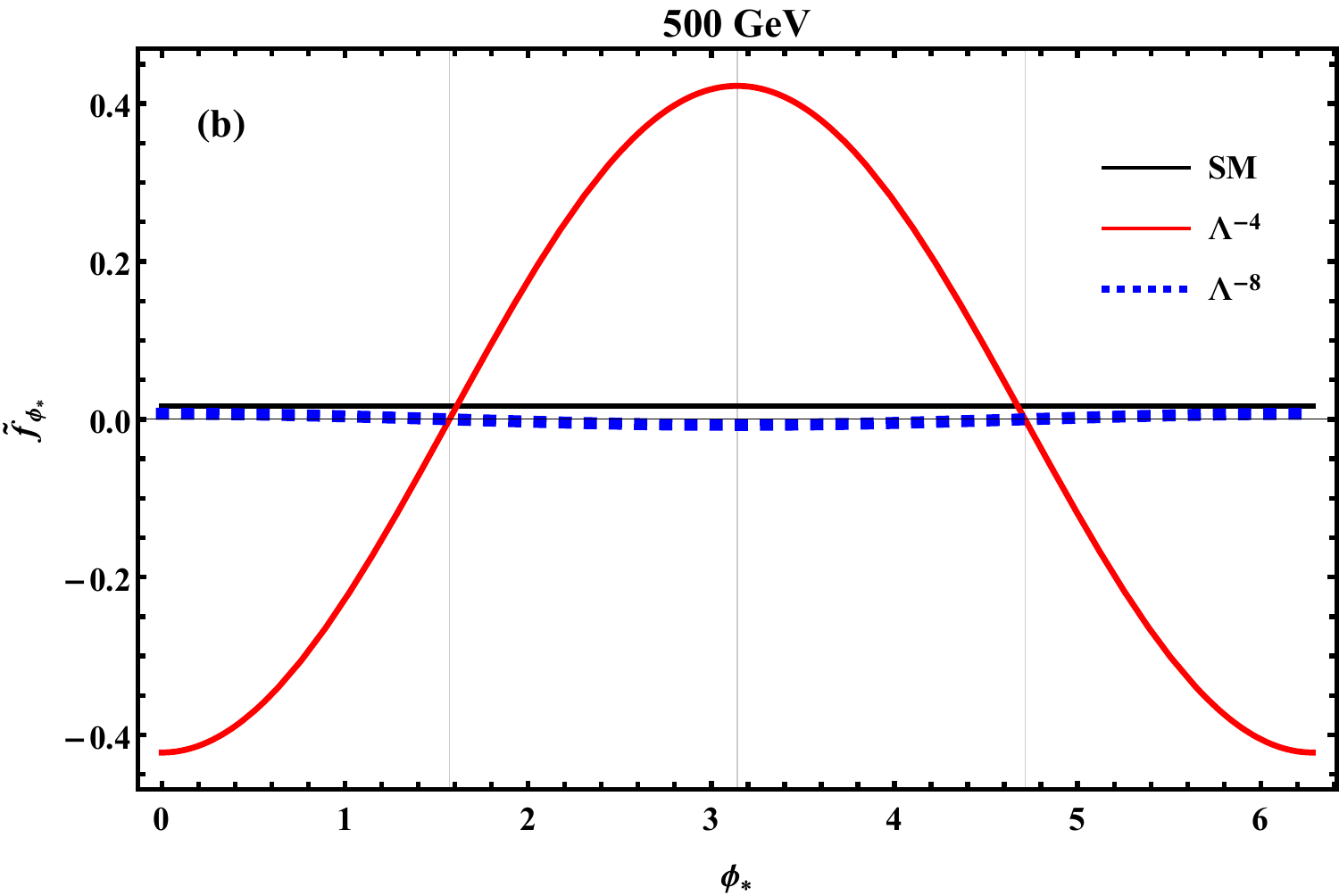}
\\[3mm]
\includegraphics[width=7.7cm,height=5.5cm]{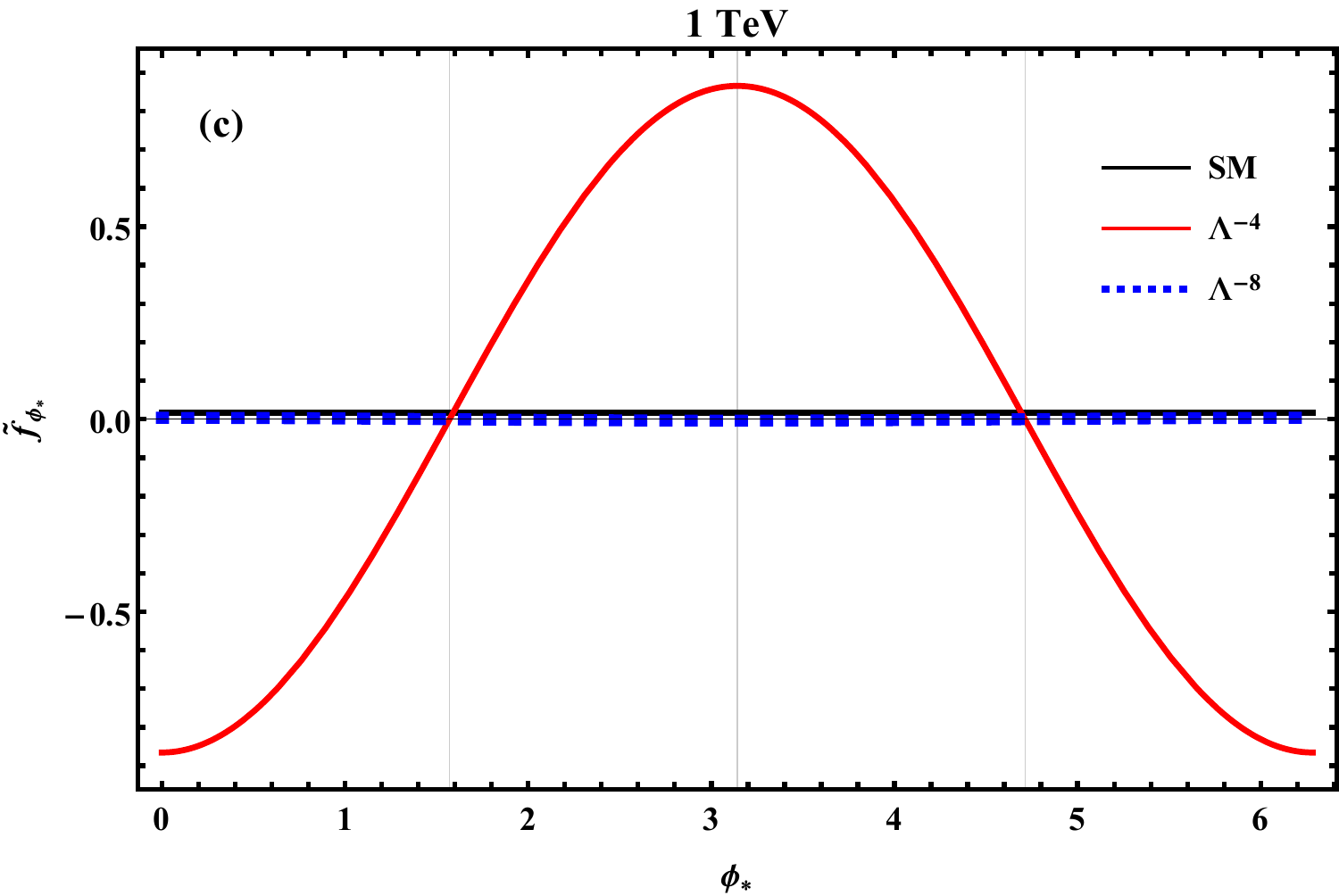}
\includegraphics[width=7.7cm,height=5.5cm]{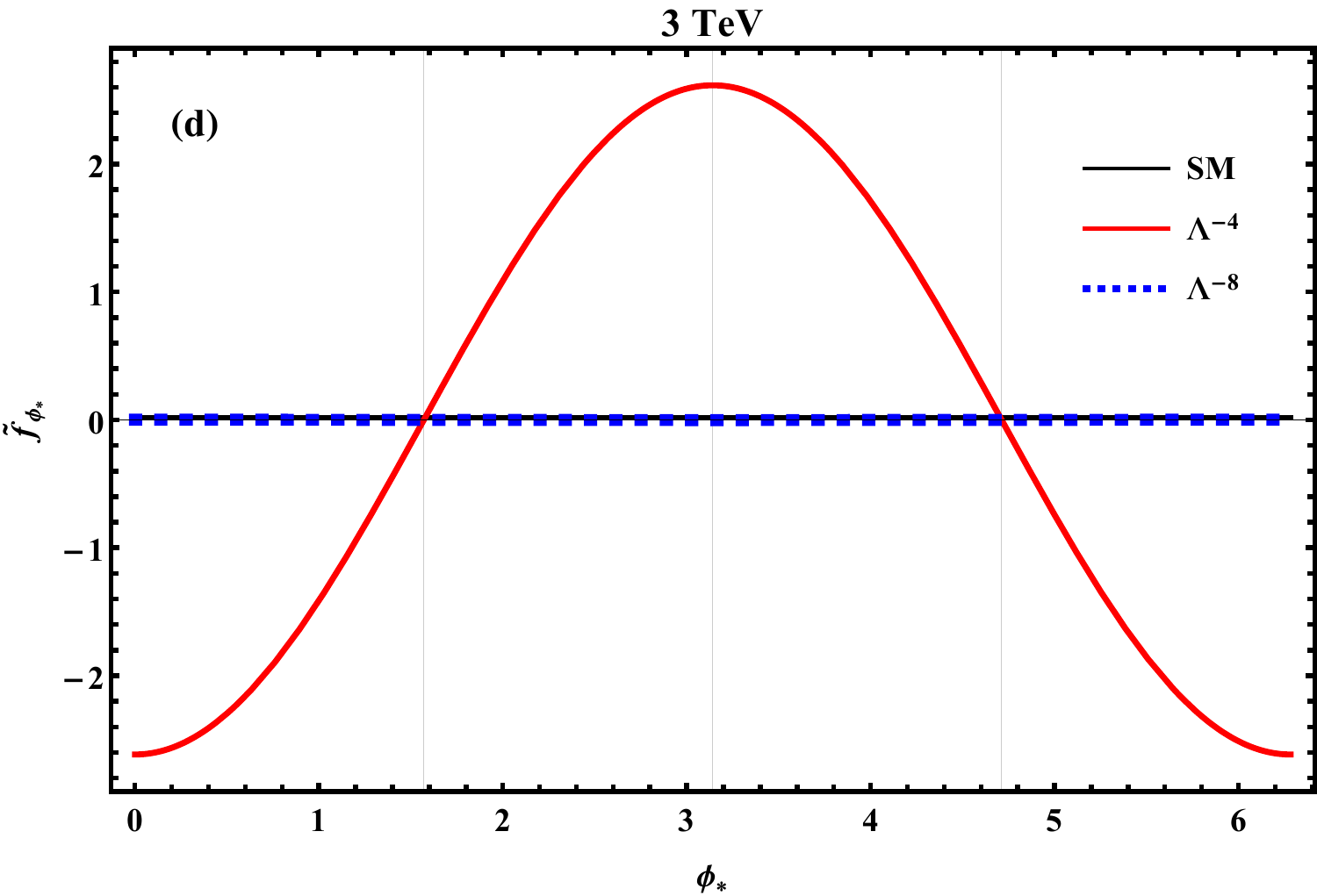}
\vspace*{-2mm}
\caption{\small{%
Normalized angular distributions in the azimuthal angle
$\phi_*^{}$ for $e^-e^+\!\!\to Z\ga$ followed by $Z\!\to\! d\bar d$ decays,
as generated by $\,\mO_{G-}^{}$\! at collision energies}
$\sqrt{s}=(0.25,\, 0.5,\, 1,\, 3)$\,TeV, {respectively.
In each plot, the black, red, and blue curves denote the contributions from
the SM, the interference term of ${O}(\cut^{-4})$, and the quadratic term
of ${O}(\cut^{-8})$, respectively,
where we note that the blue and black curves almost coincide.
We have imposed a basic cut on the polar scattering angle,
$\sin\theta>\sin\delta$,
with $\delta=0.2$ for illustration.
}}
\label{fig:4FF}
\vspace*{2mm}	
\end{figure}

We find that the above distribution $f_{\phi_*^{}}^j$
is not optimal for analyzing the operator $\mathcal{O}_{G-}^{}$.
Instead, we construct the following
angular distributions for $\mathcal{O}_{G-}^{}$:
\beqa
\tilde{f}_{\phi_*^{}}^j =\,
\frac{1}{\,\sigma_j^{}\,}\!\!\int\!\!\di\theta\di\theta_*^{}\,
\text{sign}(\cos\theta)\text{sign}(\cos\theta_*^{})
\frac{\di^3\sigma_{\!j}^{}}{\,\di\theta\,\di\theta_*^{}\di\phi_*^{}\,}\,.
\eeqa
With this, we derive the following angular distributions:
\beqs
\label{eq:fjG-}
\beqa
\tilde{f}_{\phi_*}^{0} \!\!\!&=&\!\!
\frac{3(c_L^2\!-\!c_R^2)(q_L^2\!-\!q_R^2)}
{\,8\pi (c_L^2\!+\!c_R^2)(q_L^2\!+\!q_R^2)\,}+O(\delta)\,,
\label{eq:f0G-}
\\[1mm]
\tilde{f}_{\phi_*-}^{1} \!\!\!&=&\!\!
\frac{3(c_L^{}x_L^{}\!\!+\!c_R^{}x_R^{})(q_L^2\!-\!q_R^2)M_Z^2}
{16\pi (c_L^{}x_L^{}\!\!-\!c_R^{}x_R^{})(q_L^2\!+\!q_R^2)(s\!+\!M_Z^2)}
-\frac{s^{\frac{3}{2}}_{}\cos\phi_*^{}}{\,4\pi M_Z^{}(s\!+\!M_Z^2)\,} \,,
\hspace*{10mm}
\label{eq:f1G-}
\\[1mm]
\tilde{f}_{\phi_*-}^{2} \!\!\!&=&\!\!
\frac{\,M_Z^{}\sqrt{s\,}\cos\phi_*^{}\,}{8\pi(s+\!M_Z^2)}  \,.
\label{eq:f2G-}
\eeqa
\eeqs
where the functions
$\,\tilde{f}_{\phi_*-}^{1,2}$
are contributed by the operator $\OGM$, as indicated by their subscripts
``$-$''.
The motivation and use of the above distributions are explained further
in Section\,\ref{sec:4.2}.
Here we present the angular distributions \eqref{eq:fjG-} in Fig.\ref{fig:4FF}.
We see from Eq.\eqref{eq:f1G-} that the distribution $\,f_{\phi_*^{}}^1$
from the ${O}(\cut^{-4})$ interference contribution is dominated by the
term  $\,\propto\! -\cos\phi_*^{}$\, with energy enhancement $\sqrt{s}/M_Z^{}$\,.
On the other hand, Eq.\eqref{eq:f0G-} shows that the leading SM contribution
to $\,f_{\phi_*^{}}^0$ is a constant and independent of the collider energy
$\sqrt{s}$\,, while the distribution $\,f_{\phi_*^{}}^2$ from the squared
${O}(\cut^{-8})$ contribution in Eq.\eqref{eq:f2G-}
is proportional to $\cos\phi_*^{}$ and
suppressed by $M_Z^{}/\!\sqrt{s}$ at high energies.
These analytical features of $\,f_{\phi_*^{}}^j$ enable us
to understand why the red curve in each plot of Fig.\,\ref{fig:4FF}
shows clear behaviour $\,\propto\! -\cos\phi_*^{}$\,
for the ${O}(\cut^{-4})$ interference
contribution.
Moreover, we see that the SM contributions (black curves) are essentially flat,
whereas the ${O}(\cut^{-8})$ squared contributions (blue curves) are highly suppressed
by $M_Z^{}/\!\sqrt{s}$\,, so as to be nearly flat except in the case of the relatively
low collider energy $\sqrt{s\,}=250$\,GeV, which exhibits minor fluctuations
in accord with the analytic behaviour $\propto\cos\phi_*^{}$\,.

\begin{figure}[t]
\includegraphics[width=7.7cm,height=5.5cm]{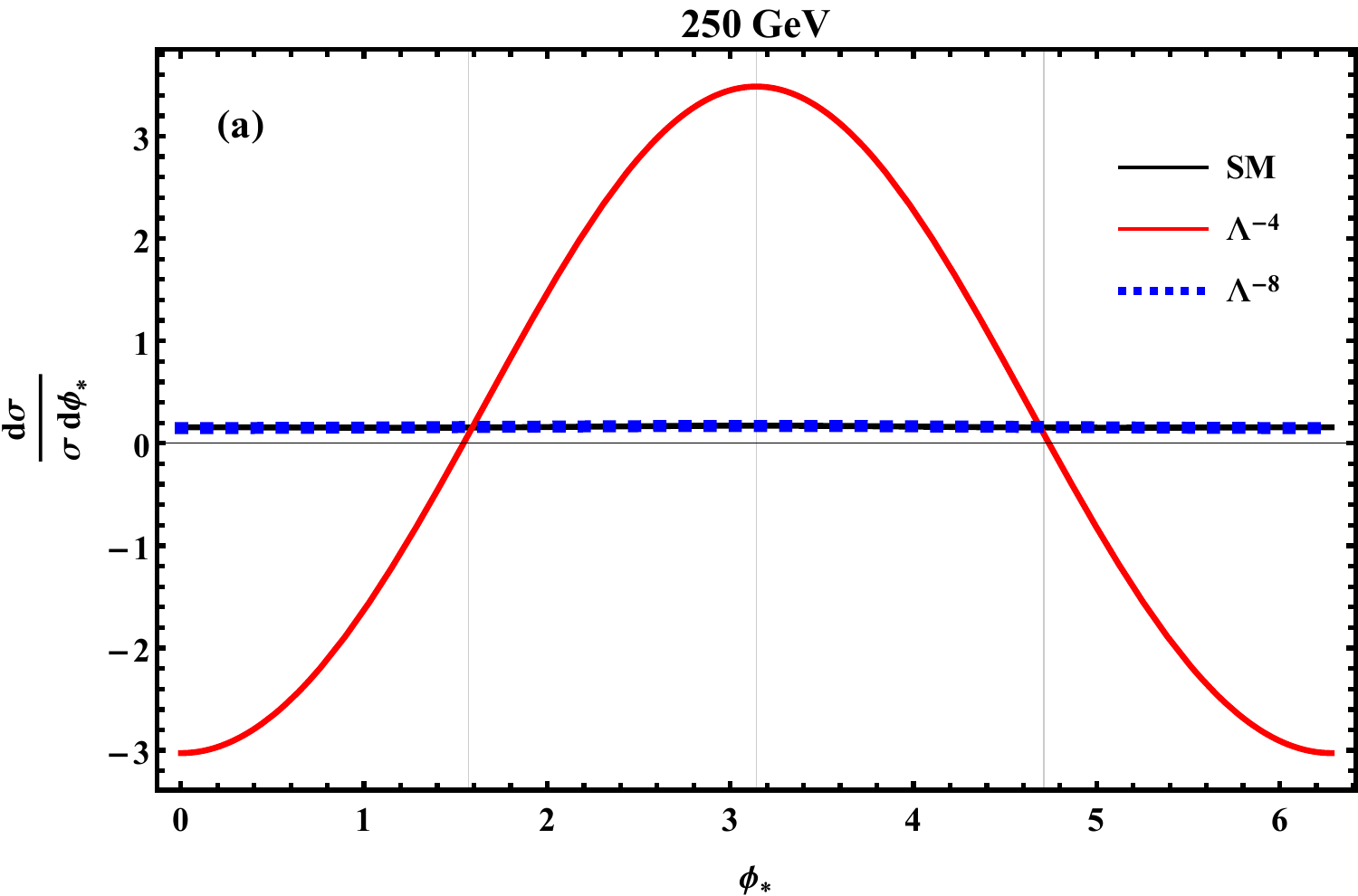}
\includegraphics[width=7.7cm,height=5.5cm]{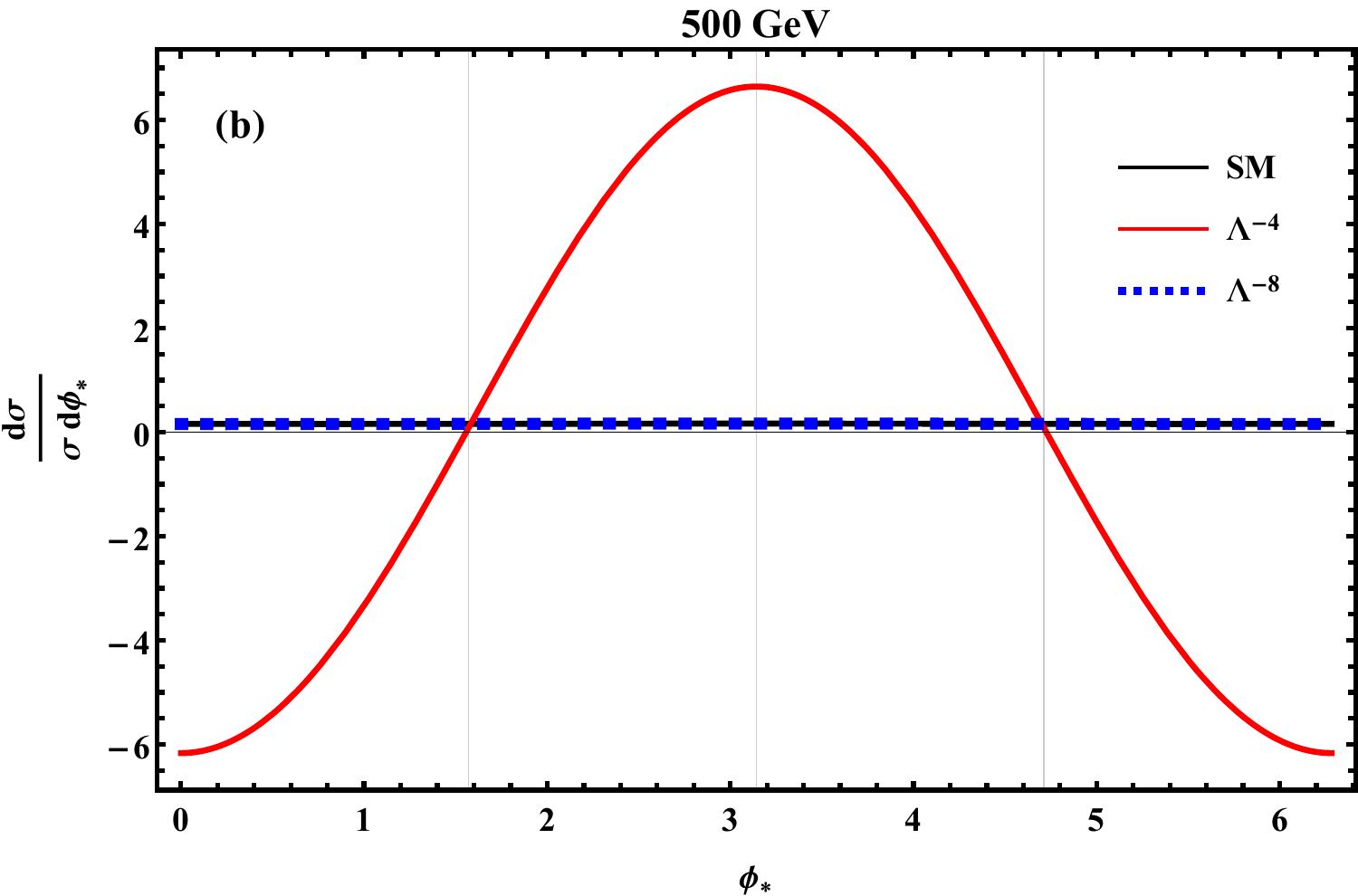}
\\[3mm]
\includegraphics[width=7.7cm,height=5.5cm]{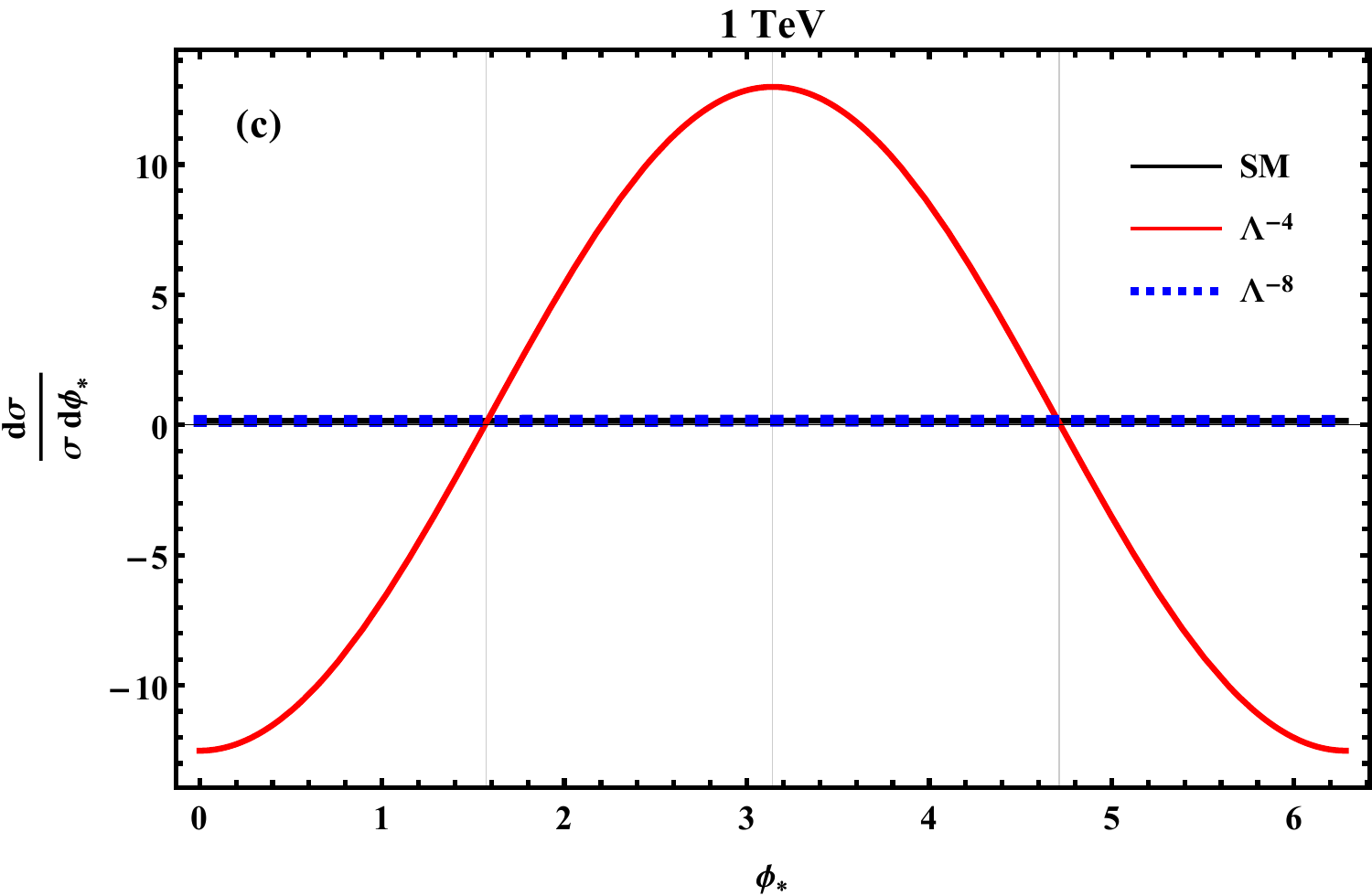}
\includegraphics[width=7.7cm,height=5.5cm]{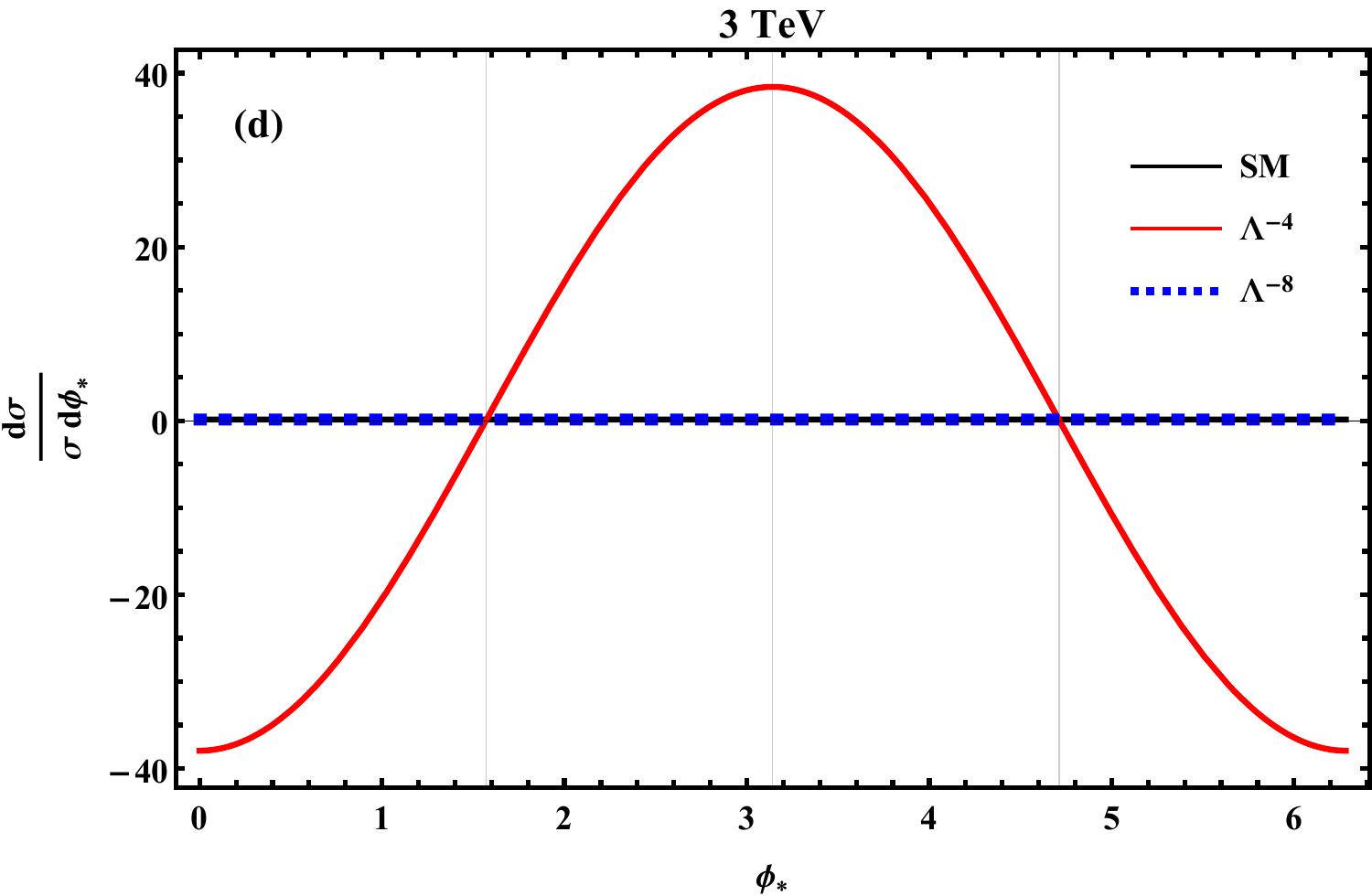}
\vspace*{-2mm}
\caption{\small{%
Normalized angular distributions in the azimuthal angle
$\phi_*^{}$ for $e^-e^+\!\!\to\! Z\ga$ followed by $Z\!\to d\bar{d}$\, decays,
as generated by $\,\mO_{\widetilde{B}W}^{}$\!
at the collision energies}
$\sqrt{s}=(0.25,\, 0.5,\, 1,\, 3)$\,TeV, {respectively.
In each plot, the black, red, and blue curves denote the contributions from
the SM, the interference term of $\,{O}(\cut^{-4})$, and the quadratic term
of $\,{O}(\cut^{-8})$, respectively,
where we note that the blue and black curves almost coincide.
We have imposed a basic cut on the polar scattering angle,
$\sin\theta>\sin\delta$,
with $\,\delta=0.2\,$ for illustration.}}
\label{fig3}
\label{fig:5F}
\end{figure}

\begin{figure}[t]
\includegraphics[width=7.7cm,height=5.5cm]{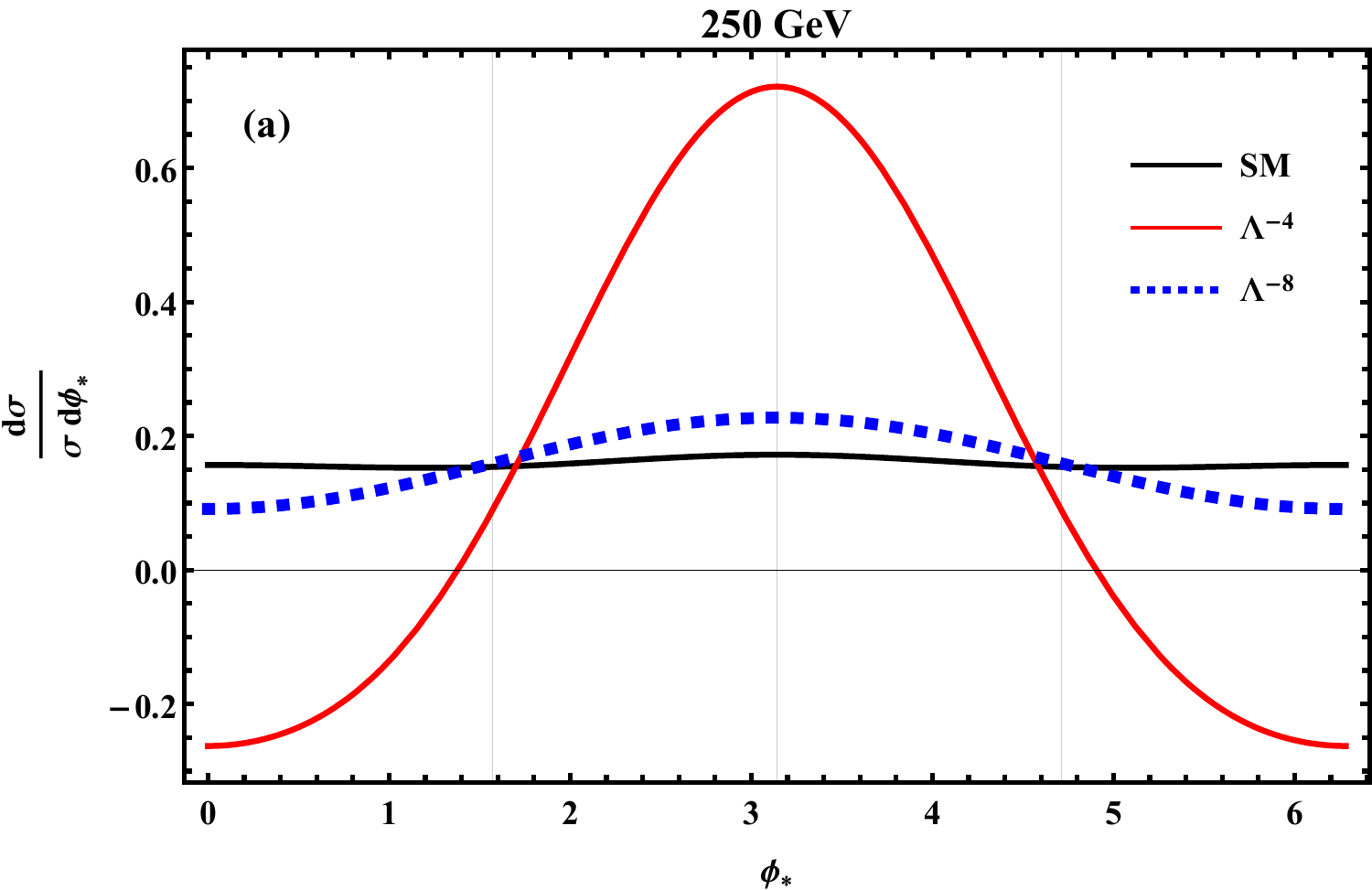}
\includegraphics[width=7.7cm,height=5.5cm]{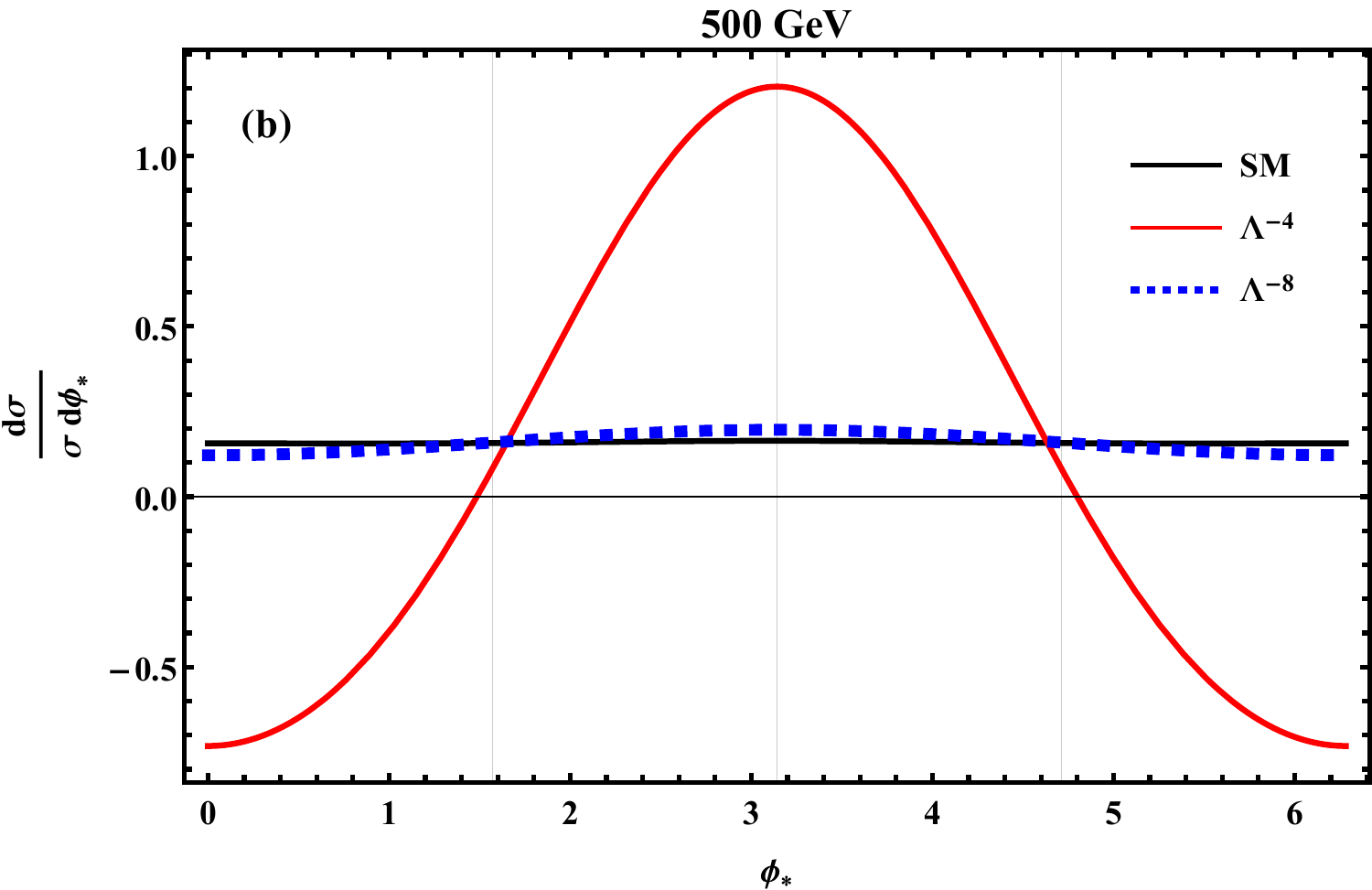}
\\[3mm]
\includegraphics[width=7.7cm,height=5.5cm]{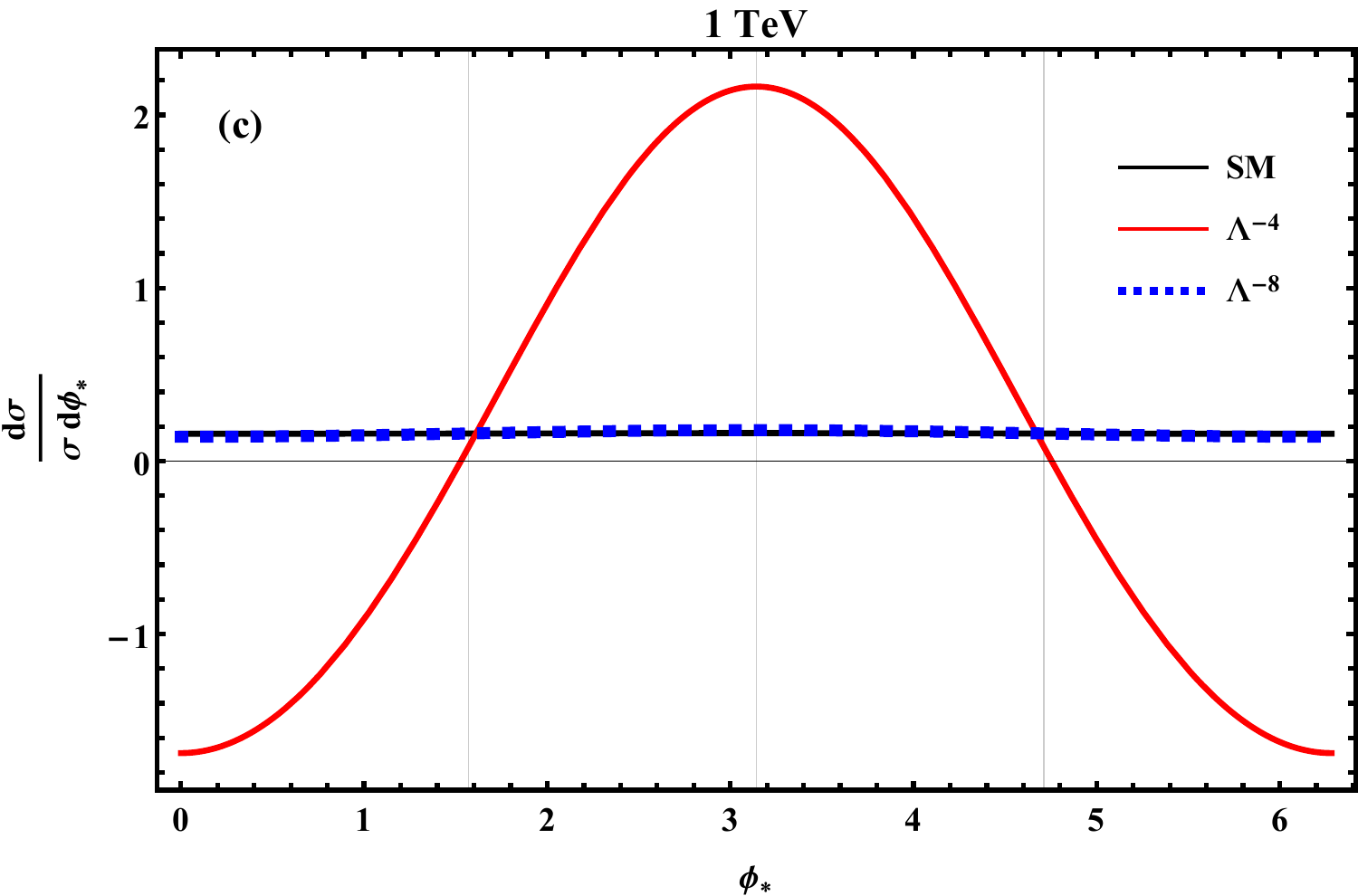}
\includegraphics[width=7.7cm,height=5.5cm]{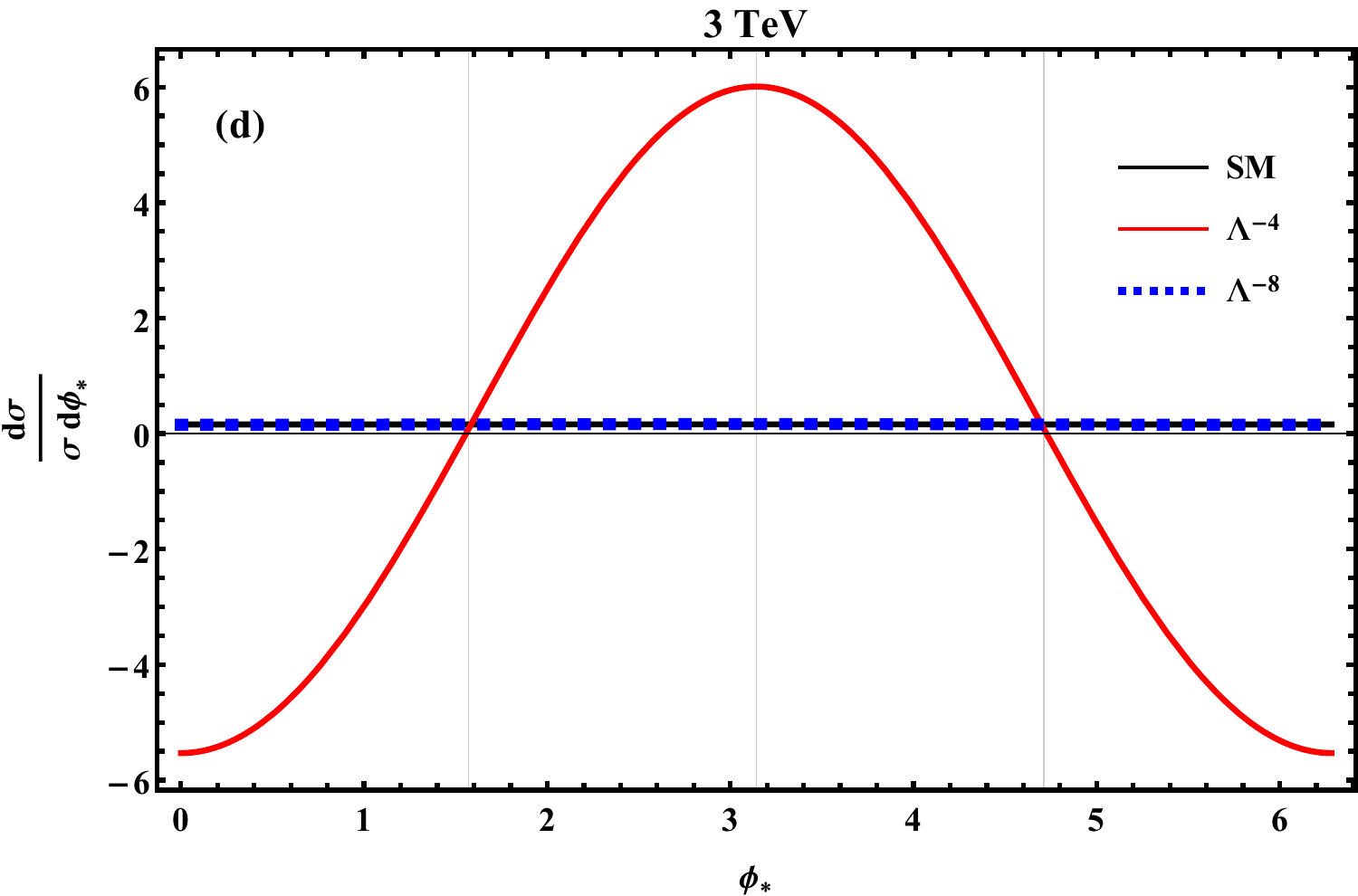}
\vspace*{-2mm}
\caption{\small{%
Normalized angular distributions in the azimuthal angle
$\phi_*^{}$ for $e^-e^+\!\!\to Z\ga$ followed by $Z\to d\bar d$ decays,
as generated by $\,\mO_{C+}^{}$\! at the collision energies
$\sqrt{s}=(0.25,\, 0.5,\, 1,\, 3)$\,TeV, respectively.
In each plot, the black, red, and blue curves denote the contributions from
the SM, the interference term of ${\cal O}(\cut^{-4})$, and the quadratic term
of ${\cal O}(\cut^{-8})$, respectively,
where we note that the blue and black curves almost coincide.
We have imposed a simple cut on the scattering angle, $\sin\theta>\sin\delta$,
with $\delta=0.2$ for illustration.}}
\label{fig:6FF}
\end{figure}

\vspace*{1mm}

The angular distributions $\,f^j_{\phi_*^{}}$ of
the operators
$\mathcal{O}_{\widetilde{B}W}^{}$ and $\mathcal{O}_{C+}^{}$
are given in Eq.\eqref{eq:f-phi*q}. We present these distributions
for $\mathcal{O}_{\widetilde{B}W}^{}$ and $\mathcal{O}_{C+}^{}$
in Figs.\,\ref{fig:5F} and \ref{fig:6FF}, respectively.
We see that the distributions have rather similar
shapes in the two figures.
This is because the distribution $\,f^1_{\phi_*^{}}\,$
of the interference contribution of $\,{O}(\cut^{-4})$\,
is dominated by the $\cos\phi_*^{}$ term, which has the enhanced
energy factor $\sqrt{s\,}/M_Z^{}$, while the
distributions $\,f^0_{\phi_*^{}}\,$ and $\,f^2_{\phi_*^{}}\,$
are mainly dominated by the constant term $\,\fr{1}{\,2\pi\,}\,$.\,
This explains why in Figs.\ref{fig:5F} and \ref{fig:6FF}
the $\,f^1_{\phi_*^{}}\,$ distributions (red curves)
have shapes similar to $\cos\phi_*^{}$,\,
and the distributions $\,f^0_{\phi_*^{}}\,$ and $\,f^2_{\phi_*^{}}\,$
are nearly flat except the case of the relatively low collider energy
$\sqrt{s}=250$\,GeV in Fig.\ref{fig:6FF}(a)
where $\,f^2_{\phi_*^{}}\,$ (blue dashed curve) shows some small deviations.
In fact, the {\it major difference} between
Figs.\ref{fig:5F} and \ref{fig:6FF}
is only in the overall magnitudes of the distributions $\,f^1_{\phi_*^{}}\,$
(red curves). We can understand this difference by inspecting
the leading terms $\,\propto\!\cos\phi_*^{}$ in
the formula \eqref{f1q} for the operators
$\mathcal{O}_{\widetilde{B}W}^{}$ and $\mathcal{O}_{C+}^{}$\,.
Using the couplings given in Eq.\eqref{eq:xLR-OBWC+G-},
we can readily estimate the ratio of their couplings appearing in the
$\cos\phi_*^{}$ term of Eq.\eqref{f1q}:

\beqs
\label{eq:R-qq}
\beqa
\label{eq:R-qq(d)}
\mathcal{R}_{q=d}^{}\,=\,
\frac{\,(c_L^{}x_L^{}\!+\!c_R^{}x_R^{})(q_L^2\!-\!q_R^2)\,}
{\,(c_L^{}x_L^{}\!-\!c_R^{}x_R^{})(q_L^2\!+\!q_R^2)\,}
\,\simeq\, \left\{
\ba{ll}
\! 6.26,    & ~~~(\text{for}~\mathcal{O}_{\widetilde{B}W}^{}) ,
\hspace*{8mm}
\\[3mm]
\! 0.936,   & ~~~(\text{for}~\mathcal{O}_{C+}^{}) ;
\ea
\right.
\\[1mm]
\mathcal{R}_{q=u}^{}\,=\,
\frac{\,(c_L^{}x_L^{}\!+\!c_R^{}x_R^{})(q_L^2\!-\!q_R^2)\,}
{\,(c_L^{}x_L^{}\!-\!c_R^{}x_R^{})(q_L^2\!+\!q_R^2)\,}
\,\simeq\, \left\{
\ba{ll}
\! 4.47,    & ~~~(\text{for}~\mathcal{O}_{\widetilde{B}W}^{}) ,
\hspace*{8mm}
\\[3mm]
\! 0.669,   & ~~~(\text{for}~\mathcal{O}_{C+}^{}) ;
\ea
\right.
\eeqa
\eeqs
where the couplings depend on the weak mixing angle $\theta_W^{}$
and we have input the $\overline{\text{MS}}$ value
$\,s_W^2=0.23122\pm 0.00003\,$ ($\mu=M_Z^{}$) \cite{PDG}.
This immediately explains why the overall size of the $\,f^1_{\phi_*^{}}\,$ distribution
for $\mathcal{O}_{\widetilde{B}W}^{}$ is larger than that for
$\mathcal{O}_{C+}^{}$ by about a factor of $\,6.26/0.936\!\simeq\!6.7$\,
for down-type quarks ($q\!=\!d,s,b$), and
$\,4.47/0.669\!\simeq\! 6.7$\, for up-type quarks ($q\!=\!u,c$),
as seen in Figs.\ref{fig:5F} and \ref{fig:6FF}.
For the operator $\mathcal{O}_{\widetilde{B}W}^{}$,\,
we can further compare the $\,f^1_{\phi_*^{}}\,$ distribution
in Fig.\ref{fig:5F} for the hadronic decay channel
$\,Z\!\to\! d\bar{d}$\,
with the same $\,f^1_{\phi_*^{}}\,$ distribution computed
for the leptonic decay channel $\,Z\!\to\!\ell\bar{\ell}$\,
in our previous study\,\cite{Ellis:2019zex} (see its Fig.4).
In the case of the $\,Z\!\to\!\ell\bar{\ell}$\, channel,
for the operator $\mathcal{O}_{\widetilde{B}W}^{}$,
the corresponding coupling ratio factor
appearing in the leading term $\propto \cos\phi_*^{}$
of $\,f^1_{\phi_*^{}}\,$ becomes
\beqa
\label{eq:R-ell}
\mathcal{R}_{\ell}^{}\,=\,
\frac{\,(c_L^{2}\!+\!c_R^{2})(c_L^2\!-\!c_R^2)\,}
{\,(c_L^{2}\!-\!c_R^{2})(c_L^2\!+\!c_R^2)\,}
\,=\, 1\,.
\eeqa
This explains why the overall size of the $\,f^1_{\phi_*^{}}\,$
distribution in hadronic $Z$ decays in the current
Fig.\ref{fig:5F} is larger than that of the analogous $f^1_{\phi_*^{}}$
distribution in leptonic $Z$ decays
(cf.\ Fig.4 of Ref.\,\cite{Ellis:2019zex})
by a significant factor
$\,\mathcal{R}_q^{}/\mathcal{R}_{\ell}^{}\!\approx\! 6.3$\,
for down-type quarks and
$\,\mathcal{R}_q^{}/\mathcal{R}_{\ell}^{}\!\approx\! 4.5$\,
for up-type quarks.

\vspace*{1mm}

In general, we find that our current study of the hadronic decay channels
$\,Z\!\!\to\!q\bar{q}$\, can give bounds on the new physics scale $\cut$
that are substantially
stronger than the leptonic and invisible $Z$-decay channels
studied previously\,\cite{Ellis:2019zex}.
The reason for this can be traced back to the fact
that the $f\bar{f}Z$ coupling combination $\,(q_L^2\!-q_R^2)$\, for quarks,
which appears in the interference term of $O(\cut^{-4})$,
is much larger than the coupling combination $\,(c_L^2\!-\!c_R^2)$\,
for the leptonic channels.
For up- and down-type quarks,
their gauge couplings
$\,(u_L^2,\, u_R^2)\simeq (0.1197,\, 0.0237)$\,
and
$\,(d_L^2,\, d_R^2)\simeq (0.1789,\, 0.0059)$,\,
so the quark final states are mostly left-handed.
We note also that for down-type quarks,
$\,q_-^2/q_+^2 \!=\! (q_L^2\!-\!q_R^2)/(q_L^2\!+\!q_R^2)\!\approx\!0.94$\,,
unlike the case of the leptonic channel which has
$\,c_-^2/c_+^2 \!= (c_L^2\!-\!c_R^2)/(c_L^2\!+\!c_R^2)
\approx 0.15\ll 1$\,.

\vspace*{1.5mm}
\subsection{\hspace{-2mm}Including the Dijet Angular Resolution}
\label{sec:3.2}
\vspace*{1mm}

As we will show in Section\,\ref{sec:4},
our current study of discriminating the signals from backgrounds will
depend on the $\phi_*^{}$ distributions (cf.\ Section\,\ref{sec:3.1})
and the imposed angular cuts. This requires a precise determination of
the azimuthal angle $\phi_*^{}$\,.
But, the accuracy of the $\phi_*^{}$ measurement depends on the
jet angular resolution $\,\delta\phi_j^{}$\,.
This is because the $\phi_*^{}$ measurement depends on the determination
of the decay plane of $\,Z\!\!\to\! q\bar{q}$\, and thus is sensitive to the
jet angular resolution $\delta\phi_j^{}$\,.
The $Z$ boson energy is determined by
$\,E_Z^{}\!=\!\fr{\sqrt{s\,}}{2}\!\!\(\!1\!+\!\fr{M_Z^2}{s}\!\)$,
which increases with the collider energy.
For the energetic fast-moving $Z$\,, the dijets from $Z$ decays
tend to be colinear and their opening angle
$\Delta_{jj}^{}$ becomes smaller for larger $\!\sqrt{s\,}$\,.
From the kinematics of $\,Z\!\to\! q\bar{q}$\,,
we derive a bound:
\beqa
\cos\Delta_{jj}^{} \,\leqq\,
1\!-\! \frac{\,8M_Z^2\,}{\,s\,}\!
\(\!1\!+\!\frac{M_Z^2}{s}\!\)^{\!\!\!-2}.
\eeqa
For a small opening angle $\Delta_{jj}^{}\!\ll 1$\,,\,
this results in a lower bound:
\beqa
\Delta_{jj}^{} \gtrsim\,
\frac{\,4M_Z^{}\,}{\,\sqrt{s\,}\,}\!
\(\!1\!+\!\frac{M_Z^2}{s}\!\)^{\!\!\!-1}.
\eeqa
For a jet with azimuthal angle $\phi_j^{}$ and angular resolution
$\delta\phi_j^{}$, the largest effect of $\delta\phi_j^{}$
on $\phi_*^{}$ is to have the variation $\delta\phi_j^{}$ perpendicular to
the $Z$ decay plane and thus we obtain an upper bound on the resultant
uncertainty of $\phi_*^{}$, namely,
$\,\delta\phi_*^{}\!\lesssim \delta\phi_j^{}/\Delta_{jj}^{}\,$.\,

\begin{figure*}[t]
\centering
\hspace*{-5mm}
\includegraphics[height=6cm,width=7.9cm]{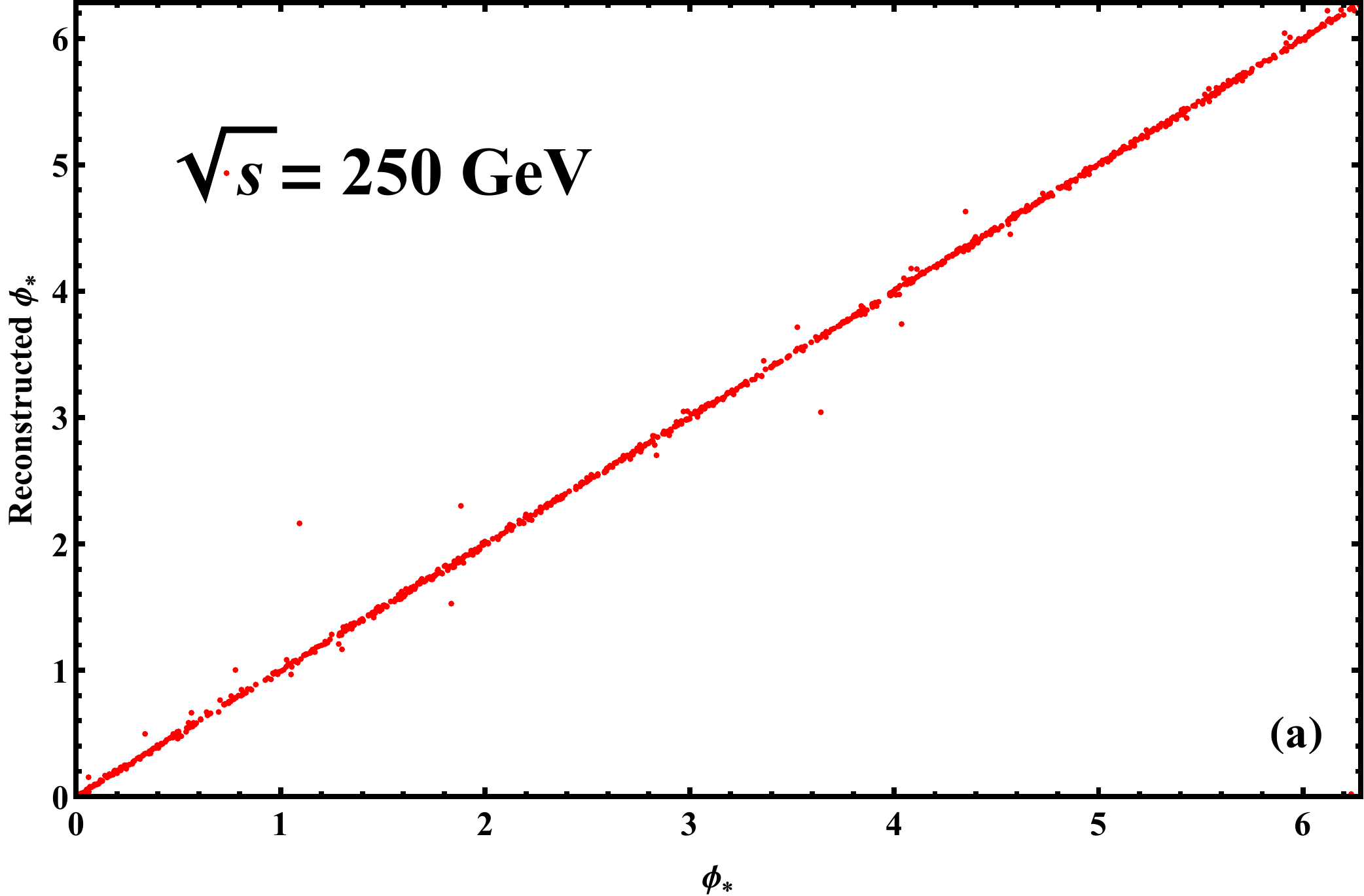}
\hspace*{2mm}
\includegraphics[height=6cm,width=7.9cm]{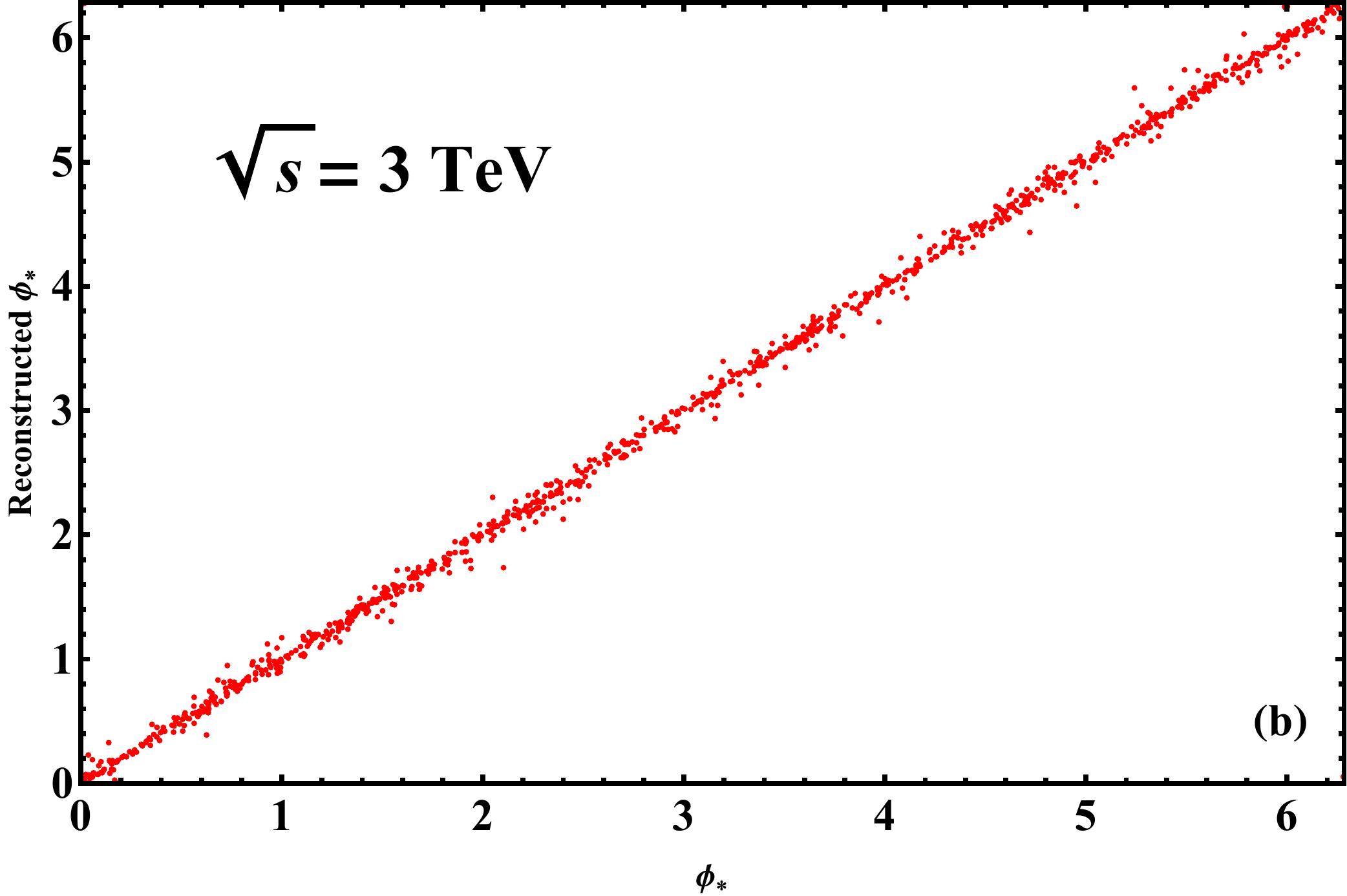}
\\
\includegraphics[height=6cm,width=8.4cm]{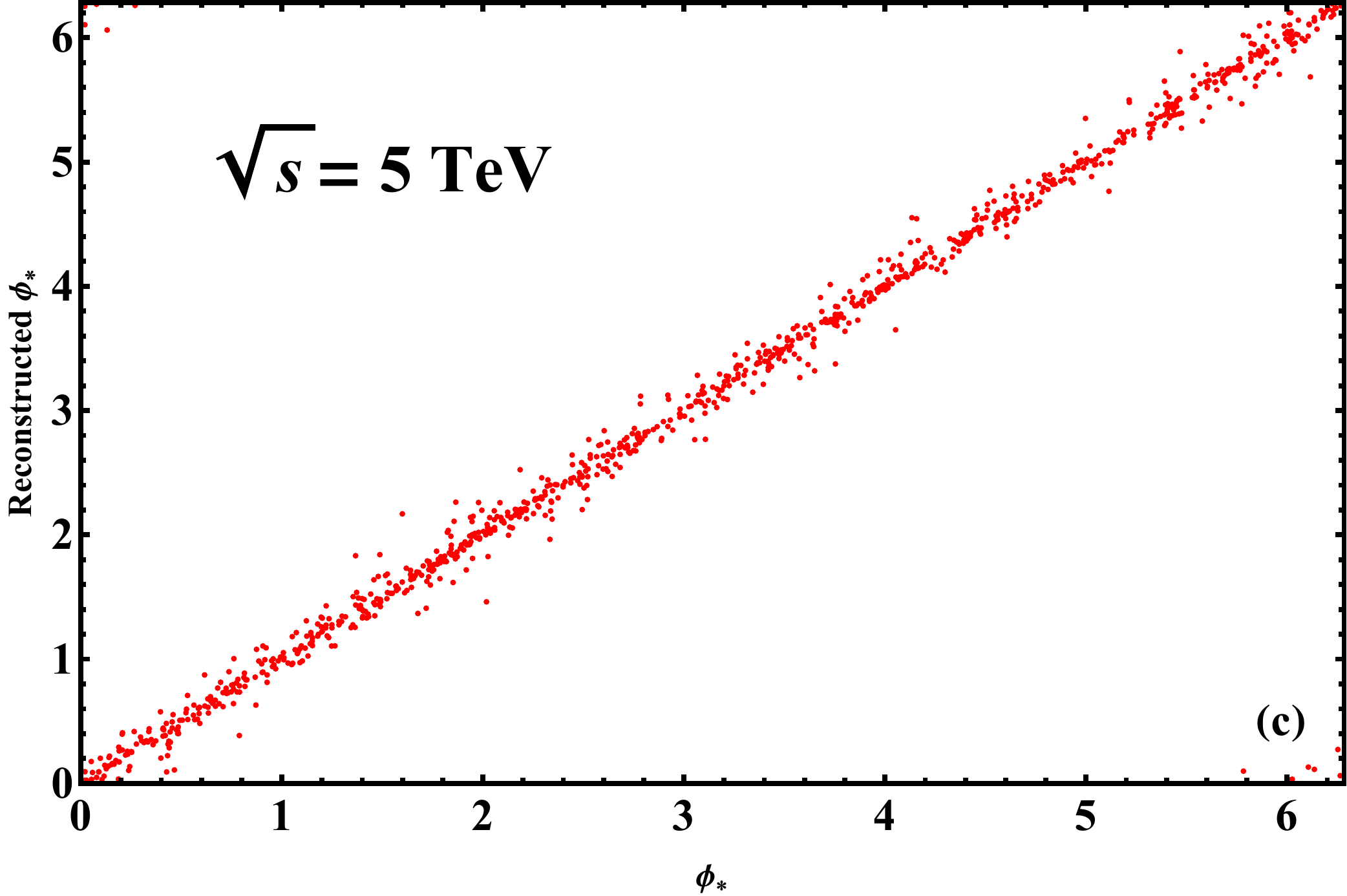}
\vspace*{-1mm}
\caption{\small{%
Comparison between the true value and reconstructed value of the azimuthal angle
$\phi_*^{}$ at} $\sqrt{s} = 250$\,GeV {[plot\,(a)]},
$\sqrt{s}=3$\,TeV {[plot\,(b)]}, {and} $\sqrt{s}=5$\,TeV {[plot\,(c)].}
}
\label{fig:phi}
\label{fig:5}
\label{fig:7F}
\end{figure*}

\vspace*{1mm}

We assume that the jet angular resolution at $e^+e^-$ colliders
is the same as that of the current LHC CMS
detector\,\cite{Sirunyan:2017ulk}, namely,
$\delta\phi_j^{}\!=\!0.01$\, and
$\delta\eta_j^{}\!=\!0.01$\, for both the jet azimuthal angle $\phi_j^{}$ and
the jet rapidity $\eta_j^{}$\,.
We input the angular parameters $(\theta,\,\theta_*^{},\,\phi_*^{})$
to determine the jet momenta
$\,\vec{k}_j^{}\!=\!(k_x^j,\, k_y^j,\, k_z^j)$\,
(for $j=1,2$) which can be re-expressed as functions of the jet
transverse momentum, rapidity and azimuthal angle
$\vec{k}_j^{}\!=\vec{k}_j^{}(k_T^j,\, \eta^{}_j,\, \phi^{}_j)$.\,
With these we can compute the true value of $\eta_j^{}$ and $\phi_j^{}$
in the collision frame, and then calculate the reconstructed momenta
by including the angular resolutions in
$\,\vec{k}_j^{}\!=\vec{k}_j^{}(k_T^j,\, \eta^{}_j\!+\delta\eta_j^{},\, \phi^{}_j\!+\delta\phi_j^{})$.\,
Here we take the uncertainties
$\,\delta\eta_j^{}\,$ and $\,\delta\phi_j^{}\,$
as random numbers obeying normal distributions
with standard deviations $0.01$.
Thus, we can use $\eta_j^{}\!+\!\delta\eta_j^{}$ and
$\phi_j^{}\!+\!\delta\phi_j^{}$ to reconstruct $\phi_*^{}$,
where the scattering plane is well determined by the directions of the incident $e^-$
and outgoing $\gamma$ (along the scattering angle $\theta$) with negligible errors.
Naively, the jet momenta $(\vec{k}_1^{},\,\vec{k}_2^{})$ and the
$\gamma$ momentum $\vec{q}_2^{}$ should be
in the same plane in the collision frame,
but they are not measured to be exactly in the same plane, because of
the soft gluon radiations from the jets and the measurement errors.
The 3-momentum of $Z$ can be fixed by that of the final state photon,
$\vec{q}_1^{}\!=\!-\vec{q}_2^{}$\,, because the photon 3-momentum $\vec{q}_2^{}$
can be measured accurately.  Thus, we can compare the angles,
$\Delta_1^{}\!\equiv
\arccos\,[\vec{q}_1^{}\!\cdot\vec{k}_1^{}/(|\vec{q}_1^{}||\vec{k}_1^{}|)]$
and
$\Delta_2^{}\!\equiv
\arccos\,[\vec{q}_1^{}\!\cdot\vec{k}_2^{}/(|\vec{q}_1^{}||\vec{k}_2^{}|)]$.\,
If $\Delta_1^{}\!>\!\Delta_2^{}$,
we choose $(\vec{q}_1^{},\,\vec{k}_1^{})$
to reconstruct the $Z$-decay plane and thus \,$\phi_*^{}$\,,
otherwise we choose $(\vec{q}_1^{},\,\vec{k}_2^{})$.
We present in Fig.\ref{fig:phi} a comparison of the true and reconstructed
values of $\phi_*^{}$ for three sample cases of collider energies
$\!\sqrt{s\,}\!=\!250$\,GeV [plot\,(a)], $\sqrt{s\,}\!=\!3$\,TeV [plot\,(b)],
and $\!\sqrt{s\,}\!=\!5$\,TeV [plot\,(c)].
It shows that the deviation $\delta\phi_*^{}\!\approx\! 0.04$\, for
$\!\sqrt{s} \!=\!250$\,GeV, while
$\delta\phi_*^{}\!\approx\!0.06$\, for $\!\sqrt{s}\!=\!3$\,TeV,
and $\delta\phi_*^{}\!\approx 0.1$ for $\!\sqrt{s}\!=\!5$\,TeV.
The reason that a higher collision energy $\!\sqrt{s\,}$ causes a larger
$\delta\phi_*^{}$ is because a higher $\!\sqrt{s\,}$ will
generate a larger $Z$ boson energy $E_Z^{}$
and thus a smaller dijet opening angle $\Delta_{jj}^{}$ in
the hadronic decays $Z\!\!\to\! q\bar{q}$\,.
This in turn will cause a larger uncertainty
in determining the $Z$-decay plane and thus a larger error
$\delta\phi_*^{}$ in the azimuthal angle $\phi_*^{}$\,.

\vspace*{1mm}

When we compute the cross sections and the observable $O_1^c$
(cf.\ Section\,\ref{sec:4}),
we note that the {azimuthal angle smearing} $\delta\phi_*^{}$ can play a role
only for values of $\phi_*^{}$ near the boundaries of the $\phi_*^{}$ cuts,
because the points far away from the cut boundaries cannot cross them
due to the small uncertainty $\delta\phi_*^{}$\,.
As we will show further in Section\,\ref{sec:4.1}, for
the observable $O_1^c$ or the cross sections (after $\phi_*^{}$ cuts),
the relative error is actually
$\propto\! \delta\phi_*^2$\,
and thus is much smaller.
We can make numerical simulations on such an error in $O_1^c$ based on the
uncertainty $\delta\phi_*^{}$ caused by the finite jet angular resolution.
We find that at $\sqrt{s} = (0.25,\, 0.5,\, 1,\, 3,\, 5)$\,TeV,
the uncertainty $\delta\phi_*^{}$ could cause a relative error of the order of
$\(10^{-5}\!,\, 10^{-5}\!,\, 10^{-4}\!,\, 10^{-3}\!,\, 10^{-2}\)$
in the observable $O_1^c$, respectively.
Moreover, we find that
the error of the SM background cross section $\sigma_0^c$ under the $\phi_*^{}$ cuts
(cf.\ Section\,\ref{sec:4}) is $\,O(10^{-4})$,
so the effect due to the $\delta\phi_*^{}$ error is negligible in most cases.
For our practical analyses of the sensitivities to the new physics scales $\cut$
in the next Section, we include the effects of $\delta\phi_*^{}$ uncertainty
in our simulations.

\vspace*{1mm}
\section{\large\hspace{-2mm}Probing nTGCs by
\boldmath{$Z\ga$} Production with Hadronic \boldmath{$Z$} Decays}
\label{sec:4}

We analyze in this Section the sensitivities of probing the contributions of
dimension-8 operators to nTGCs in the reaction
$\,e^-e^+ \!\!\to\! Z\ga$\,
followed by hadronic $Z$ decays.
We also compare them with the sensitivities by using
the leptonic $Z$-decay channels.
As we discussed in Section\,\ref{sec:2},
there are only 3 independent operators under the EOM.
We first present our analysis of the sensitivities to each of the four operators $(\mathcal{O}_{G+}^{},\, \mathcal{O}_{G-}^{},\, \mathcal{O}_{\widetilde{B}W}^{},\,
\mathcal{O}_{C+}^{})$
following the conventional approach of one operator at a time,
where we keep in mind the possibility that the dynamics beyond the SM might
generate any one of these operators by itself.
Finally, in the last part of this Section we present our analysis of fitting
pairs among these four operators and study the correlations
between each pair of operators.

\vspace*{-2mm}
\subsection{\hspace*{-2mm}Probing the Dimension-8 Pure Gauge Operator~\boldmath{$\mO_{G+}^{}$}}
\vspace*{1.5mm}
\label{sec:4.1}

In this Subsection, we analyze the contribution of the pure gauge operator
$\mathcal{O}_{G+}^{}$ to the reaction
$\,e^-e^+\!\!\to\! Z\ga$\,
using the hadronic decays $\,Z\!\!\to\! q\bar{q}$\,.
We find that the sensitivity to $\Lambda$ is greatly enhanced
as compared to our previous study via the leptonic decay channels
$\,Z\!\!\to\!\ell\bar{\ell}$ \cite{Ellis:2019zex}.
In consequence, the squared contribution
of $O(\cut^{-8})$ is severely suppressed relative to the interference term
of $O(\cut^{-4})$, and thus can be safely neglected.
Hence, we can focus on the interference contributions
in the following analysis.

\vspace*{1mm}

From Eq.(\ref{eq:phi-}), we note
that the $\cos2\phi_ *$ term dominates $f_{\phi_*}^1$.
Thus, we can construct the following observable $\mathbb{O}_1^c$\,:
\begin{eqnarray}
\mathbb{O}_1^c & \equiv &
\left|\sigma_1^{}\!\int\!\! \di\theta \di\theta_*^{}\di\phi_*^{} \di M_*^{}\,
f_j^{(4)} \text{sign}(\cos\!2\phi_*^{})
\right|,
\label{eq:O1-G+}
\end{eqnarray}
where
\begin{eqnarray}
f_j^{(4)} \,=\, \dis
	\frac{\di^4\sigma_j^{}}
	{\,\sigma_j^{}\,\di\theta\, \di\theta_*^{} \di\phi_*^{}\di M_*^{}\,}\,.
	\end{eqnarray}
For other kinematic variables, we impose cuts on the transverse momentum
of each final state quark
$\,P_{qT}\!>0.2P_q^{}\,$
and on the invariant-mass
$\,M(q\gamma)\!>\!0.1\sqrt{s}\,$
to remove the soft or collinear divergences due to the diagram
in Fig.\,\ref{feyndiag}(c), and
$|M(q\bar q)\!-\!M_Z^{}|\!<\!10$\,GeV
to satisfy the nearly on-shell condition for the $Z$ boson.
We also place cuts on the polar scattering angle $\theta$\,
that are the same as in \cite{Ellis:2019zex}.
For the $\phi_*$ distribution,
we require $|\cos2\phi_*|\!>0.394$\,.
We use the Monte Carlo method to compute numerically
the contributions of all the relevant diagrams in Fig.\ref{feyndiag}.
In the case of hadronic decays $\,Z\!\to\!q\bar{q}$\,
with up-type-quark final states $q=u,c$, we obtain the following results for
the SM cross section $\sigma_0^c$
and the interference contribution $\mathbb{O}_1^c$
of the nTGC operator $\mathcal{O}_{G+}^{}$,
\beqs
\begin{eqnarray}
\vspace*{-1mm}
\label{eq:Z4-u-}
\sqrt{s}=250\,\text{GeV}, &~~~&
\(\sigma_0^c,\, \mathbb{O}_1^c\)
=\left(\!371,\, 0.68\!\left(\!\frac{\text{TeV}}{\Lambda}\right)^{\!\!4}\right)
\!\text{fb}\,,
\hspace*{12mm}
\\
\sqrt{s}=500\,\text{GeV}, &~~~&
\(\sigma_0^c,\,\mathbb{O}_1^c\)
=\left(\!81.1,\, 3.30\!\left(\!\frac{\text{TeV}}{\Lambda}\right)^{\!\!4}\right)
\!\text{fb}\,,
\\
\sqrt{s}=1\,\text{TeV}, &~~~&
(\sigma_0^c,\,\mathbb{O}_1^c)
= \left(\!20.0,\, 13.9\!\left(\!\frac{\text{TeV}}{\Lambda}\right)^{\!\!4}\right)
\!\text{fb}\,,
\\
\sqrt{s}=3\,\text{TeV}, &~~~&
(\sigma_0^c,\,\mathbb{O}_1^c)
= \left(2.30,\, 128\!\left(\!\frac{\text{TeV}}{\Lambda}\right)^{\!\!4}\right)\!
\text{fb}\,,
\\
\sqrt{s}=5\,\text{TeV}, &~~~&
(\sigma_0^c,\, \mathbb{O}_1^c)
= \left(\!0.838,\, 355\!\left(\!\frac{\text{TeV}}{\Lambda}\right)^{\!\!4}\right)
\!\text{fb}\,.
\end{eqnarray}
\eeqs
In the above and following numerical analyses, we have computed
$\,\mathbb{O}_1^c\,$ and $\,\sigma_0^c\,$
for the reaction $\,e^-e^+\!\to\!q\,\bar{q}\,\ga\,$,\,
as shown in Fig.\,\ref{fig:2}.
For hadronic decays $\,Z\!\to q\bar{q}$\, with the down-type-quark
final states $q=d,s,b$,
we derive the following results:
\beqs
\begin{eqnarray}
\sqrt{s}=250\,\text{GeV}, &~~~&
(\sigma_0^c,\, \mathbb{O}_1^c)
=\left(\!472,\, 0.878\!\left(\!\frac{\text{TeV}}{\Lambda}\right)^{\!\!4}\right)
\!\text{fb}\,,
\hspace*{12mm}
\\
\sqrt{s}=500\,\text{GeV}, &~~~&
(\sigma_0^c,\,\mathbb{O}_1^c)
=\left(\!103,\, 4.24\!\left(\!\frac{\text{TeV}}{\Lambda}\right)^{\!\!4}\right)
\!\text{fb}\,,
\\
\sqrt{s}=1\,\text{TeV}, &~~~&
(\sigma_0^c,\,\mathbb{O}_1^c)
= \left(\!25.5,\, 18\!\left(\!\frac{\text{TeV}}{\Lambda}\right)^{\!\!4}\right)
\!\text{fb}\,,
\\
\sqrt{s}=3\,\text{TeV}, &~~~&
(\sigma_0^c,\,\mathbb{O}_1^c)
= \left(\!2.95,\, 165\!\left(\!\frac{\text{TeV}}{\Lambda}\right)^{\!\!4}\right)\!
\text{fb}\,,
\\
\sqrt{s}=5\,\text{TeV}, &~~~&
(\sigma_0^c,\, \mathbb{O}_1^c)
= \left(\!1.07,\, 458\!\left(\!\frac{\text{TeV}}{\Lambda}\right)^{\!\!4}\right)
\!\text{fb}\,.
\end{eqnarray}
\eeqs
Using the above results, we derive the signal significances
$\SZZ_u^{}\,(\SZZ_d^{})$ for the two types of quark final state
and their combined signal significance $\SZZ$\,,
\beqs
\begin{eqnarray}
\label{eq:Zq}
\SZZ_{u,d} &\!\!=\!\!&
\frac{S_{}}{\sqrt{B\,}\,}=
\frac{\,\mathbb{O}_{1(u,d)}^c\,}
{\sqrt{\sigma_{0(u,d)}^{c}\,}\,}\!\times\!
\sqrt{\mL\!\times\!\epsilon\,}\,,
\\[1mm]
\label{eq:Z-combined}
\SZZ &\!\!\simeq\!\!& \sqrt{2\SZZ_u^2\!+3\SZZ_d^2\,}\,,
\end{eqnarray}
\eeqs
where $\,\epsilon\,$ denotes the detection efficiency and the signal significance
$\SZZ_u$\,($\SZZ_d$) corresponds to the contribution from each
final state of the up-type quarks (down-type quarks).
In Eq.\eqref{eq:Z-combined}, the coefficient~2\, for
$\SZZ_u^2$ denotes the contributions of the two up-type quarks
$\,q=u,c\,$ and the coefficient 3 for
$\SZZ_d^2$ denotes the contributions of three down-type quarks
$\,q\!=\!d,s,b$\,.
The combined signal significance $\SZZ$
has the following values for each $e^+e^-$ collider energy, with
a sample integrated luminosity of $\,\mL\!= 2$\,ab$^{-1}$ at each energy:
\beqs
\label{eq:Z4-q}
\begin{eqnarray}
\sqrt{s}=250\,\text{GeV}, &~~~&
	\mathcal{Z} =\,
	3.85\!\(\!\frac{\,\text{TeV}\,}{\Lambda}\!\)^{\!\!4}\!\times
	\!\sqrt{\epsilon\,}\,,
	\hspace*{12mm}
	\\
	\sqrt{s}=500\,\text{GeV}, &~~~&
	\mathcal{Z} =\,
	2.49\!\left(\!\frac{\,2\text{TeV}\,}{\Lambda}\!\)^{\!\!4}\! \times
	\!\sqrt{\epsilon\,}\,,
	\\
	\sqrt{s}=1\,\text{TeV}, &~~~&
	\mathcal{Z} =\,
	4.19\!\(\!\frac{\,3\text{TeV}\,}{\Lambda}\!\)^{\!\!4}\!\times
	\!\sqrt{\epsilon\,}\,,
	\\
	\sqrt{s}=3\,\text{TeV}, &~~~&
	\mathcal{Z} \,=\,
	2.24\!\left(\!\frac{\,8\text{TeV}\,}{\Lambda}\!\right)^{\!\!4}\!\times
	\!\sqrt{\epsilon\,}\,,
	\\
	\sqrt{s}=5\,\text{TeV}, &~~~&
	\mathcal{Z} =\,
	4.22\!\(\!\frac{\,10\text{TeV}\,}{\Lambda}\!\)^{\!\!4}\! \times
	\!\sqrt{\epsilon\,}\,.
	\end{eqnarray}
\eeqs
%

\vspace*{1mm}

\begin{figure*}[t]
\vspace*{-10mm}
\centering
\hspace*{-5mm}
\includegraphics[width=7.8cm,height=7cm]{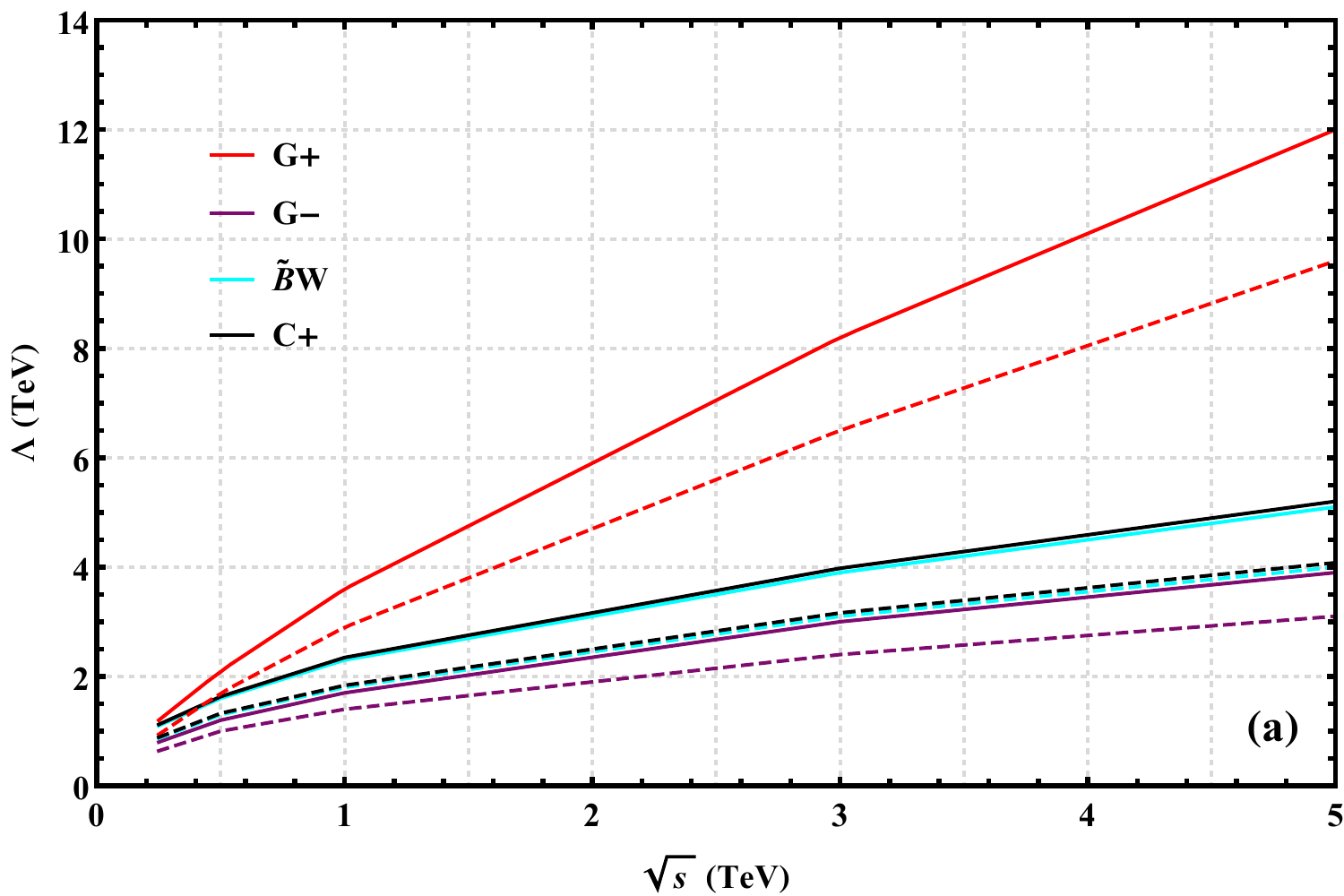}
\hspace*{2mm}
\includegraphics[width=7.8cm,height=7cm]{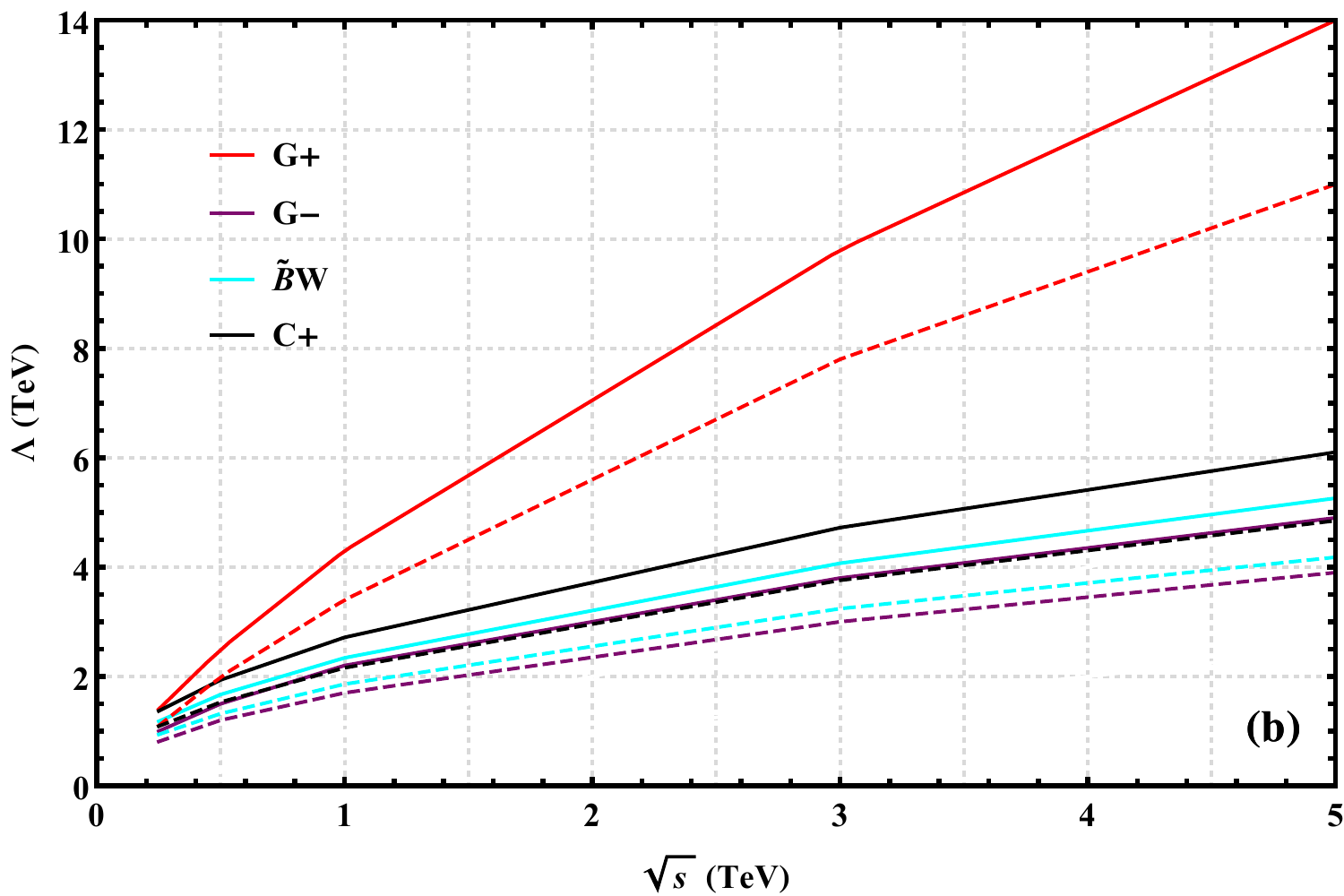}
\\[1.5mm]
\includegraphics[width=7.8cm,height=7cm]{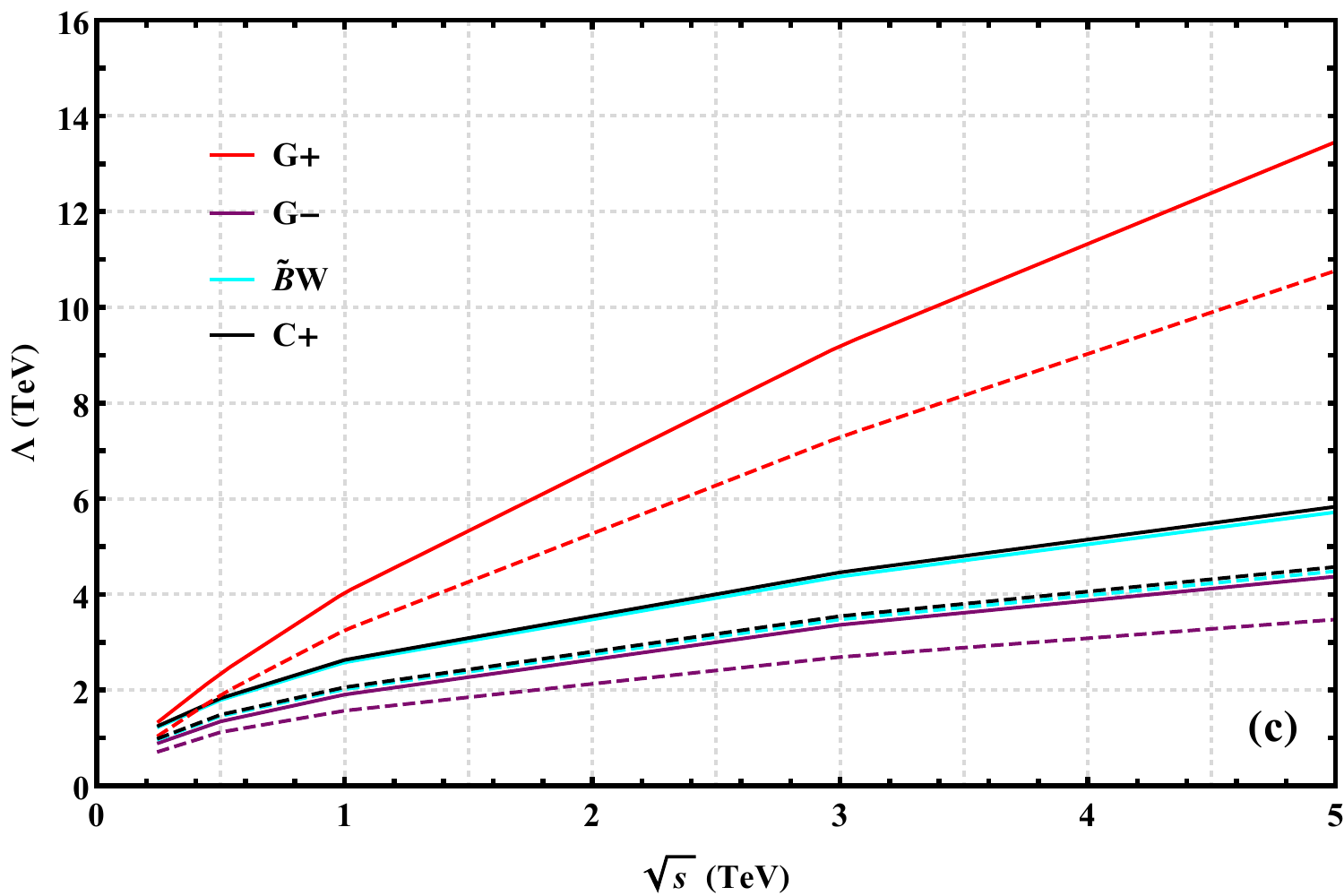}
\hspace*{2mm}
\includegraphics[width=7.8cm,height=7cm]{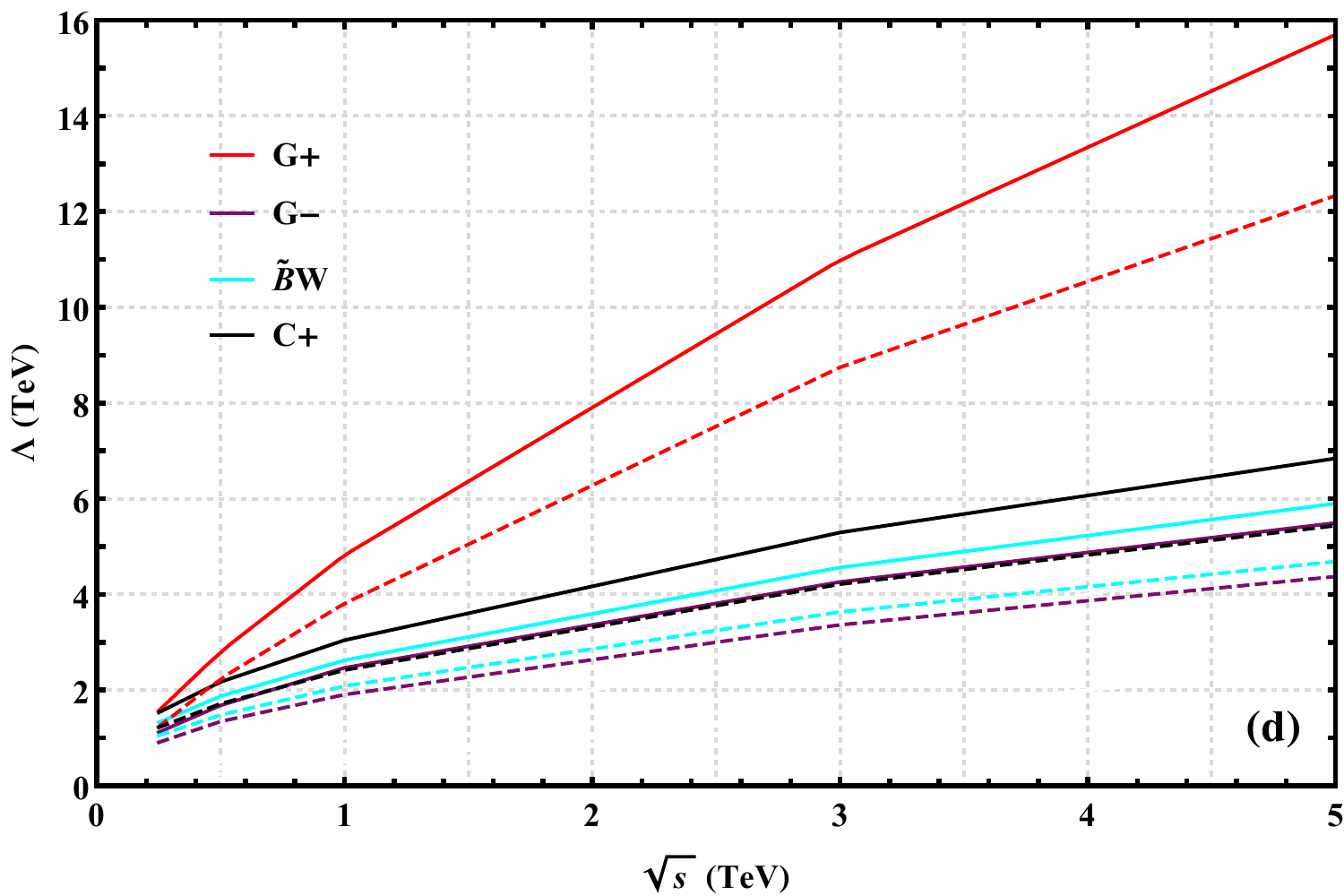}	
\vspace*{-3mm}
\caption{\small Reaches of the new physics scale $\cut$ as functions
of the $e^+e^-$ collision energy $\sqrt{s\,}$\,.
In each plot, the combined sensitivities are presented at
$\,2\sigma$ (solid curves) and $5\sigma$ (dashed curves) levels,
and for each individual dimension-8 operator among
$(\OGP,\,\OGM,\,\OBW,\OCP)$ which correspond to
the (red,\,purple,\,blue,\,black) curves, respectively.
Plots\,(a) and (c) are for unpolarized $e^\mp$ beams, whereas
plots\,(b) and (d) are for polarized $e^\mp$ beams with
$(P_L^e,\, P_R^{\bar e})=(0.9,\,0.65)$.
The plots\,(a) and (b) have input a sample
integrated luminosity $\,\mL\!=\!2\,\text{ab}^{-1}$,
whereas the plots\,(c) and (d) have used as input a sample of
$\,\mL\!=\!5\,\text{ab}^{-1}$.
}
\label{fig:Z4}
\label{fig:8}
\end{figure*}

At high energies $s\!\gg\! M_Z^2$\,,\,
we can deduce the following scaling relation
for the signal significance,
\begin{eqnarray}
\label{eq:Z-scaling}
\mathcal{Z}^{}\propto
\frac{~s^{\fr{3}{2}}\,}{\,\Lambda^4\,}
\sqrt{\text{Br}(q)\!\times\!\LL\!\times\!\epsilon\,} \,,
\end{eqnarray}
Thus, for a given signal significance $\SZZ^{}$, the corresponding reach of
the new physics scale $\Lambda$ scales as follows:
\begin{eqnarray}
\cut ~\propto~
\(\!\sqrt{s\,}\)^{\!\!\fr{3}{4}}_{}\!\times\!
\frac{\,\left[\,\text{Br}(q)\!\times\!\mathcal{L}\!\times\!\epsilon\,
        \right]^{\frac{1}{8}}_{}\,}
{\,\mathcal{Z}^{\frac{1}{4}}_{}} \,,
%
\label{eq:lam4}
\end{eqnarray}	
where we denote the hadronic branching fraction
$\,\text{Br}(q)\!\equiv\!\text{Br}[Z\!\!\to\! q\bar{q}]\,$.\,
For the dependence on $\,\text{Br}(q)\,$ in Eq.\eqref{eq:lam4},
we have assumed that the SM backgrounds are dominated by the irreducible
background of the diagram\,(b) in Fig.\,\ref{fig:2}.
Under kinematical cuts to single out the on-shell $Z\ga$ final state,
we find that the other background of diagram\,(c) can be sufficiently suppressed,
and thus the scaling relation $\,\Lambda\!\propto\! [\text{Br}(q)]^{1/8}\,$
works well.
In addition, we also note that Eq.\eqref{eq:lam4} exhibits a scaling relation
$\,\Lambda\!\propto\!\ep^{1/8}$.
This means that the sensitivity reach of $\cut$,
as shown in Tables\,\ref{tab:1}-\ref{tab:3} and Fig.\,\ref{fig:8},
should be rather insensitive to the detection efficiency $\,\ep\,$,
where we input an ideal detection efficiency $\,\ep =\!100\%$\,
for illustration.
The overall efficiency of detecting the final state $\gamma$
and $Z$ (with $Z\to q\bar{q}$ or $Z\to \ell\bar{\ell}$)
is expected to be at the level of $95\%$ \cite{efficiency}.
As an illustration, if we reduce the assumed detection efficiency from the ideal value
$\,\ep =100\%\,$ to a lower value $\,\ep \!\in\! (90\!-\!95)\%$\,,\,
the reach for $\Lambda$ would be reduced only slightly,
by about $(1.3-0.6)\%$\,. Hence the sensitivity reaches
shown in Tables\,\ref{tab:1}-\ref{tab:3} and Fig.\,\ref{fig:8}
would be largely unchanged.

\begin{table}
\begin{center}
\begin{tabular}{cccccccccc}
\hline\hline
&&&&&&&&& \\[-3.5mm]
$\sqrt{s\,}$ & $\mL$\,(ab$^{-1}$) & $\Lambda^{2\sigma}_{G+}$&$\Lambda^{5\sigma}_{G+}$&$\Lambda^{2\sigma}_{G-}$&$\Lambda^{5\sigma}_{G-}$&$\Lambda^{2\sigma}_{\!\widetilde{B}W}$&$\Lambda^{5\sigma}_{\!\widetilde{B}W}$&$\Lambda^{2\sigma}_{C+}$&$\Lambda^{5\sigma}_{C+}$\\
&&&&&&&&& \\[-3.7mm]
\hline
&&&&&&&&& \\[-4mm]
250\,GeV & 2 &	1.2 & 0.94 & 0.8 & 0.64 & 1.1 & 0.87 & 1.1 & 0.87 \\
    & 5 &	1.3 & 1.0 & 0.9 & 0.72 & 1.2 & 0.97 & 1.2 & 0.97 \\
&&&&&&&&& \\[-3.8mm]
\hline
&&&&&&&&& \\[-3.8mm]
500\,GeV & 2 &	2.1 & 1.7 & 1.2 & 1.0 & 1.6 & 1.3 & 1.6 & 1.3 \\
    & 5 &	2.3 & 1.9 & 1.3 & 1.1 & 1.8 & 1.4 & 1.8 & 1.4 \\
&&&&&&&&& \\[-3.8mm]
\hline
&&&&&&&&& \\[-3.8mm]
1\,TeV & 2	&3.6 & 2.9 & 1.7 & 1.4 & 2.3 & 1.8 & 2.3 & 1.8 \\
  & 5	&3.9 & 3.2 & 1.9 & 1.6 & 2.6 & 2.0 & 2.6 & 2.0 \\
&&&&&&&&& \\[-3.8mm]
\hline
&&&&&&&&& \\[-3.8mm]
3\,TeV & 2	&8.2 & 6.5 & 3.0 & 2.4 & 3.9 & 3.1 & 3.9 & 3.1 \\
  & 5	&9.2 & 7.2 & 3.3 & 2.7 & 4.3 & 3.5 & 4.4 & 3.4 \\
&&&&&&&&& \\[-3.8mm]
\hline
&&&&&&&&& \\[-3.8mm]
5\,TeV & 2	&12.0 & 9.6 & 3.9 & 3.1 & 5.1 & 4.0 & 5.1 & 4.0 \\
  & 5	&13.4 & 10.8 & 4.4 & 3.4 & 5.7 & 4.5 & 5.7 & 4.5 \\
&&&&&&&&& \\[-4.4mm]
\hline\hline
\end{tabular}
\end{center}
\vspace*{-3mm}
\caption{\small %
Sensitivity reaches of the new physics scale $\,\cut\,$ (in TeV)
for each of the dimension-8 nTGC operators or related contact operator
$(\OGP,\,\OGM,\,\OBW,\OCP)$,
at the $2\sigma$ (exclusion) and $5\sigma$ (discovery) levels,
as obtained from the reaction $\,e^-e^+\!\!\to\!Z\ga\!\to\! q\bar{q}\ga$\,
at different collider energies with unpolarized $e^\mp$ beams.	
For illustration, we have input two sample representative integrated luminosities
$\,\LL\!=\!2\,\text{ab}^{-1}\!$ {and}
$5\,\text{ab}^{-1}\!$.
}
\vspace*{2mm}
\label{tab:all}
\label{tab:1}
\end{table}

\vspace*{1mm}

Using the combined signal significances given in Eq.\eqref{eq:Z4-q},
we can evaluate numerically the sensitivity reaches on the new physics scale
$\,\cut\,$ associated with the dimension-8 nTGC operator $\,\mO_{G+}^{}\,$
assuming a sample integrated luminosity $\,\mathcal{L}=2\,\text{ab}^{-1}$.
We present in Table\,\ref{tab:1}
our findings for the sensitivity reaches of $\,\cut\,$ (in TeV)
at the $2\sigma$ level (3rd column) and $5\sigma$ level (4th column),
respectively.
It is very impressive to see that the new physics scale can be probed up to
$\,\cut = 1.2\,(0.94)$\,TeV\, at the $\,2\sigma\,(5\sigma)$ level
for the collider energy $\sqrt{s\,}=250$\,GeV, and
$\,\cut = 8.2\,(6.5)$TeV at the $\,2\sigma\,(5\sigma)$ level
for $\sqrt{s\,}=3$\,TeV.
At a collision energy $\sqrt{s\,}=5$\,TeV,
we see that the new physics scale can be probed up to
$\,\cut = 12\,(9.6)$TeV
at the $\,2\sigma\,(5\sigma)$ level,
which is $\,\cut = O(10\,\text{TeV})$.
Eq.\eqref{eq:lam4} shows that the new physics scale $\,\cut$\, has
a rather weak dependence on the signal significance $\mathcal{Z}$ via
$\,\cut\propto \mathcal{Z}^{-1/4}$.\,
Thus, the lower bounds on $\cut$ at $2\sigma$ and $5\sigma$ levels
are connected by
\beqa
\label{eq:Lambda-5/2sigma}
\frac{\,\cut_{5\sigma}\,}{\cut_{2\sigma}}
\,= \(\frac{2}{5}\)^{\!\!1/4}\! \simeq 0.80\,.
\eeqa
These two bounds differ by only 20\%, so they are quite close,
as expected.

\vspace*{1mm}

\begin{table}[t]
\begin{center}
\begin{tabular}{cccccccccc}
\hline\hline
&&&&&&&&&\\[-3.5mm]
$\sqrt{s\,}$ & $\mL$\,(ab$^{-1}$)
& $\Lambda^{2\sigma}_{G+}$ &$\Lambda^{5\sigma}_{G+}$ & $\Lambda^{2\sigma}_{G-}$
& $\Lambda^{5\sigma}_{G-}$ & $\Lambda^{2\sigma}_{\!\widetilde{B}W}$
& $\Lambda^{5\sigma}_{\!\widetilde{B}W}$ &
$\Lambda^{2\sigma}_{C+}$&$\Lambda^{5\sigma}_{C+}$\\
&&&&&&&&& \\[-3.7mm]
\hline
&&&&&&&&& \\[-3.8mm]
250\,GeV & 2 &	1.4 & 1.1 & 1.0 & 0.81 & 1.2 & 0.94 & 1.4 & 1.1 \\
         & 5 &	1.6 & 1.2 & 1.1 & 0.89 & 1.3 & 1.0 & 1.6 & 1.2 \\
&&&&&&&&& \\[-3.8mm]
\hline
&&&&&&&&& \\[-3.8mm]
500\,GeV & 2 &	2.5 & 2.0 & 1.5 & 1.2 & 1.7 & 1.3 & 2.0 & 1.5 \\
         & 5 &	2.7 & 2.2 & 1.7 & 1.3 & 1.9 & 1.4 & 2.2 & 1.7 \\
&&&&&&&&& \\[-3.8mm]
\hline
&&&&&&&&& \\[-3.8mm]
1\,TeV   & 2 &	4.3 & 3.4 & 2.2 & 1.7 & 2.3 & 1.9 & 2.6 & 2.2 \\
         & 5 &	4.7 & 3.7 & 2.4 & 1.9 & 2.6 & 2.1 & 2.9 & 2.4 \\
&&&&&&&&& \\[-3.8mm]
\hline
&&&&&&&&& \\[-3.8mm]
3\,TeV   & 2 &	9.8 & 7.8 & 3.8 & 3.0 & 4.1 & 3.2 & 4.8 & 3.7 \\
         & 5 &	11.0 & 8.6 & 4.2 & 3.3 & 4.5 & 3.6 & 5.2 & 4.1 \\
&&&&&&&&& \\[-3.8mm]
\hline
&&&&&&&&& \\[-3.8mm]
5\,TeV   & 2 &	14.2 & 11.3 & 4.9 & 3.9 & 5.3 & 4.2 & 6.1 & 4.9 \\
         & 5 &	15.9 & 12.7 & 5.5 & 4.4 & 5.9 & 4.7 & 6.8 & 5.5 \\
&&&&&&&&& \\[-4.4mm]
\hline\hline
\end{tabular}
\end{center}
\vspace*{-3mm}
\caption{{\small
Sensitivity reaches of the new physics scale $\,\cut\,$ (in TeV)
for each of the dimension-8 nTGC operators or related contact operator
$(\OGP,\,\OGM,\,\OBW,\OCP)$,
at the $2\sigma$ (exclusion) and $5\sigma$ (discovery) levels,
as obtained from the reaction $\,e^-e^+\!\!\to\!Z\ga\!\to\! q\bar{q}\ga$\,
at different collider energies. 
This is for {\it polarized $e^\mp$ beams}
with $(P_L^e,\, P_R^{\bar e})=(0.9,\,0.65)$.
All other inputs are the same as in Table\,\ref{tab:1}.
}}
\vspace*{2mm}
\label{tab:allpol}
\label{tab:2}
\end{table}

Next, we study the effects of $e^\mp$ beam polarizations.
We define $P_L^e$ ($P_R^{\bar e}$) as
the fractions of left-handed (right-handed) electrons (positrons)
in the beam,$\!$\footnote{%
The degree of longitudinal beam polarization for $e^-$ or $e^+$
is defined as
$\,\widehat{P}\!=\!P_R^{}\!-\!P_L^{}$ \cite{eePol}.
Since the sum of left-handed and right-handed fractions
equals one ($P_L^{}\!+\!P_R^{}=1$),
we can express the left-handed and right-handed fractions of
$e^-$ and $e^+$ as,
$P_{L,R}^e\!=\!\fr{1}{2}(1\!\mp\!\widehat{P}^e)$\, and
$P_{L,R}^{\bar{e}}\!=\!\fr{1}{2}(1\!\mp\!\widehat{P}^{\bar{e}})$,\,
respectively. For instance, the unpolarized $e^-$ or $e^+$ beam has
a vanishing degree of polarization $\,\widehat{P}\!=\!0$\,,
while a polarized $e^-$ beam with a fraction $P_L^e=90\%$
has $\,\widehat{P}^e\!=\!-0.8$\,
and a polarized $e^+$ beam with a fraction $P_R^{\bar{e}}=0.65\%$
has $\,\widehat{P}^{\bar{e}}\!=\!0.3$\,.
}
where the unpolarized $e^-$ ($e^+$)
beam has 50\% left-handed $e^-$  (right-handed $e^+$).
Thus, we can substitute
$\,(c_L^2,\, c_R^2)\to
\(4P_L^e P_R^{\bar e}c_L^2,\, 4(1\!-\!P_L^e) (1\!-\!P_R^{\bar e})c_R^2\)$\,
in the above formulae for unpolarized beams
to obtain the results for partially-polarized beams.
With these, we derive the following formulae
for the case of the partially-polarized $e^\mp$ beams:
\beqs
\begin{eqnarray}
\mathbb{O}_1^c(P_L^e,P_R^{\bar e}) &\!\!=\!\!&
4\,P_L^e P_R^{\bar e}\,\mathbb{O}_1^c(0.5,0.5) \,,
\\[2mm]
\sigma_0^c(P_L^e,P_R^{\bar e}) &\!\!=\!\!&
4\frac{\,P_L^e P_R^{\bar e}c_L^2\!+\!
(1\!-\!P_L^e)(1\!-\!P_R^{\bar e})c_R^2\,}{c_L^2\!+c_R^2}
\,\sigma_0^c(0.5,0.5) \,.
\end{eqnarray}
\eeqs
Using the above equations,
we can estimate the sensitivity reaches for the case of
the partially polarized $e^\mp$ beams.
We input the nominal polarizations
$(P_L^e,\,P_R^{\bar e}) = (0.9,\,0.65)$\,
for the numerical analyses.

{\small
\tabcolsep 1pt
\begin{table}[t]
{\small
\begin{center}
\begin{tabular}{ccccccccc}
\hline\hline
&&&&&&&&\\[-3.5mm]
$\sqrt{s\,}\,$
& $\Lambda^{2\sigma}_{G+}$ &$\Lambda^{5\sigma}_{G+}$ & $\Lambda^{2\sigma}_{G-}$
& $\Lambda^{5\sigma}_{G-}$ & $\Lambda^{2\sigma}_{\!\widetilde{B}W}$
& $\Lambda^{5\sigma}_{\!\widetilde{B}W}$ &
$\Lambda^{2\sigma}_{C+}$&$\Lambda^{5\sigma}_{C+}$\\
&&&&&&&& \\[-3.7mm]
\hline
&&&&&&&& \\[-3.8mm]
0.25  & ({\blu 1.3},\,{\red 1.6})
& ({\blu 1.0},\,{\red 1.2})
& ({\blu 0.9},\,{\red 1.1})
& ({\blu 0.72},\,{\red 0.89})
& ({\blu 1.2},\,{\red 1.3})
& ({\blu 0.97},\,{\red 1.0})
& ({\blu 1.2},\,{\red 1.6})
& ({\blu 0.97},\,{\red 1.2}) \\
&&&&&&&& \\[-3.8mm]
\hline
&&&&&&&& \\[-3.8mm]
0.5  &	({\blu 2.3},\,{\red 2.7})  & ({\blu 1.9},\,{\red 2.2}) & ({\blu 1.3},\,{\red 1.7})& ({\blu 1.1},\,{\red 1.3})&({\blu 1.8},\,{\red 1.9}) & ({\blu 1.4},\,{\red 1.4}) & ({\blu 1.8},\,{\red 2.2}) &({\blu 1.4},\,{\red 1.7}) \\
&&&&&&&& \\[-3.8mm]
\hline
&&&&&&&& \\[-3.8mm]
1    &	({\blu 3.9},\,{\red  4.7}) & ({\blu 3.2},\,{\red  3.7})& ({\blu 1.9},\,{\red 2.4})& ({\blu 1.6},\,{\red 1.9})& ({\blu 2.6},\,{\red 2.6})&({\blu 2.0},\,{\red  2.1})& ({\blu 2.6},\,{\red 2.9})& ({\blu 2.0},\,{\red 2.4}) \\
&&&&&&&& \\[-3.8mm]
\hline
&&&&&&&& \\[-3.8mm]
3   &({\blu 9.2},\,{\red 11.0})&({\blu 7.2},\,{\red  8.6})&({\blu 3.3},\,{\red  4.2})& ({\blu 2.7},\,{\red 3.3})& ({\blu 4.3},\,{\red 4.5}) & ({\blu 3.5},\,{\red 3.6}) & ({\blu 4.4},\,{\red 5.2}) & ({\blu 3.4},\,{\red 4.1}) \\
&&&&&&&& \\[-3.8mm]
\hline
&&&&&&&& \\[-3.8mm]
5   &\,	({\blu 13.4},\,{\red 15.9}) \,&\, ({\blu 10.8},\,{\red 12.7}) \,&\,
({\blu 4.4},\,{\red 5.5}) \,&\, ({\blu 3.4},\,{\red 4.4}) \,&\, ({\blu 5.7},\,{\red 5.9})
\,&\, ({\blu 4.5},\,{\red 4.7}) \,&\, ({\blu 5.7},\,{\red  6.8}) \,&
({\blu 4.5},\,{\red  5.5})\\
&&&&&&&& \\[-4.4mm]
\hline\hline
\end{tabular}
\end{center}
\vspace*{-3mm}
\caption{{\small
Sensitivity reaches of the new physics scale $\,\cut\,$ (in TeV)
for each of the dimension-8 nTGC operators or related contact operator
$(\OGP,\,\OGM,\,\OBW,\,\OCP )$,
at the $2\sigma$ (exclusion) and $5\sigma$ (discovery) levels,
as obtainable from the reaction $\,e^-e^+\!\!\to\!Z\ga\!\to\! q\bar{q}\ga$\,
at different collider energies $\sqrt{s\,}$ (in TeV),
with (unpolarized,\,polarized) $e^\mp$ beams as marked by (blue, red) colors
in each entry.	
We choose a sample integrated luminosity
$\,\LL\!=\!5\,\text{ab}^{-1}\!$ and the $e^\mp$ beam
polarizations $(P_L^e,\, P_R^{\bar e})=(0.9,\,0.65)$.
}}
\label{tab:3new}
}
\end{table}
}

\vspace*{1mm}

With these, we derive in Table\,\ref{tab:2}
the sensitivity reaches of $\,\cut\,$ (in TeV)
at the $2\sigma$ level (3rd column) and $5\sigma$ level (4th column),
respectively, for a polarized electron beam
($P_L^e\!=\!0.9$) and positron beam
($P_R^{\bar e}\!=\!0.65$).
Table\,\ref{tab:2} shows that for the polarized $e^\mp$ beams,
the new physics scale can be probed up to
$\,\cut = 1.4\,(1.1)$\,TeV\, at $\,2\sigma\,(5\sigma)$ level
for the collider energy $\sqrt{s\,}=250$\,GeV, and
$\,\cut = 9.8\,(7.8)$TeV at $\,2\sigma\,(5\sigma)$ level
for the collider energy $\sqrt{s\,}=3$\,TeV.
In comparison with the unpolarized case (Table\,\ref{tab:1}),
we see that when using the polarized $e^\mp$ beams, the sensitivities
to the new physics scale $\,\cut\,$ are increased by about $20\%$
for most cases except for the operator $\OBW$ whose bounds are
raised within a few percent.
For comparison, we further summarize the new physics sensitivities for
the (unpolarized,\,polarized) $e^\mp$ beams in Table\,\ref{tab:3new}
at various collider energies and with a common integrated luminosity
$\,\LL\!=\!5\,\text{ab}^{-1}\!$.

\vspace*{1mm}

We present our findings graphically in Fig.\,\ref{fig:8}, where
the reaches of the new physics scale $\cut$ are plotted as
functions of the collider energy $\sqrt{s\,}$ (in TeV)
for each individual dimension-8 operator among
$(\OGP,\,\OGM,\,\OBW,\OCP)$.
We need not to study the contact operator $\OCM$ separately
because it is equivalent to the pure gauge operator $\OGP$ under the EOM
\eqref{eq:OG+QBW-F} and for the reaction $\,e^+ e^-\!\!\to\! Z\ga\,$,
as we noted before.
In each plot, the $2\sigma$ ($5\sigma$) sensitivity reach
of the new physics scale $\cut$ is depicted by the solid (dashed) curve
for each operator,
while the (red,\,purple,\,blue,\,black) curves correspond to
operators $(\OGP,\,\OGM,\,\OBW,\OCP)$, respectively.
The plots\,(a) and (c) give results for the case of unpolarized $e^\mp$ beams,
whereas the plots\,(b) and (d) show results for the case of
polarized $e^\mp$ beams with
$(P_L^e,\, P_R^{\bar e})=(0.9,\,0.65)$.
In addition, the plots\,(a) and (b) input a sample
integrated luminosity $\,\mL\!=\!2$\,ab$^{-1}$,
while the plots\,(c) and (d) adopt another sample
input $\,\mL\!=\!5$\,ab$^{-1}$.
From the scaling relation \eqref{eq:lam4}, we see that the new physics
scale has a rather weak dependence on the integrated luminosity,
$\,\cut\propto \mathcal{L}^{1/8}\,$,
which holds for all dimension-8 effective operators.
Thus, we deduce that increasing the integrated luminosity from
$\,2$\,ab$^{-1}$ to $\,5$\,ab$^{-1}$ could enhance the sensitivity reach
of the new physics scale $\,\cut\,$ by a factor
\beqa
\label{eq:Lambda-Lumi=5/2/ab}
\frac{~\cut (5\text{ab}^{-1})~}{\cut (2\text{ab}^{-1})}
\,=\, \(\frac{\,5\,}{2}\)^{\!\!\!1/8}\simeq 1.12\,,
\eeqa
which is about 12\% improvement for both the unpolarized and polarized
$e^\mp$ beams.
This improvement is reflected in plots\,(c) and (d).
In Fig.\,\ref{fig:8}, we also present the sensitivity reaches
of $\,\cut\,$ for other nTGC operators,
which are discussed in the following subsections.

\vspace*{1mm}

In passing, we clarify the possible contributions of dimension-6
operators to the reaction $e^+e^-\!\!\to\!Z\ga$\,.
As we discussed in \cite{Ellis:2019zex} [see its Eq.\,(3.41)],
there are three Higgs-related dimension-6
operators ($\mathcal{O}_L^{(3)}$,\,$\mathcal{O}_L^{}$,\,$\mathcal{O}_R^{}$)
that contribute to this reaction via the $e^+e^-Z$ vertex.
The cutoff scales of these operators can be sensitively probed independently
by existing electroweak precision data and projected $Z$-pole measurements
at a future $Z$ factory such as CEPC, yielding
$\,\cut_6^{}\!>\! (19.2\!-\!34.5)$\,GeV
at the $2\sigma$ level in the event of a null measurement
(cf.\ Table\,4 of \cite{Ellis:2019zex}).
This can be compared to the sensitivity reaches ($\cut_8^{}$) in the present
Tables\,\ref{tab:1} and \ref{tab:2} for dimension-8 operators such as
$\OGP$, from which we see that $\,\cut_8^{}/\cut_6^{}\approx 1/(2\!-\!3.5)$\,
for $\sqrt{s}=(3\!-\!5)$~TeV, and
$\,\cut_8^{}/\cut_6^{}\approx 1/(12\!-\!22)$\,
for $\sqrt{s}=250$~GeV.
We can estimate the leading contributions of the relevant dimension-6 and dimension-8
operators to the cross section for $e^+e^-\!\!\to\!Z\ga$\, by power counting.
For instance, taking ($\mathcal{O}_L^{(3)}$,\,$\mathcal{O}_L^{}$,\,$\mathcal{O}_R^{}$)
and $\OGP$ as examples, we estimate their leading-order cross sections to be
$\,\sigma_6^{}\!\sim\! M_Z^2/(E^2\cut_6^2)\,$ and
$\,\sigma_8^{}[\mathbb{O}_1^c]\!\sim\! E^2\!/\cut_8^4\,$,
where $\,\sigma_8^{}[\mathbb{O}_1^c]$ is based on
Eqs.\eqref{eq:O1-G+} and \eqref{eq:f1-phi*}.
Thus, we estimate the ratio of their cross sections as follows:
\beqa
\frac{~\sigma_6^{}~}{\,\sigma_8^{}[\mathbb{O}_1^c]\,}
\sim \(\!\frac{M_Z^{}}{E}\!\)^{\!\!\!2}\!
\(\!\frac{\,\cut_8^{}\,}{\cut_6^{}}\!\)^{\!\!\!2}\!
\(\!\frac{\,\cut_8^{}\,}{E}\!\)^{\!\!\!2} \, ,
\label{eq:sigma6/sigma8}
\eeqa
where  $E=\!\sqrt{s}$\,.
From the above, we can estimate
$\,{\sigma_6^{}}/{\sigma_8^{}}\!\sim\!(1\!-\!4)\%$
for $\sqrt{s}=250$\,GeV,
$\,{\sigma_6^{}}/{\sigma_8^{}}\!\sim\!(0.6\!-\!2)\%$
for $\sqrt{s}=500$\,GeV,  and
$\,{\sigma_6^{}}/{\sigma_8^{}}\!\sim\!(0.1\!-\!0.4)\%$
for $\sqrt{s}=3$\,TeV.
For the dimension-8 operators other than $\OGP$,
Tables\,\ref{tab:1}-\ref{tab:3new}
show that their sensitivity bounds ($\cut_8^{}$) are generally lower,
so the ratio \eqref{eq:sigma6/sigma8} is even smaller.
Hence, we conclude that the effects of the dimension-6 operators
are negligible for our present nTGC study via the reaction
$e^+e^-\!\!\to\!Z\ga$\,.

\vspace*{1.5mm}
\subsection{\hspace*{-2mm}Probing~the~Dimension-8~Pure~Gauge~Operator~\boldmath{$\mO_{G-}^{}$}}
\vspace*{1.5mm}
\label{sec:4.2}

In this case, the leading term in the differential cross section at
$O(\Lambda^{\!-4})$\,
is proportional to
%
\beqa
\label{eq:dSigmaZA}
	&\hspace*{-5mm}&
	\Re\texttt{e}\!\!
	\left[\mathcal{T}^L_{(8)}(0\pm)\mathcal{T}^{T*}_{\sm}(\mp\pm)\right]
	\!\sin\theta\sin\theta_*^{}
\nn\\ 
	&\hspace*{-5mm}&
	\propto
	\dis\frac{v^2\!\sqrt{s\,}}{\,\Lambda^4 M_Z^{}\,}\!
	\!\left[C_1(1\!+\!\cos^2\theta)
	+C_2\cos\!\theta\cos\!\theta_*^{}
	\right]\sin^2\!\theta_*^{}\cos\phi_*^{}\,,~~~
\eeqa
where the coefficients
$\,C_1\!=\!({c_L x_L}\!+\!c_R x_R)(q_L^2\!-\!q_R^2)$\, and
$\,C_2\!=\!2(c_L x_L\!-\!c_R x_R)(q_L^2\!+\!q_R^2)$\,.
For the pure gauge operator $\mathcal{O}_{G-}$,
we find the coupling combinations
${c_L x_L}\!+\!c_R x_R\!\simeq\! -0.0088$\,
and
${c_L x_L}\!-\!c_R x_R\!\simeq\!-0.1155$\,.
We then construct the following observable $\mathbb{O}_1^c$
for the effective operator $\,\mO_{G-}^{}$\,:
%
%
\begin{eqnarray}
	\mathbb{O}_1^c \,=\,
	\left|\sigma_1^{}\!\int\!\! \di\theta \di\theta_*^{}\di\phi_*^{} \di M_*^{}\,
	f_1^{(4)}
	\text{sign}(\cos\!\theta)\,\text{sign}(\cos\!\theta_*^{})\,
	\text{sign}(\cos\!\phi_*^{})
	\right|.
\label{eq:O1cG-}
\end{eqnarray}

For the $\phi_*^{}$ distribution, we impose a cut
$\,|\cos\phi_*|>0.394$\,,
while for the other kinematic variables we place the same cuts
as in Sec.\,\ref{sec:4.1}.
With these we compute the values of the observable $\mathbb{O}_1^c$ for
the final state $\,u\bar u\ga\,$ at different collider energies,
\beqs
\vspace*{-1mm}
\begin{eqnarray}
	\sqrt{s}=250\,\text{GeV}, &~~~&
	(\sigma_0^c,\, \mathbb{O}_1^c)
	=\left(\!379,\, 0.149\!\left(\!\frac{\text{TeV}}{\Lambda}\right)^{\!\!4}\right)
	\!\text{fb}\,,
	\hspace*{12mm}
	\\
	\sqrt{s}=500\,\text{GeV}, &~~~&
	(\sigma_0^c,\,\mathbb{O}_1^c)
	=\left(\!81.2,\, 0.354\!\left(\!\frac{\text{TeV}}{\Lambda}\right)^{\!\!4}\right)
	\!\text{fb}\,,
	\\
	\sqrt{s}=1\,\text{TeV}, &~~~&
	(\sigma_0^c,\,\mathbb{O}_1^c)
	= \left(\!20.0,\, 0.728\!\left(\!\frac{\text{TeV}}{\Lambda}\right)^{\!\!4}\right)
	\!\text{fb}\,,
	\\
	\sqrt{s}=3\,\text{TeV}, &~~~&
	(\sigma_0^c,\,\mathbb{O}_1^c)
	= \left(2.31,\, 2.32\!\left(\!\frac{\text{TeV}}{\Lambda}\right)^{\!\!4}\right)\!
	\text{fb}\,,
	\\
	\sqrt{s}=5\,\text{TeV}, &~~~&
	(\sigma_0^c,\, \mathbb{O}_1^c)
	= \left(\!0.838,\, 3.89\!\left(\!\frac{\text{TeV}}{\Lambda}\right)^{\!\!4}\right)
	\!\text{fb}\,,
	\end{eqnarray}
	\eeqs
and derive the corresponding signal significances $\SZZ^{}$ as follows,
	\beqs
	\vspace*{-1mm}
	\label{eq:LZ4-ee}
	\begin{eqnarray}
	\sqrt{s}=250\,\text{GeV}, &~~~&
	\mathcal{Z}_u =\,
	5.51\!\(\!\frac{\,0.5\text{TeV}\,}{\Lambda}\!\)^{\!\!\!4}\!\times
	\!\sqrt{\epsilon\,}\,,
	\hspace*{12mm}
	\\
	\sqrt{s}=500\,\text{GeV}, &~~~&
	\mathcal{Z}_{u} =\,
	4.28\!\left(\!\frac{\,0.8\text{TeV}\,}{\Lambda}\!\)^{\!\!\!4}\! \times
	\!\sqrt{\epsilon\,}\,,
	\\
	\sqrt{s}=1\,\text{TeV}, &~~~&
	\mathcal{Z}_{u} =\,
	7.29\!\(\!\frac{\,\text{TeV}\,}{\Lambda}\!\)^{\!\!\!4}\!\times
	\!\sqrt{\epsilon\,}\,,
	\\
	\sqrt{s}=3\,\text{TeV}, &~~~&
	\mathcal{Z}_{u} \,=\,
	4.28\!\left(\!\frac{\,2\text{TeV}\,}{\Lambda}\!\right)^{\!\!\!4}\!\times
	\!\sqrt{\epsilon\,}\,,
	\\
	\sqrt{s}=5\,\text{TeV}, &~~~&
	\mathcal{Z}_{u} =\,
	4.87\!\(\!\frac{\,2.5\text{TeV}\,}{\Lambda}\!\)^{\!\!\!4}\! \times
	\!\sqrt{\epsilon\,}\,.
	\end{eqnarray}
\eeqs
In the above and following numerical analyses, we have computed
$\,\mathbb{O}_1^c\,$ and $\,\sigma_0^c\,$
for the reaction $\,e^-e^+\!\!\to\!q\,\bar{q}\,\ga\,$,\,
as shown in Fig.\,\ref{fig:2}.

\vspace*{1mm}

In the case of the final state $d\bar d\ga$ with down quarks, we obtain
the following results:
\beqs
	\begin{eqnarray}
	\sqrt{s}=250\,\text{GeV}, &~~~&
	(\sigma_0^c,\, \mathbb{O}_1^c)
	=\left(\!479,\, 0.192\!\left(\!\frac{\text{TeV}}{\Lambda}\right)^{\!\!4}\right)
	\!\text{fb}\,,
	\hspace*{12mm}
	\\
	\sqrt{s}=500\,\text{GeV}, &~~~&
	(\sigma_0^c,\,\mathbb{O}_1^c)
	=\left(\!103,\, 0.454\!\left(\!\frac{\text{TeV}}{\Lambda}\right)^{\!\!4}\right)
	\!\text{fb}\,,
	\\
	\sqrt{s}=1\,\text{TeV}, &~~~&
	(\sigma_0^c,\,\mathbb{O}_1^c)
	= \left(\!25.6,\, 0.945\!\left(\!\frac{\text{TeV}}{\Lambda}\right)^{\!\!4}\right)
	\!\text{fb}\,,
	\\
	\sqrt{s}=3\,\text{TeV}, &~~~&
	(\sigma_0^c,\,\mathbb{O}_1^c)
	= \left(2.94,\, 2.98\!\left(\!\frac{\text{TeV}}{\Lambda}\right)^{\!\!4}\right)\!
	\text{fb}\,,
\eeqa
\beqa
	\sqrt{s}=5\,\text{TeV}, &~~~&
	(\sigma_0^c,\, \mathbb{O}_1^c)
	= \left(\!1.07,\, 5.06\!\left(\!\frac{\text{TeV}}{\Lambda}\right)^{\!\!4}\right)
	\!\text{fb}\,,
	\end{eqnarray}
	\eeqs
Accordingly, we derive the corresponding signal significances,
with a sample integrated luminosity of $\,\mL = 2$\,ab$^{-1}$ at
each collision energy:
\beqs
\label{eq:LZ4-mumu}
\begin{eqnarray}
	\sqrt{s}=250\,\text{GeV}, &~~~&
	\mathcal{Z}_d =\,
	6.28\!\(\!\frac{\,0.5\text{TeV}\,}{\Lambda}\!\)^{\!\!\!4}\!\times
	\!\sqrt{\epsilon\,}\,,
	\hspace*{12mm}
	\\
	\sqrt{s}=500\,\text{GeV}, &~~~&
	\mathcal{Z}_{d} =\,
	4.87\!\left(\!\frac{\,0.8\text{TeV}\,}{\Lambda}\!\)^{\!\!\!4}\! \times
	\!\sqrt{\epsilon\,}\,,
\\
	\sqrt{s}=1\,\text{TeV}, &~~~&
	\mathcal{Z}_{d} =\,
	8.36\!\(\!\frac{\,\text{TeV}\,}{\Lambda}\!\)^{\!\!\!4}\!\times
	\!\sqrt{\epsilon\,}\,,
	\\
	\sqrt{s}=3\,\text{TeV}, &~~~&
	\mathcal{Z}_{d} \,=\,
	4.87\!\left(\!\frac{\,2\text{TeV}\,}{\Lambda}\!\right)^{\!\!\!4}\!\times
	\!\sqrt{\epsilon\,}\,,
	\\
	\sqrt{s}=5\,\text{TeV}, &~~~&
	\mathcal{Z}_{d} =\,
	5.60\!\(\!\frac{\,2.5\text{TeV}\,}{\Lambda}\!\)^{\!\!\!4}\! \times
	\!\sqrt{\epsilon\,}\,.
\end{eqnarray}
\eeqs

With the above, we first derive in Table\,\ref{tab:1}
the sensitivity reaches of $\,\cut\,$ (in TeV) for the
pure gauge operator $\mO_{G-}^{}$
at the $2\sigma$ level (5th column) and $5\sigma$ level (6th column),
respectively, for the unpolarized $e^\mp$ beams.
Table\,\ref{tab:1} shows that in this case
the new physics scale can be probed up to
$\,\cut = 0.80\,(0.64)$\,TeV\, at $\,2\sigma\,(5\sigma)$ level
for the collider energy $\sqrt{s\,}=250$\,GeV, and
$\,\cut = 3.0\,(2.4)$TeV at $\,2\sigma\,(5\sigma)$ level
for the collider energy $\sqrt{s\,}=3$\,TeV.
Then, we obtain in Table\,\ref{tab:2}
the sensitivity reaches of $\,\cut\,$ (in TeV)
at the $2\sigma$ level (5th column) and $5\sigma$ level (6th column),
respectively, for a polarized electron beam
($P_L^e\!=\!0.9$) and positron beam
($P_R^{\bar e}\!=\!0.65$).
From Table\,\ref{tab:2}, we find that in the case of
polarized $e^\mp$ beams,
the new physics scale can be probed up to
$\,\cut = 1.0\,(0.81)$\,TeV\, at $\,2\sigma\,(5\sigma)$ level
for the collider energy $\sqrt{s\,}=250$\,GeV, and
$\,\cut = 3.8\,(3.0)$TeV at $\,2\sigma\,(5\sigma)$ level
for the collider energy $\sqrt{s\,}=3$\,TeV.

\vspace*{1mm}

Finally, Fig.\,\ref{fig:8} presents the reaches of the new physics scale
$\,\cut\,$ as functions of the collider energy $\sqrt{s\,}$ (in TeV)
for the dimension-8 pure gauge operator $\mO_{G-}^{}$.\,
These are depicted by the green solid and dashed curves
at the $2\sigma$ and $5\sigma$ levels, respectively.
In this Figure, we show the results for the case of unpolarized $e^\mp$ beams
in plots\,(a) and (c); whereas the results for the case of
polarized $e^\mp$ beams with $(P_L^e,\, P_R^{\bar e})=(0.9,\,0.65)$
are given in the plots\,(b) and (d).
For comparison,  we have input a sample integrated luminosity
$\,\mL\!=\!2$\,ab$^{-1}$ in the plots\,(a) and (b),
and another sample input $\,\mL\!=\!5$\,ab$^{-1}$
in the plots\,(c) and (d).

\vspace*{1.5mm}
\subsection{\hspace{-2mm}Probing~the~Higgs-Related~Dimension-8~Operator \boldmath{$\mO_{\!\widetilde{B}W}^{}$}}
\vspace*{1.5mm}
\label{sec:4.3}

To analyze the contributions of $\mathcal{O}_{\widetilde{B}W}^{}$
via hadronic $Z$ decays, we use the same signal observable
$\,\mathbb{O}_1^c\,$ and the background fluctuation $\sqrt{\sigma_0^c\,}\,$
as we introduced in Ref.\,\cite{Ellis:2019zex}.
A major difference is that we will study the probe of
$\mathcal{O}_{\widetilde{B}W}^{}$
via hadronic decay channels of the final state $Z$ boson.
As we will show, this increases the sensitivity to a much larger cutoff scale
$\,\cut\,$, and consequently the squared contribution of $O(\cut^{-8})$
becomes negligible. Thus, we include the new physics contributions
up to $O(\cut^{-4})$ in the present analysis.

\vspace*{1mm}

For the reaction $\,e^-e^+\!\!\to\!Z\ga\,$ with hadronic decays
$\,Z\!\to\!q\bar{q}\,$,
we derive the following analytical formulae for the observable
$\,\mathbb{O}_1^c\,$
and the SM background cross section $\sigma_0^c$\,:
\beqs
\label{eq:qO1c-sigma0c}
\begin{eqnarray}	
\mathbb{O}_1^c &=&
|\sigma_1^{}|\!\(
\int_{\pi-\phi_c^{}}^{\pi+\phi_c^{}}
- \int_0^{\phi_c^{}} - \int_{2\pi-\phi_c^{}}^{2\pi}
\) \!f^1_{\phi_*}\di\phi_*^{}
\nn\\[1mm]
&\simeq&
\frac{\,3\alpha {c_X^{}}q_-^2M_Z^{}(s\!-\!M_Z^2)\!
	\sin\!\phi_c^{}\,}
{\,64s_W^{} c_W^{}q_+^2\, \cut^4\, s^{\frac 3 2}_{}}\!
\left[3 (\pi \!-\!2 \delta )(s\!+\!M_Z^2)\!
- \!(s\!-\!3 M_Z^2)\sin\!2\delta\right]\!\!\times\!\text{Br}(q) \,,
\hspace*{15mm}
\label{eq:O1c-q}
\\[1.5mm]
\sigma_0^c
&\simeq& \frac{\,2\phi_c^{}\,}{\pi}\sigma^0
\nn\\[1mm]
&=&
\frac{\,4\alpha^2c_+^2\!
	\left[2(s^2\!+\!M_Z^4)
	\ln\cot\!\frac{\delta}{2}
    -\cos\!\delta\,(s\!-\!M_Z^2)^2\right]\!\phi_c^{}\,}
{c_W^2s_W^2(s\!-\!M_Z^2)s^2}\!\times\!\text{Br}(q)\,,
\hspace*{30mm}
\end{eqnarray}
\eeqs
where for convenience we denote the coupling combination
$\,c_X^{}\!\equiv c_L^{}x_L^{}\!+c_R^{}x_R^{}\,$
and the decay branching fraction
$\,\text{Br}(q)\!\equiv\!\text{Br}[Z\!\!\to\! q\bar{q}]\!\simeq\! 69.9\%\,$.
Because the quark-related gauge coupling ratio
$\,q_-^2/q_+^2$\, is closer to 1 and
the hadronic decay branching fraction
Br$[Z\!\!\to\! q\bar q]$\, is roughly 7 times larger
than the leptonic one
\,Br$[Z\!\!\to\!\ell\bar{\ell}]\!\simeq\! 10.1\%$\,,\,
we find that $\,\mathbb{O}_1^c\,$ becomes much larger
in the hadronic channel than that in the leptonic channel.

\vspace*{1mm}

Since the operators $\OBW$ and $\OGM$ have the same $\phi_*^{}$ dependence
in their leading energy terms $\,f^1_{\phi_*^{}}$,
we may apply the same kinematic cuts as in Sec.\,\ref{sec:4.2}.
For each of the up-type quarks ($u,c$),
we compute the SM cross section $\sigma_0^c$ and the observable
$\mathbb{O}_1^c$ for different collison energies as follows:
\beqs
	\vspace*{-1mm}
	\label{eq:Z4-u}
	\begin{eqnarray}
	\sqrt{s}=250\,\text{GeV}, &~~~&
	(\sigma_0^c,\, \mathbb{O}_1^c)
	=\left(\!379,\, 0.392\!\left(\!\frac{\text{TeV}}{\Lambda}\right)^{\!\!4}\right)
	\!\text{fb}\,,
	\hspace*{12mm}
	\\
	\sqrt{s}=500\,\text{GeV}, &~~~&
	(\sigma_0^c,\,\mathbb{O}_1^c)
	=\left(\!81.2,\, 0.794\!\left(\!\frac{\text{TeV}}{\Lambda}\right)^{\!\!4}\right)
	\!\text{fb}\,,
	\\
	\sqrt{s}=1\,\text{TeV}, &~~~&
	(\sigma_0^c,\,\mathbb{O}_1^c)
	= \left(\!20.0,\, 1.65\!\left(\!\frac{\text{TeV}}{\Lambda}\right)^{\!\!4}\right)
	\!\text{fb}\,,
	\\
	\sqrt{s}=3\,\text{TeV}, &~~~&
	(\sigma_0^c,\,\mathbb{O}_1^c)
	= \left(2.31,\, 5.15\!\left(\!\frac{\text{TeV}}{\Lambda}\right)^{\!\!4}\right)\!
	\text{fb}\,,
	\\
	\sqrt{s}=5\,\text{TeV}, &~~~&
	(\sigma_0^c,\, \mathbb{O}_1^c)
	= \left(\!0.838,\, 8.60\!\left(\!\frac{\text{TeV}}{\Lambda}\right)^{\!\!4}\right)
	\!\text{fb}\,.
	\end{eqnarray}
	\eeqs
In the above and following numerical analyses, we have computed
$\,\mathbb{O}_1^c\,$ and $\,\sigma_0^c\,$
for the reaction $\,e^-e^+\!\to\!q\,\bar{q}\,\ga\,$,\,
as shown in Fig.\,\ref{fig:2}.

\vspace*{1mm}

Then, we can derive the following estimated signal significances
$\mathcal{Z}_4^u$ at each given collision energy with
an integrated luminosity $\mL = 2$\,ab$^{-1}$,
\beqs
	\vspace*{-1mm}
	\label{eq:Z4-U}
	\begin{eqnarray}
	\sqrt{s}=250\,\text{GeV}, &~~~&
	\mathcal{Z}_u =\,
	14.4\!\(\!\frac{\,0.5\text{TeV}\,}{\Lambda}\!\)^{\!\!4}\!\times
	\!\sqrt{\epsilon\,}\,,
	\hspace*{12mm}
	\\
	\sqrt{s}=500\,\text{GeV}, &~~~&
	\mathcal{Z}_{u} =\,
	9.59\!\left(\!\frac{\,0.8\text{TeV}\,}{\Lambda}\!\)^{\!\!4}\! \times
	\!\sqrt{\epsilon\,}\,,
	\\
	\sqrt{s}=1\,\text{TeV}, &~~~&
	\mathcal{Z}_{u} =\,
	16.5\!\(\!\frac{\,\text{TeV}\,}{\Lambda}\!\)^{\!\!4}\!\times
	\!\sqrt{\epsilon\,}\,,
	\\
	\sqrt{s}=3\,\text{TeV}, &~~~&
	\mathcal{Z}_{u} \,=\,
	9.47\!\left(\!\frac{\,2\text{TeV}\,}{\Lambda}\!\right)^{\!\!4}\!\times
	\!\sqrt{\epsilon\,}\,,
	\\
	\sqrt{s}=5\,\text{TeV}, &~~~&
	\mathcal{Z}_{u} =\,
	10.8\!\(\!\frac{\,2.5\text{TeV}\,}{\Lambda}\!\)^{\!\!4}\! \times
	\!\sqrt{\epsilon\,}\,.
	\end{eqnarray}
	\eeqs
For each of the down-type quarks ($d, s, b$), we arrive at
\beqs
\label{eq:Z4-d}
\begin{eqnarray}
	\sqrt{s}=250\,\text{GeV}, &~~~&
	(\sigma_0^c,\, \mathbb{O}_1^c)
	=\left(\!479,\, 0.702\!\left(\!\frac{\text{TeV}}{\Lambda}\right)^{\!\!4}\right)
	\!\text{fb}\,,
	\hspace*{12mm}
	\\
	\sqrt{s}=500\,\text{GeV}, &~~~&
	(\sigma_0^c,\,\mathbb{O}_1^c)
	=\left(\!103,\, 1.42\!\left(\!\frac{\text{TeV}}{\Lambda}\right)^{\!\!4}\right)
	\!\text{fb}\,,
\\
	\sqrt{s}=1\,\text{TeV}, &~~~&
	(\sigma_0^c,\,\mathbb{O}_1^c)
	= \left(\!25.6,\, 2.98\!\left(\!\frac{\text{TeV}}{\Lambda}\right)^{\!\!4}\right)
	\!\text{fb}\,,
   \\
	\sqrt{s}=3\,\text{TeV}, &~~~&
	(\sigma_0^c,\,\mathbb{O}_1^c)
	= \left(2.94,\, 9.28\!\left(\!\frac{\text{TeV}}{\Lambda}\right)^{\!\!4}\right)\!
	\text{fb}\,,
	\\
	\sqrt{s}=5\,\text{TeV}, &~~~&
	(\sigma_0^c,\, \mathbb{O}_1^c)
	= \left(\!1.07,\, 15.5\!\left(\!\frac{\text{TeV}}{\Lambda}\right)^{\!\!4}\right)
	\!\text{fb}\,.
	\end{eqnarray}
	\eeqs
Accordingly, we derive the signal significances $\mathcal{Z}^d$
at these collision energies with
a sample integrated luminosity $\mL = 2$\,ab$^{-1}$,
\beqs
	\label{eq:Z4-D}
	\begin{eqnarray}
	\sqrt{s}=250\,\text{GeV}, &~~~&
	\mathcal{Z}_d =\,
	23.0\!\(\!\frac{\,0.5\text{TeV}\,}{\Lambda}\!\)^{\!\!4}\!\times
	\!\sqrt{\epsilon\,}\,,
	\hspace*{12mm}
	\\
	\sqrt{s}=500\,\text{GeV}, &~~~&
	\mathcal{Z}_{d} =\,
	15.3\!\left(\!\frac{\,0.8\text{TeV}\,}{\Lambda}\!\)^{\!\!4}\! \times
	\!\sqrt{\epsilon\,}\,,
	\\
	\sqrt{s}=1\,\text{TeV}, &~~~&
	\mathcal{Z}_{d} =\,
	26.4\!\(\!\frac{\,\text{TeV}\,}{\Lambda}\!\)^{\!\!4}\!\times
	\!\sqrt{\epsilon\,}\,,
	\\
	\sqrt{s}=3\,\text{TeV}, &~~~&
	\mathcal{Z}_{d} \,=\,
	15.1\!\left(\!\frac{\,2\text{TeV}\,}{\Lambda}\!\right)^{\!\!4}\!\times
	\!\sqrt{\epsilon\,}\,,
	\\
	\sqrt{s}=5\,\text{TeV}, &~~~&
	\mathcal{Z}_{d} =\,
	17.2\!\(\!\frac{\,2.5\text{TeV}\,}{\Lambda}\!\)^{\!\!4}\! \times
	\!\sqrt{\epsilon\,}\,.
	\end{eqnarray}
\eeqs
With the these, we can deduce the combined signal significance
for the hadronic decay channels of $Z$ boson,
$\,\SZZ_q^{}\!=\!\sqrt{2\SZZ_u^2\!+\!3\SZZ_d^2\,}\,$.

\vspace*{1mm}

From the above, we first compute in Table\,\ref{tab:1}
the sensitivity reaches of $\,\cut\,$ (in TeV) for the
Higgs-related operator $\mO_{\widetilde{B}W}^{}$
at the $2\sigma$ level (7th column) and $5\sigma$ level (8th column),
respectively, for the unpolarized $e^\mp$ beams.
Table\,\ref{tab:1} shows that in this case
the new physics scale can be probed up to
$\,\cut = 1.1\,(0.87)$\,TeV\, at $\,2\sigma\,(5\sigma)$ level
for the collider energy $\sqrt{s\,}=250$\,GeV, and
$\,\cut = 3.9\,(3.1)$TeV at $\,2\sigma\,(5\sigma)$ level
for $\sqrt{s\,}\!=\!3$\,TeV.
Then, we obtain in Table\,\ref{tab:2}
the sensitivity reaches of $\,\cut\,$ (in TeV)
at the $2\sigma$ level (7th column) and $5\sigma$ level (8th column),
respectively, for a polarized electron beam
($P_L^e\!=\!0.9$) and positron beam
($P_R^{\bar e}\!=\!0.65$).
From Table\,\ref{tab:2}, we find that in the case of
polarized $e^\mp$ beams,
the new physics scale can be probed up to
$\,\cut = 1.2\,(0.94)$\,TeV\, at $\,2\sigma\,(5\sigma)$ level
for the collider energy $\sqrt{s\,}=250$\,GeV, and
$\,\cut = 4.1\,(3.2)$TeV at $\,2\sigma\,(5\sigma)$ level
for $\sqrt{s\,}=3$\,TeV.

\vspace*{1mm}

Finally, we present in Fig.\,\ref{fig:8} the reaches of the new physics scale
$\,\cut\,$ as functions of the collider energy $\sqrt{s\,}$ (in TeV)
for the dimension-8 Higgs-related nTGC operator $\mO_{\widetilde{B}W}^{}$.\,
These are shown by the blue solid and dashed curves
at the $2\sigma$ and $5\sigma$ levels, respectively.
In this figure, we show the results for the case of unpolarized $e^\mp$ beams
in plots\,(a) and (c), whereas the results for the case of
polarized $e^\mp$ beams with $(P_L^e,\, P_R^{\bar e})=(0.9,\,0.65)$
are given in the plots\,(b) and (d).
For comparison,  we have input a sample integrated luminosity
$\,\mL\!=\!2$\,ab$^{-1}$ in the plots\,(a) and (b),
and another sample integrated luminosity $\,\mL\!=\!5$\,ab$^{-1}$
in the plots\,(c) and (d).

\vspace*{1mm}

We note in passing that the SM contributes to the nTGC vertex
at the one-loop and higher-loop levels.
These contributions are automatically finite
because the SM is renormalizable and it does not contain any nTGC
vertex at the tree-level.
Such electroweak loop corrections are expected
to modify the SM amplitude for $e^+e^-\!\!\to\!Z\ga$ only at
${\cal O}(1\!-\!10)\%$ level.
We find from Eq.\eqref{eq:Zq} that the sensitivity reach of $\cut$
scales as $\,\cut\!\propto\! (S')^{1/4}/B^{1/8}\,$, where
the signal rate $\,S'\equiv S\cut^4\,$ is independent of the cutoff $\cut$\,.
Thus, a SM loop correction to $S'$ by an amount of ${\cal O}(\pm 10\%)$
could only cause a rather minor effect on the $\cut$ reach by about
$\pm 2.5\%$\,.
The new physics scale $\cut$ depends on the SM background rate $B$
via  $\,\cut\!\propto\! B^{-1/8}\,$,
so a SM loop corection to $B$ by a fraction of ${\cal O}(\pm 10\%)$
could affect the $\cut$ reach only by about $\pm 1.2\%$\,.
Thus, if the SM loop corrections to both the signal rate $S$
and background rate $B$ amount to about ${\cal O}(\pm 10\%)$,
it could cause a total change of the $\cut$ reach by about
$\pm(1.2\!-\!3.8)\%$, depending on whether the two corrections are combined
destructively or constructively.
Such loop effects on the sensitivity reaches of $\cut$ are rather small
and negligible for the current study.

\vspace*{1.5mm}
\subsection{\hspace{-2mm}Comparison with Probing Fermionic Contact Operators}
\vspace*{1.5mm}
\label{sec:4.4}

In the case of the fermionic contact operator $\,\mO_{C+}^{}\!=\widetilde{B}_{\!\mu\nu}^{}W^{a\mu\rho}\!
\left[D_{\!\rho}^{}(\overline{\psi_{\!L}^{}}T^a\!\gamma^\nu\!\psi_{\!L}^{}\!)
+\!D^\nu(\overline{\psi_{\!L}^{}}T^a\!\gamma_\rho^{}\psi_{\!L}^{})\right]$
as shown in Eq.\eqref{C+},
we can obtain the expression of the observable $\mathbb{O}_1^c$
by replacing the coupling factor $\,c_L^2\!+\!c_R^2$\,
(related to $\mO_{\!\widetilde B W}^{}$)
by $\,{c_L x_L}\,$ (related to $\mO_{C+}^{}$)
in Eq.(\ref{eq:qO1c-sigma0c}), where we have used the couplings defined
in Eqs.\eqref{eq:xLR-OBW}-\eqref{eq:xLR-OC+}.
Thus, for the unpolarized case, we can derive the following ratio
between the two operators
as follows:
\begin{eqnarray}
\frac{\,\mathcal{Z}_{C+}^{}\cut_{C+}^4\,}
{\,\mathcal{Z}_{\widetilde B W}^{}\cut_{\widetilde B W}^4\,}
\,=\,
\left|\frac{c_L x_L}{\,c_L^2\!+\!c_R^2\,}\right|
\simeq 1.07\,,
\end{eqnarray}
\noindent
with $\,x_L\!=\!1/2$\,.
Requiring the same signal significance
$\,\mathcal{Z}_{\widetilde B W}^{}\!\!=\!\mathcal{Z}_{C+}^{}\,$,
we deduce a relation between the two cutoff scales,
$\,\Lambda_{C+}^{}/\Lambda_{\widetilde B W}^{}\!\simeq 1.02$\,.
This shows that the sensitivity reaches of
$\,\Lambda_{\widetilde B W}^{}$ and $\,\Lambda_{C+}^{}\,$
are equal within about 2\%\,.
This feature holds very well in Table\,\ref{tab:1} for the
operators $\,\mO_{\!\widetilde B W}^{}$ and $\,\mO_{C+}^{}\,$,
where we have computed directly the $2\sigma$ and $5\sigma$ limits
for these two operators separately.

\vspace*{1mm}

For the polarized case, we derive the following ratio accordingly:
\begin{eqnarray}
\frac{\,\mathcal{Z}_{C+}^{}\cut_{C+}^4\,}
{\,\mathcal{Z}_{\widetilde B W}^{}\cut_{\widetilde B W}^4\,}
\,=\,
\left|\frac{\,P_L^eP_R^{\bar e} c_L x_L\,}
{\,P_L^eP_R^{\bar e}c_L^2\!+\!(1\!-\!P_L^e)(1\!-\!P_R^{\bar e})c_R^2\,}\right|.
\end{eqnarray}
For the given $e^\mp$ beam polarizations
$(P_L^e,\,P_R^{\bar{e}})=(0.9,\,0.65)$,
we evaluate numerically the above ratio
$\,(\mathcal{Z}_{C+}^{}\cut_{C+}^4)/
(\mathcal{Z}_{\widetilde B W}^{}\cut_{\widetilde B W}^4) =1.78\,$.
Requiring the same signal significance
$\,\mathcal{Z}_{\widetilde B W}^{}\!\!=\!\mathcal{Z}_{C+}^{}\,$,
we derive a relation between the two cutoff scales,
$\,\Lambda_{C+}^{}/\Lambda_{\widetilde B W}^{}\!\simeq 1.16$\,.
Thus, the sensitivity reach of
$\,\Lambda_{\widetilde B W}^{}$ is higher than that of $\,\Lambda_{C+}^{}$
by about 16\%\,.
Inspecting the $2\sigma$ and $5\sigma$ limits in Table\,\ref{tab:2},
we see that this relation
$\,\Lambda_{C+}^{}/\Lambda_{\widetilde B W}^{}\!\!\simeq\! 1.16$\,
indeed holds rather well.
Hence, the case with polarized $e^\mp$ beams provides a more sensitive probe of
the new physics scale of $\,\mO_{C+}^{}$
than that of $\,\mO_{\!\widetilde B W}^{}$.

\vspace*{1mm}

According to Eq.\eqref{eq:OG+QBW-F},
we see that the other combination of contact operators $\mO_{C-}$
has the same effect as $\mO_{G+}$
in the reaction $\,e^-e^+\!\to\!Z\gamma$\,,
because the other related operators with the Higgs fields
do not contribute to this reaction.
Hence we need not to discuss it separately, but just refer to our study of
$\mO_{G+}$ in Section\,\ref{sec:4.1}.

\vspace*{1mm}
\subsection{\hspace*{-2mm}Comparison with Leptonic Channels of \boldmath{$Z$} Decays}
\vspace*{1.5mm}
\label{sec:4.5}

In our previous study\,\cite{Ellis:2019zex},
we analyzed the sensitivities for probes of the nTGC operator
$\mO_{\widetilde B W}$
via leptonic $Z$ decay channels.
By replacing the coupling coefficients
$(q_L^{},\,q_R^{})$ by $(c_L^{},\,c_R^{})$,
we can obtain all the formulae for the leptonic $Z$ channels
from the corresponding formulae in the case of the hadronic $Z$ decays.
For comparison, we summarize in Table\,\ref{tab:3}
the sensitivities for probing $\mO_{G+}^{}$ and $\mO_{\widetilde B W}^{}$ in
the leptonic $Z$ decay channels,
for the case of unpolarized $e^\mp$ beams (in black color) and
the case of polarized $e^\mp$ beams (in red color)
with $(P_L^e,\, P_R^{\bar e})=(0.9,\,0.65)$.
We also assume an ideal detection efficiency
$\,\ep\!=\!100\%$ for illustration, but a lower detection efficiency has
little effect on our sensitivity reaches of $\cut$,\, as we discussed
below Eq.\eqref{eq:lam4}.

\vspace*{1mm}

The sensitivities for probing the $\mO_{G+}^{}$ and $\mO_{\widetilde B W}^{}$
in the leptonic channels are generally
lower than that in the hadronic channels.
For the pure gauge operator $\mO_{G+}^{}$,
we can derive the ratio of the two signal significances as follows,
\begin{eqnarray}
\label{eq:Zq/Zl}
\frac{\,\Lambda^q_{G+}\,}{\,\Lambda^\ell_{G+}\,} \,=\,
\(\!\frac{\,\SZZ^{\ell}_{G+}\,}{\SZZ^{q}_{G+}}\!\)^{\!\!\!\fr{1}{4}}
\!\!\(\!\frac{\,\text{Br}(Z\!\!\to\!q\bar{q})\,}
{\,\text{Br}(Z\!\!\to\!\ell\bar{\ell})\,}
\!\)^{\!\!\!\fr{1}{8}}\!\simeq 1.27 \,,
\end{eqnarray}
where we have required the same signal significance
$\,\SZZ^{\ell}_{G+}\!\!=\!\SZZ^{q}_{G+}$.\,
This is because the observable
$\,\mathbb{O}_1^c(G+)\propto f_L^2\!+\!f_R^2\,$,
where $(f_L,\,f_R)$ denote the coupling coefficients of
the left- and right-handed fermions to the weak gauge boson $Z$.
For comparison, the corresponding ratio for the Higgs-related
nTGC operator $\,\mO_{\widetilde B W}^{}$
was given in Eq.\eqref{eq:Lambda-q/Lambda-lep}.
Because the corresponding observable
$\,\mathbb{O}_1^c(\widetilde{B}W)\propto f_L^2\!-\!f_R^2$\,,\,
it leads to a much weaker sensitivity for $\mO_{\widetilde B W}^{}$ in
the leptonic channels, as shown in Table\,\ref{tab:3}.
Inspecting the $2\sigma$ and $5\sigma$ limits on the new physics scale
of $\OGP$ via hadronic and leptonic $Z$ decays in
Tables\,\ref{tab:1} and \ref{tab:3}, we find that the above relation
\eqref{eq:Zq/Zl} indeed holds rather well.

\vspace*{1mm}

Next, we compare the sensitivity reaches for the operator
$\mO_{\widetilde{B}W}^{}$ via $Z$ decays
into the hadronic and leptonic channels.
We note that the coupling coefficient of the observable $\mathbb{O}_1^c$
for the final-state quarks is much larger than that for the final-state leptons.
The case with the final state $Z\!\to\!q\bar{q}$\,
has much larger signal significance than that
of $Z\!\to\!\ell\bar{\ell}$\, that we studied previously\,\cite{Ellis:2019zex}.
For the sake of comparison, we rederive in Table\,\ref{tab:3} the limits on the
operator $\OBW$ by including its contributions up to $O(\cut^{-4})$,
which is the same as we do for the pure gauge operator $\OGP$\,.

%
\begin{table}[t]
\begin{center}
\begin{tabular}{cccccc}
\hline\hline
&&&&&\\[-3.5mm]
$\sqrt{s\,}$ & $\mL$
& $\Lambda^{\ell,2\sigma}_{G+}$ & $\Lambda^{\ell,5\sigma}_{G+}$
& $\Lambda^{\ell,2\sigma}_{\widetilde{B}W}$
& $\Lambda^{\ell,5\sigma}_{\widetilde{B}W}$
\\[1.5mm]
(energy) & (ab$^{-1}$) & (unpol,\,pol) & (unpol,\,pol)
& (unpol,\,pol) & (unpol,\,pol)
\\
&&&&&\\[-3.8mm]
\hline	
&&&&&\\[-3.8mm]
250\,GeV & 2 &	({\blu 0.93},\,\,{\red 1.1}) & ({\blu 0.74},\,\,{\red 0.87})
& ({\blu 0.56},\,\,{\red 0.65}) & ({\blu 0.44},\,\,{\red 0.51})
\\
& 5 &	({\blu 1.0},~{\red 1.2}) & ({\blu 0.83},~{\red 0.97})
& ({\blu 0.63},~{\red 0.73}) & ({\blu 0.49},~{\red 0.57})
\\
&&&&&\\[-4mm]
\hline	
&&&&&\\[-3.8mm]
500\,GeV & 2 & ({\blu 1.7},~{\red 2.0}) & ({\blu 1.3},\,{\red 1.5})
& ({\blu 0.8},~{\red 1.0}) & ({\blu 0.64},~{\red 0.78})
\\
& 5 & ({\blu 1.9},~{\red 2.2}) & ({\blu 1.4},~{\red 1.7})
& ({\blu 0.90},~{\red 1.1}) & ({\blu 0.72},~{\red 0.87})
\\
&&&&&\\[-4mm]
\hline	
&&&&&\\[-3.8mm]
1\,TeV & 2 &	({\blu 2.8},~{\red 3.3}) & ({\blu 2.3},~{\red 2.7})
& ({\blu 1.2},~{\red 1.4}) & ({\blu 0.91},~{\red 1.1})
\\
& 5 &	({\blu 3.1},~{\red 3.7}) & ({\blu 2.6},~{\red 3.0})
& ({\blu 1.3},~{\red 1.6}) & ({\blu 1.0},~{\red 1.2})
\\
&&&&&\\[-4mm]
\hline	
&&&&&\\[-3.8mm]
3\,TeV & 2 &	({\blu 6.5},~{\red 7.7}) & ({\blu 5.1},~{\red 6.0})
&  ({\blu 2.0},~{\red 2.5}) & ({\blu 1.6},~{\red 2.0})
\\
& 5 &	({\blu 7.3},~{\red 8.6}) & ({\blu 5.7},~{\red 6.7})
& ({\blu 2.2},~{\red 2.8}) & ({\blu 1.8},~{\red 2.2})
\\
&&&&&\\[-4mm]
\hline	
&&&&&\\[-3.8mm]
5\,TeV & 2 & ({\blu 9.5},~{\red 11.2}) & ({\blu 7.5},~{\red 8.8})
& ({\blu2.6},~{\red3.2}) & ({\blu2.0},~{\red2.6})
\\
& 5 &	({\blu 10.6},~{\red 12.5}) & ({\blu 8.4},~{\red 9.9})
& ({\blu 2.9},~{\red 3.6}) & ({\blu 2.2},~{\red 2.9})
\\
&&&&&\\[-4.4mm]
\hline\hline
\end{tabular}
\end{center}
\vspace*{-3mm}
\caption{\small{%
Sensitivities for probes of the new physics scale $\cut$} (in TeV)
{of the nTGC pure gauge operator $\,\mO_{G+}^{}$
in comparison with that of the Higgs-related nTGC operator
$\,\mO_{\widetilde{B}W}^{}$,
at the $2\sigma$ (exclusion) and $5\sigma$ (discovery) levels
for different dimension-8 operators,
as obtained from the reaction
$\,e^-e^+\!\!\!\to\!Z\ga\!\to\! \ell\bar{\ell}\ga$\,
at different collider energies, for}
$\,\LL\!=\!2\,\text{ab}^{-1}\!$ {and}
$\,\LL\!=\!5\,\text{ab}^{-1}\!$, respectively.
The sensitivity limits on $\cut$ are shown in pair inside the
parentheses of each entry
and correspond to the cases with (unpolarized,\,polarized) $e^\mp$
beams, which are marked with (blue,\,red) colors and under the abbreviations
(unpol,\,pol) in the second row.	
The polarized $e^\mp$ beams correspond to
$(P_L^e,\, P_R^{\bar e})=(0.9,\,0.65)$.
}
\vspace*{2mm}
\label{tab:3}
\end{table}

\vspace*{1mm}

Then, we can estimate the ratio of the significances 
for the hadronic and leptonic $Z$ decay channels as follows:
\begin{eqnarray}
\label{eq:Z4q/Z4lep}
\frac{\,\bar{\SZZ}^q_{\widetilde{B}W}\,}{\,\SZZ^{\ell}_{\widetilde{B}W}\,}
=\,
\frac{\,\mathbb{O}_1^{c,q}(\widetilde{B}W)\,}
{\mathbb{O}_1^{c,\ell}(\widetilde{B}W)}
\sqrt{\!\frac{\sigma_0^{c,\ell}}{\,\sigma_0^{c,q}\,}}
\,\simeq
\(\!\!\frac{\cut^{\ell}_{\widetilde{B}W}}
{\,\cut^q_{\widetilde{B}W}\,}\!\!\)^{\!\!\!4}
\!\(\!\frac{\,q_-^2\,}{\,c_-^2\,}\!
\sqrt{\frac{\,N_{\!c}^{}\,c_+^2\,}{\,N_{\ell}^{}\,q_+^2\,}}\,\) \!,
\end{eqnarray}
where $\,\bar{\SZZ}_q^{}\!=\!\SZZ_u^{}\,$ for up-type quarks or
$\,\bar{\SZZ}_q^{}\!=\!\SZZ_d^{}\,$ for down-type quarks, and
we have denoted the coupling factors
$\,c_\pm^2\!=\!c_L^2\!\pm\! c_R^2\,$
and $\,q_\pm^2\!=\!q_L^2\!\pm\! q_R^2\,$ as before.
In the above, $\,N_{\!c}^{}\!=\!3$\, is the color factor
of each final state quarks $q\bar{q}$\,, while $\,N_{\ell}^{}\!=\!3\,$
counts the contributions from the three types of the final-state leptons
$\,\ell^-\ell^+\,$ ($\ell =e,\mu,\tau$).
For our estimate in the last step of Eq.\eqref{eq:Z4q/Z4lep},
we have assumed that the SM background cross sections $\sigma_0^{c,q}$ and
$\sigma_0^{c,\ell}$ are dominated by irreducible backgrounds as generated from
the diagram-(b) of Fig.\,\ref{fig:2}.
But in general we have the scaling relation
$\,{\cut^{q}_{\widetilde{B}W}}/{\cut^{\ell}_{\widetilde{B}W}}
\propto \(\!{\sigma_0^{c,q}}/{\sigma_0^{c,\ell}}\!\)^{\!\!1/8}$,\,
so the ratio $\,{\cut^{q}_{\widetilde{B}W}}/{\cut^{\ell}_{\widetilde{B}W}}$\,
is rather insensitive to a small change in the SM background cross sections.
Hence we expect that our estimate in Eq.\eqref{eq:Z4q/Z4lep} should hold well,
which is indeed the case as we verify shortly.


As before, we have combined the signal significances
for the hadronic $Z$ decay channels,
\beqa
\SZZ^q_{\widetilde{B}W}\!=\!
\sqrt{2(\SZZ^u_{\widetilde{B}W}\!)^2\!+\!3(\SZZ^d_{\widetilde{B}W}\!)^2\,}\,,
\eeqa
which takes into account the contributions of the two relevant up-type quarks
$(q=u,c)$ and the three down-type quarks $(q=d,s,b)$.
With this we can extend the above formula \eqref{eq:Z4q/Z4lep} and derive
the ratio between the combined signal significance
$\,{\SZZ}^q_{\widetilde{B}W}\,$ for hadronic channels
and the signal significance
$\,\SZZ^{\ell}_{\widetilde{B}W}\,$ for leptonic channels:
\begin{eqnarray}
\label{eq:Zqtot/Zlep}
\frac{\,\SZZ^q_{\widetilde{B}W}\,}{\,\SZZ^{\ell}_{\widetilde{B}W}\,}
= \!\left[
 2\!\(\!\!\frac{\,\SZZ^u_{\widetilde{B}W}\,}{\SZZ^{\ell}_{\widetilde{B}W}}\!\!\)^{\!\!\!2}
\!+\!3\!\(\!\!\frac{\,\SZZ^d_{\widetilde{B}W}\,}{\SZZ^{\ell}_{\widetilde{B}W}}\!\!\)^{\!\!\!2}
\right]^{\!\!\fr{1}{2}}
\!\!\simeq \(\!\!\frac{\cut^{\ell}_{\widetilde{B}W}}
{\,\cut^q_{\widetilde{B}W}\,}\!\!\)^{\!\!\!4}
\!\!\(\!\!\frac{\,N_{\!c}^{}\,c_+^2\,}{\,N_{\ell}^{}\,c_-^4\,}\!\!\)^{\!\!\!\fr{1}{2}}
\!\!\(\!\!\frac{\,2u_-^4\,}{u_+^2}\!+\!\frac{\,3d_-^4\,}{d_+^2}\!\!\)^{\!\!\!\fr{1}{2}}
\!, \hspace*{15mm}
\end{eqnarray}
where $\,N_{\!c}^{}\!=\!N_\ell^{}\!=\!3\,$,
and we have used the notations
$\,u_\pm^2\!=\!\left. q^2_\pm\right|_{q=u}^{}\,$
and
$\,d_\pm^2\!=\!\left. q^2_\pm\right|_{q=d}^{}\,$
with the coupling factors $\,q^2_\pm=q_L^2\pm q_R^2\,$
defined as before.

\vspace*{1mm}

Requiring the same signal significances
$\,\SZZ^q_{\widetilde{B}W}\!\!\!=\!\SZZ^{\ell}_{\widetilde{B}W}\,$
in Eq.\eqref{eq:Zqtot/Zlep},
we can compare the sensitivity reaches
of the cutoff scales $\cut^{q}_{\widetilde{B}W}$ and
$\cut^{\ell}_{\widetilde{B}W}$
via the hadronic and leptonic $Z$ decay channels:
\beqa
\label{eq:Lambda-q/Lambda-lep}
\frac{\,\cut^{q}_{\widetilde{B}W}\,}
{\,\cut^{\ell}_{\widetilde{B}W}\,}
\,=\,
\(\!\!\frac{\,N_{\!c}^{}\,c_+^2\,}{\,N_{\ell}^{}\,c_-^4\,}\!\!\)^{\!\!\!\fr{1}{8}}
\!\!\(\!\!\frac{\,2u_-^4\,}{u_+^2}\!+\!\frac{\,3d_-^4\,}{d_+^2}\!\!\)^{\!\!\!\fr{1}{8}}
\!.
\eeqa
We have computed this ratio numerically and found a relation
$\,{\cut^{q}_{\widetilde{B}W}}/{\cut^{\ell}_{\widetilde{B}W}}
\!\simeq\! 1.96\,$.
Inspecting the sensitivity limits on $\OBW$
in Tables\,\ref{tab:1} and \ref{tab:3}
which are derived separately for the hadronic and leptonic channels of $Z$ decays,
we can readily verify that this ratio
$\,{\cut^{q}_{\widetilde{B}W}}/{\cut^{\ell}_{\widetilde{B}W}}
\!\simeq\! 1.96\,$ holds rather well.

\vspace*{1.5mm}
\subsection{\hspace*{-2mm}Correlations between Dimension-8 Operators}
\vspace*{1.5mm}
\label{sec:4.6}

In this Subsection we analyze the correlations between each pair of the
dimension-8 operators.
We construct for this purpose the following three types of
$\mathbb{O}_1^c$ observables to extract different signal terms
in the distribution $\,f_1^{(4)}$ at $O(\cut^{-4})$:
%
\beqs	
\label{eq:O-ABC}
	\begin{eqnarray}
	\mathbb{O}_A^c &\!\!=\!\!&
	\sigma_1^{}\!\int\!\! \di\theta \di\theta_*^{}\di\phi_*^{} \di M_*^{}\,
	f_1^{(4)}
	\text{sign}(\cos\!\phi_*^{})\,,
\\
	\mathbb{O}_B^c &\!\!=\!\!&
	\sigma_1^{}\!\int\!\! \di\theta \di\theta_*^{}\di\phi_*^{} \di M_*^{}\,
	f_1^{(4)}
	\text{sign}(\cos\!\phi_*^{})\text{sign}(\cos\!\theta^{})
    \text{sign}(\cos\!\theta_*^{})\,,
\\
	\mathbb{O}_C^c &\!\!=\!\!&
	\sigma_1^{}\!\int\!\! \di\theta \di\theta_*^{}\di\phi_*^{} \di M_*^{}\,
	f_1^{(4)}
	\text{sign}(\cos\!2\phi_*^{})\,.
\end{eqnarray}
\eeqs	
Using the formulae \eqref{Zgamma-},\,\eqref{eq:f-phi*},\,\eqref{Zgamma1}-\eqref{eq:f-phi*q}
and \eqref{eq:dSigmaZA}-\eqref{eq:O1cG-},
we derive expressions for these observables for each of the operators
$(\OGP,\,\OBW,\,\OGM,\,\OCP)$.
For the observables of type-A, we have
\beqs
\vspace*{-2mm}
\begin{eqnarray}
\mathbb{O}_A^c(G+) &\!\!=\!\!&
\mathbb{A}\times \fr{1}{2}c_L^{}P_L^eP_R^{\bar{e}}
(5s\!+\!M_Z^2)\Lambda^{-4}_{G+}\,,
\\[1mm]
\mathbb{O}_A^c(\widetilde{B}W) &\!\!=\!\!&
\mathbb{A}\times
3\!\left[c_L^2P_L^eP_R^{\bar{e}}\!+\!
         c_R^2(1\!-\!P_L^e)(1\!-\!P_R^{\bar{e}})\right]\!
(s\!+\!M_Z^2)\Lambda^{-4}_{\widetilde{B}W}\,,
\\[1mm]
\mathbb{O}_A^c(G-) &\!\!=\!\!&
\mathbb{A}\times 3\!\left[c_L^{}P_L^eP_R^{\bar{e}}\!+\!
                   c_R^{}(1\!-\!P_L^e)(1\!-\!P_R^{\bar{e}})\right]
s_W^2(s\!+\!M_Z^2)\Lambda^{-4}_{G-}\,, \hspace*{15mm}
\\[1mm]
\mathbb{O}_A^c(C+) &\!\!=\!\!&
\mathbb{A}\times\fr{3}{2}c_L^{}P_L^eP_R^{\bar{e}}
(s\!+\!M_Z^2)\Lambda^{-4}_{C+}\,,
\end{eqnarray}
\eeqs
whereas for the type-B observables we obtain
\beqs
\vspace*{-2mm}
\begin{eqnarray}
\mathbb{O}_B^c(G+) &\!\!=\!\!&
\mathbb{B}\times \fr 1 2c_L^{}P_L^eP_R^{\bar{e}}
(5s\!+\!M_Z^2)\Lambda^{-4}_{G+}\,,
\\[1mm]
\mathbb{O}_B^c(\widetilde{B}W) &\!\!=\!\!&
\mathbb{B}\times 3\!\left[c_L^2P_L^eP_R^{\bar{e}}\!-\!
         c_R^2(1\!-\!P_L^e)(1\!-\!P_R^{\bar{e}})\right]\!
         (s\!+\!M_Z^2)\Lambda^{-4}_{\widetilde{B}W}\,,
\\[1mm]
\mathbb{O}_B^c(G-) &\!\!=\!\!&
\mathbb{B}\times 3\left[c_L^{}P_L^eP_R^{\bar{e}}\!-\!
                   c_R^{}(1\!-\!P_L^e)(1\!-\!P_R^{\bar{e}})\right]
                   s_W^2(s\!+\!M_Z^2)\Lambda^{-4}_{G-}\,, \hspace*{15mm}
\\[1mm]
\mathbb{O}_B^c(C+) &\!\!=\!\!&
\mathbb{B}\times\fr 3 2c_L^{}P_L^eP_R^{\bar{e}}
(s\!+\!M_Z^2)\Lambda^{-4}_{C+}\,.
\end{eqnarray}
\eeqs
Finally, for the type-C observables we derive
\beqs
\begin{eqnarray}
\mathbb{O}_C^c(G+) &\!\!=\!\!&
\mathbb{C}\times \fr 1 2c_L^{}P_L^eP_R^{\bar{e}}\,s\,\Lambda^{-4}_{G+}\,,
\\[1mm]
\mathbb{O}_C^c(\widetilde{B}W) &\!\!=\!\!&
\mathbb{C}\times \!\left[c_L^2P_L^eP_R^{\bar{e}}\!-\!
         c_R^2(1\!-\!P_L^e)(1\!-\!P_R^{\bar{e}})\right]\!
M_Z^2\Lambda^{-4}_{\widetilde{B}W}\,,
\\[1mm]
\mathbb{O}_C^c(G-) &\!\!=\!\!&
\mathbb{C}\times \left[c_L^{}P_L^eP_R^{\bar{e}}\!-\!
                   c_R^{}(1\!-\!P_L^e)(1\!-\!P_R^{\bar{e}})\right]
s_W^2M_Z^2\Lambda^{-4}_{G-}\,,
\hspace*{27mm}
\\[1mm]
\mathbb{O}_C^c(C+) &\!\!=\!\!&
\mathbb{C}\times\fr{1}{2}c_L^{}P_L^eP_R^{\bar{e}}M_Z^2\Lambda^{-4}_{C+}\,.
\end{eqnarray}
\eeqs
In the above,
the values of the coefficients $\mathbb{A}$, $\mathbb{B}$ and $\mathbb{C}$
are obtained from our numerical results for the observables in Eq.\eqref{eq:O-ABC}.
The case with unpolarized $e^\mp$ beams corresponds to
$(P_L^e,\,P_R^{\bar{e}})=(0.5,\,0.5)$\,, while for the fully-polarized $e^\mp$ beams
we have $(P_L^e,\,P_R^{\bar{e}})=(1,\,1)$\,.

\vspace*{1mm}

Then, we combine contributions of the above three types
of observables $\,\mathbb{O}_\rho^c$\, ($\rho=\!A,B,C$)
and include different operators $\mO_j^{}$
($\,j\!=G+,\,\widetilde{B}W,\,G-,\,C+$)
into a global $\chi^2$ function:

\begin{eqnarray}
\label{eq:chi2}
\chi^2 \,=\, \sum_\rho\mathcal Z_\rho^2\,=\, \sum_\rho\!\frac{S_\rho^2}{B_\rho}\,=\, \sum_\rho\!\left[
2\frac{\left(\sum_\rho\mathbb{O}_{\rho}^{c(u)}(j)\right)^{\!\!2}\,}
{\sigma_{0\rho}^{c(u)}}
+3\frac{\left(\sum_\rho\mathbb{O}_{\rho}^{c(d)}(j)\right)^{\!\!2}\,}
{\sigma_{0\rho}^{c(d)}}
\right]\!\!\times\!\mL\!\times\!\ep \,,
\hspace*{6mm}
\end{eqnarray}
where we use the same notations for the signals and error bars
as in Eq.\eqref{eq:Zq}.

\begin{figure}[t]
\vspace*{-2mm}
\begin{center}
\includegraphics[height=6.8cm,width=7.5cm]{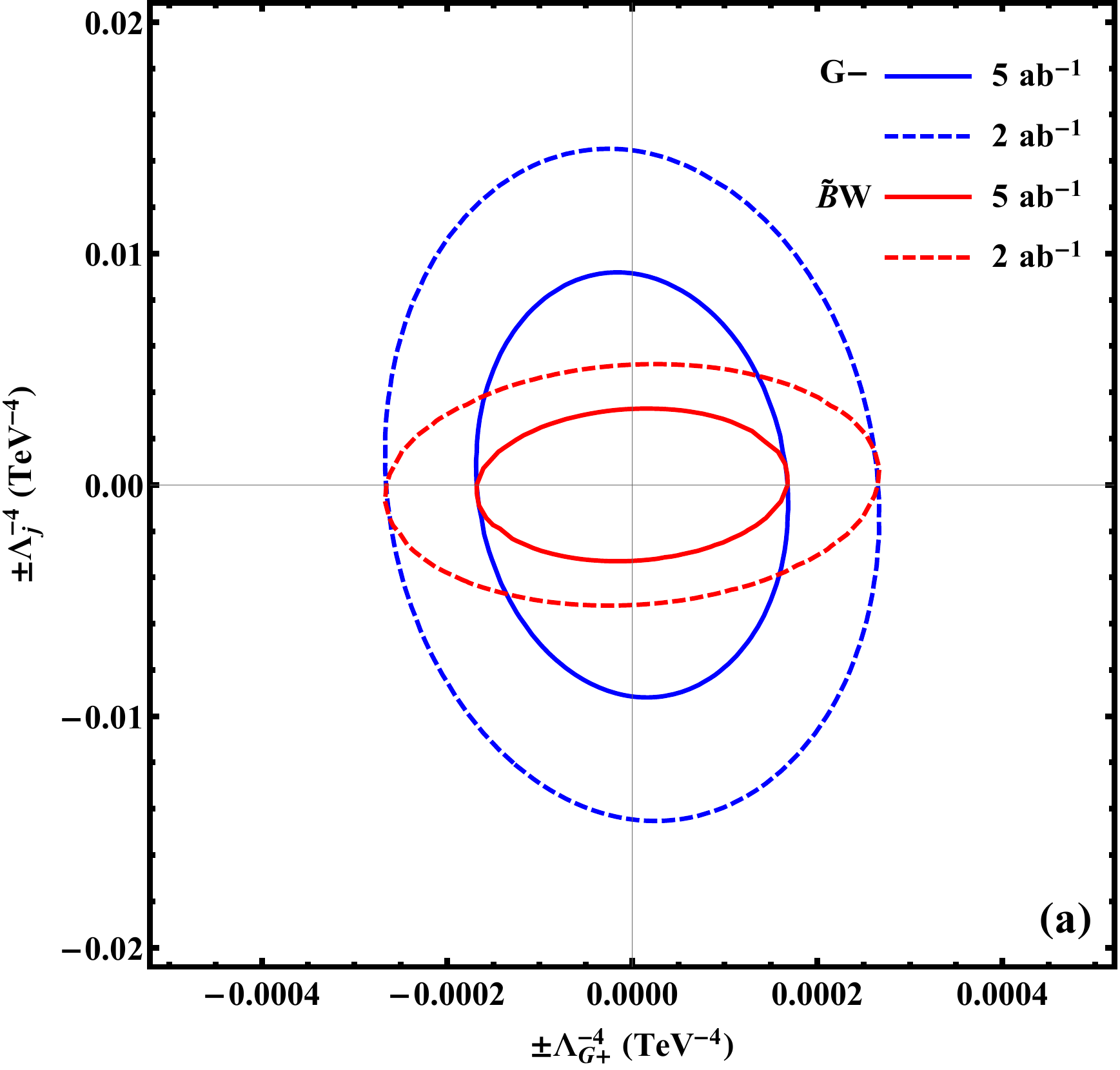}
\hspace*{1.5mm}
\includegraphics[height=6.8cm,width=7.5cm]{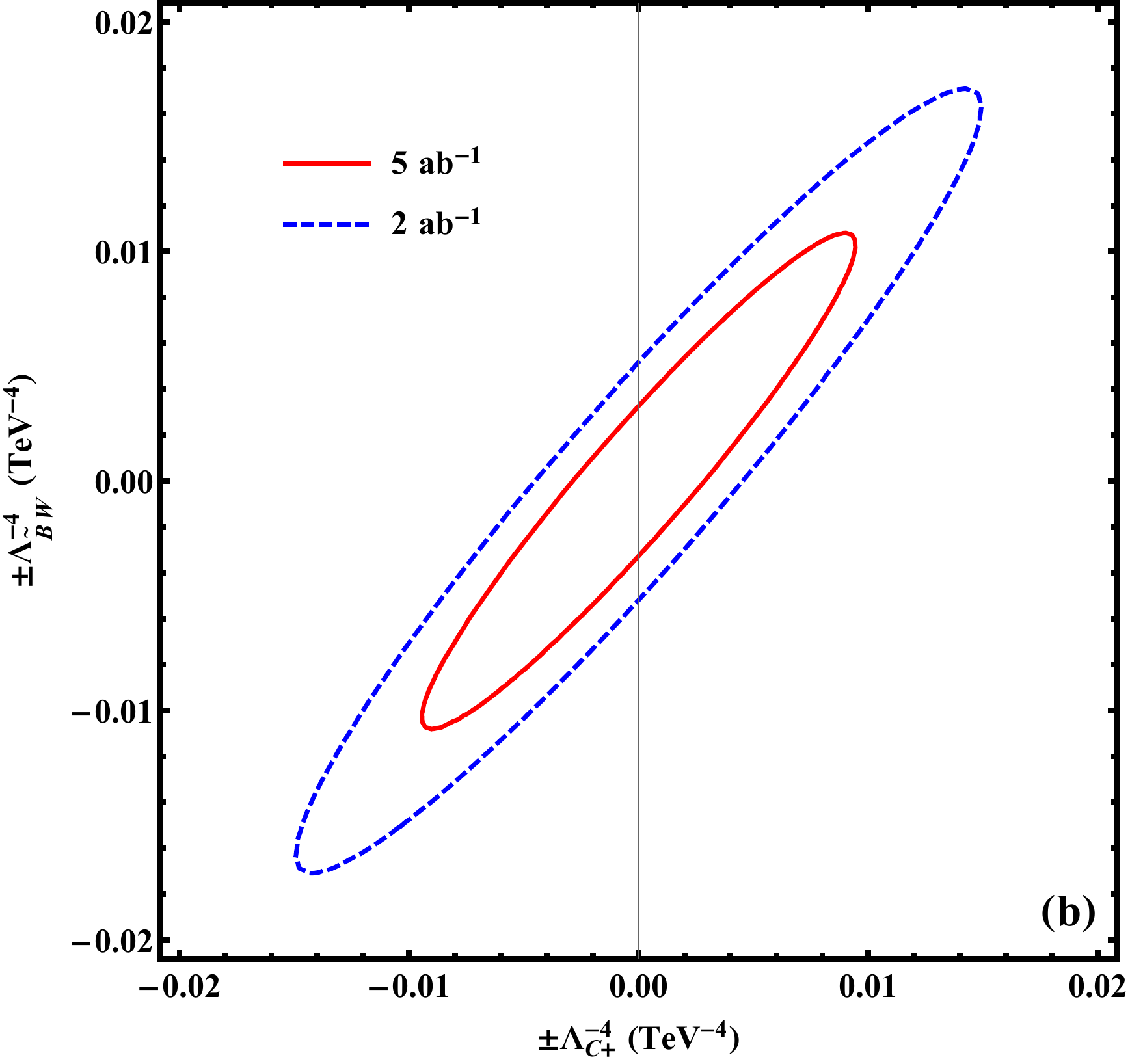}
\\[2.5mm]
\includegraphics[height=6.8cm,width=7.5cm]{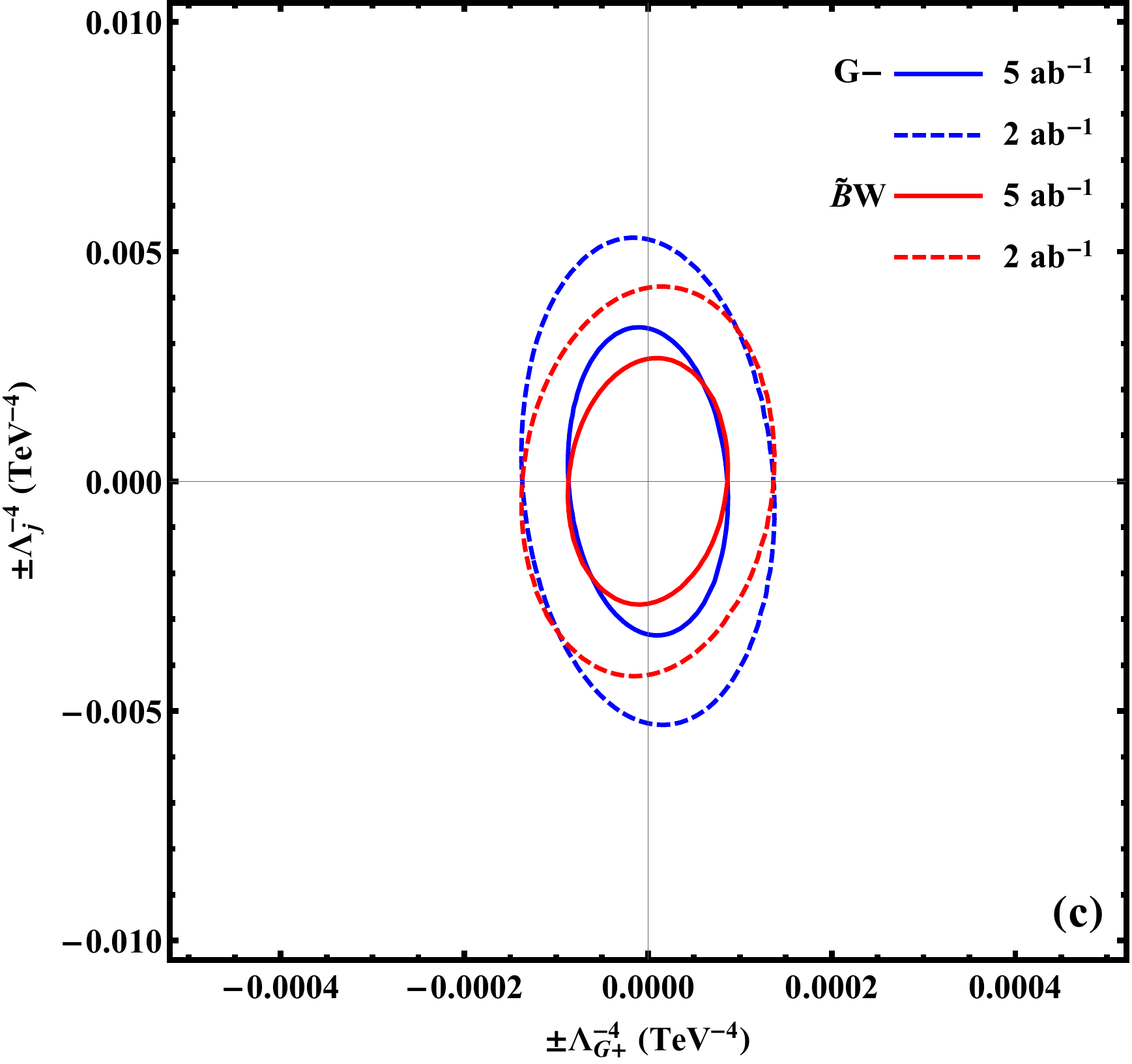}
\hspace*{1.5mm}
\includegraphics[height=6.8cm,width=7.5cm]{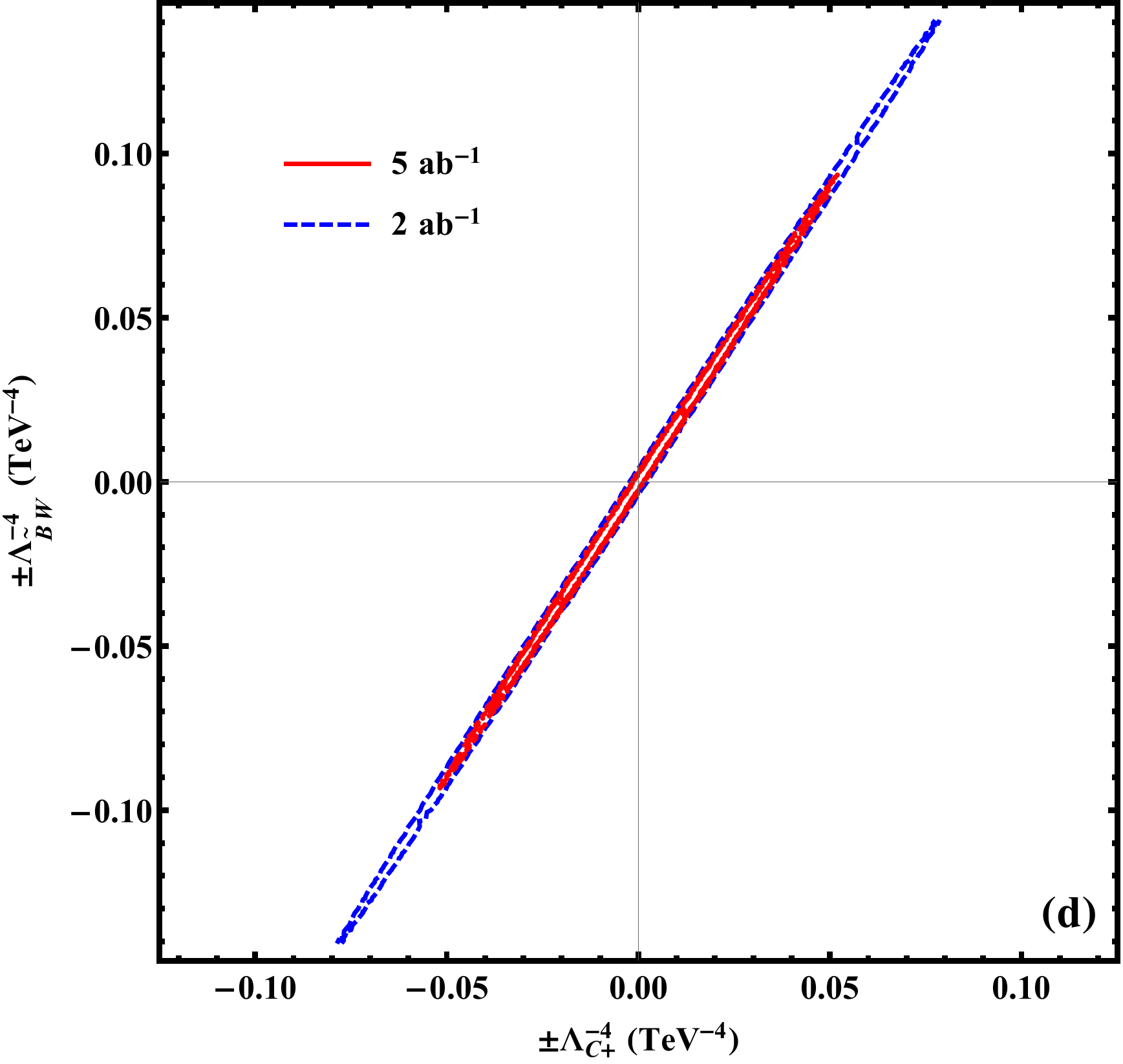}
\vspace*{-1.5mm}
\caption{\small{%
Sensitivity bounds on the new physics scales $\cut_j^{}$ at $2\sigma$ level
for each pair of the operators $(\mO_{G+}^{},\, \mO_{G-}^{})$,
\,$(\mO_{G+}^{},\, \mO_{\widetilde{B}W}^{})$,\, and
\,$(\mO_{C+}^{},\, \mO_{\widetilde{B}W}^{})$,
respectively, for the collider energy} $\sqrt{s\,}\!=\!3$\,TeV.
In each plot, the results are shown for an integrated luminosity
$\mL\!=\!2\,\text{ab}^{-1}$ (dashed curves) and
$\mL\!=\!5\,\text{ab}^{-1}$ (solid curves).
The plots\,(a) and (b) are shown for the case of unpolarized $e^\mp$ beams,
whereas the plots\,(c) and (d) depict the case with polarized beams
$(P_L^e,\,P_R^{\bar{e}})\!=\!(0.9,\,0.65)$\,.
Here the $\,\pm\,$ inside each axis label denote the sign of the coefficient
$\tilde{c}_j^{}$ of the corresponding operator.
}
\label{figchi}
\label{fig:9}
\end{center}
\vspace*{-10mm}
\end{figure}

\vspace*{1mm}

Minimizing the $\chi^2$ function \eqref{eq:chi2}, we can derive constraints
among the cutoff scales $\,\cut_\rho^{}\,$ for any given subset of effective operators.
For the current analysis, we study each pair of operators and
analyze the correlations between their associated cutoff scales.
We find that the most sensitive observable for probing the operators
$\OBW$ and $\OCP$ is $\,\mathbb{O}_A^c$,
whereas for probing the operator $\OGM$, it is $\,\mathbb{O}_B^c$,
and for probing the operator $\mO_{G+}^{}$,\, it is $\mathbb{O}_C^c$\,.

\vspace*{1mm}

We present in Fig.\,\ref{fig:9} the sensitivity bounds on the new physics
scales $\cut_j^{}$ at the $2\sigma$ level
for the pairs of the operators $(\mO_{G+}^{},\, \mO_{G-}^{})$,
\,$(\mO_{G+}^{},\, \mO_{\widetilde{B}W}^{})$,\, and
\,$(\mO_{C+}^{},\, \mO_{\widetilde{B}W}^{})$,
respectively.
For illustration, we have chosen in each plot a collider energy
$\sqrt{s\,}\!=\!3$\,TeV,
and an integrated luminosity
$\mL\!=\!2\,\text{ab}^{-1}$ (dashed curve) and
$\mL\!=\!5\,\text{ab}^{-1}$ (solid curve).
The $\,\pm\,$ in each axis label denotes the sign of the coefficient
$\tilde{c}_j^{}$ of the corresponding operator $\mO_j^{}$\,.
The plots\,(a) and (b) are for the case of unpolarized $e^\mp$ beams,
and plots\,(c) and (d) depict the case with partially-polarized
beams: $(P_L^e,\,P_R^{\bar{e}})\!=\!(0.9,\,0.65)$.
In plots\,(a) and (c) we present the correlation contours for
the operators $(\mO_{G+}^{},\, \mO_{\widetilde{B}W}^{})$ in red color,
and for the operators $(\mO_{G+}^{},\, \mO_{G-}^{})$ in blue color.
The correlation contours for the operators
$(\mO_{C+}^{},\, \mO_{\widetilde{B}W}^{})$
are shown in plots\,(b) and (d).

\vspace*{1mm}

We see in plots\,(b) and (d) of Fig.\,\ref{fig:9}
that the operators $\mO_{\widetilde{B}W}^{}$ and $\mO_{C+}^{}$
are highly correlated, unlike the cases shown in plots (a) and (c).
This is because the operators $\mO_{\widetilde{B}W}^{}$ and $\mO_{C+}^{}$
are both sensitive to the same type of observable $\mathbb{O}_A^c$.
In particular, we note that in the polarized case shown in plot\,(d)
the ellipse contour collapses towards a straight line, which corresponds to
the limit of fully-polarized $e^\mp$ beams.
In this limit we find the following $\chi^2$ function
for the contributions of the two operators
$(\mO_{\widetilde{B}W}^{},\,\mO_{C+}^{})$:
\begin{eqnarray}
\label{eq:chi2p}
\chi^2 &\!\!=\!\!& \sum_q\!\left\{\!
\frac{\left[3\mathbb{A}_q^{}(s\!+\!M_Z^2)\right]^{\!2}}{\sigma_{0A}^{c(q)}}
+\frac{\left[3\mathbb{B}_q(s\!+\!M_Z^2)\right]^{\!2}}{\sigma_{0B}^{c(q)}}
+\frac{\left[\mathbb{C}_qM_Z^2\right]^{\!2}}{\sigma_{0C}^{c(q)}}
\!\right\}\!\times\!\mL\!\times\!\!
\left[\!\frac{\,c_L^{}\!\!
\(c_L \tilde{c}_{\widetilde{B}W}^{}\!+\!\fr{1}{2}\tilde{c}_{C+}^{}\!\)\,}
     {\tilde{\Lambda}^{4}}\!\right]^{\!2}
\hspace*{10mm}
\nn\\[1mm]
&\!\!=\!\!&
\text{constant}\times
\left[\!\frac{\,c_L^{} \tilde{c}_{\widetilde{B}W}^{}\!+\!\fr{1}{2}\tilde{c}_{C+}^{}\,}
     {\tilde{\Lambda}^{\!4}_{}}\!\right]^{\!2},
\end{eqnarray}
where the summation index $q$ runs over the relevant quark flavors in the
final state $Z\!\to\!q\bar{q}$\,.
We see that the minimum value $\,\chi^2=0\,$ leads to the equation
\beqa
\label{eq:FullPolEq}
c_L^{} \frac{\,\tilde{c}_{\widetilde{B}W}^{}\,}{\tilde{\Lambda}^{\!4}_{}}
+\frac{1}{2}\frac{\,\tilde{c}_{C+}^{}\,}{\tilde{\Lambda}^{\!4}_{}} \,=\,0\,,
\eeqa
which describes a straight line whose slope is
$\,-\fr{1}{\,2c_L^{}\,}\!=\!(1\!-\!2s_W^2)^{-1}\!\simeq\! 1.86\,$.
We note that this feature is reflected in plot\,(d)
even for the partially-polarized case
$(P_L^e,\,P_R^{\bar{e}})\!=\!(0.9,\,0.65)$\,.
In the case of $100\%$ polarized $e^\mp$ beams, the operators
$\mO_{\widetilde{B}W}^{}$ and $\mO_{C+}^{}$ would be fully correlated
along the straight line determined by equation\,\eqref{eq:FullPolEq}.

\vspace*{1.5mm}
\section{\large\hspace{-2mm}Conclusions}
\label{sec:5}
\label{sec:conclusion}

In this work, we have used the SMEFT formulation to study systematically how
dimension-8 operators that contribute to neutral triple gauge couplings (nTGCs)
can be probed via the reaction $\,e^+ e^-\!\!\to\! Z\ga\,$
with hadronic decays $Z\!\!\to\!q\bar{q}$\, at future $e^+e^-$ colliders.

\vspace*{1mm}

We have constructed a new set of dimension-8 CP-conserving pure gauge operators
\eqref{eq:OG} that contribute to the nTGC vertices with a leading energy
dependence $\propto\!E^5$.\,
We paid special attention to the dimension-8 pure gauge operators \eqref{eq:OG+-}
and the related dimension-8 $\,e^-e^+Z\ga$\, contact interactions \eqref{eq:C+-},
in addition to the Higgs-related operators
that we had studied previously
via leptonic $Z$ decay channels\,\cite{Ellis:2019zex}.
We found that among the five dimension-8 operators
$(\OGP,\,\OGM,\,\OBW,\,\OCP,\,\OCM)$,
the fermionic contact operator $\OCM$ is equivalent to
the pure gauge operator $\OGP$
for the reaction $\,e^+ e^-\!\!\to\! Z\ga\,$ due to the
equation of motion (EOM) \eqref{eq:OG+QBW-F}, whereas
the fermionic contact operator $\OCP$ can be reexpressed in terms
of the pure gauge operator $\OGM$ and the Higgs-related operator
$\OBW$ due to the EOM \eqref{eq:OG-QBW-F}.
Hence, only three of the five operators are independent due to the EOMs.
We have mainly focused on the linearly-independent operators
$(\OGP,\,\OGM,\,\OBW)$ because they
contribute directly to the nTGCs and the contact operators
$\OCP$ and $\OCM$ do not.
As expected from the strong energy dependence of the contributions
to amplitudes from the type of dimension-8 operators shown in Eq.\eqref{eq:OG+-},
higher-energy colliders generally have greater sensitivity than their lower-energy
counterparts, assuming that comparable amounts of integrated luminosity can be
accumulated. This is shown in the scaling relations
\eqref{eq:Z-scaling}-\eqref{eq:lam4} for instance.
Moreover, the use of hadronic $Z$ decays with larger branching fractions
increases the sensitivity significantly,
as compared to leptonic decay channels
studied in our previous work\,\cite{Ellis:2019zex}.

\vspace*{1mm}

We have analyzed the prospective sensitivities to the new physics cutoff scales
$\cut_j^{}$  for individual dimension-8 operators
via the reaction $\,e^+ e^-\!\!\to\! Z\ga\,$ with $\,Z\!\!\to\!q\bar{q}$\,.
We found that the dimension-8 pure gauge operator $\OGP$ in Eq.\eqref{eq:OG+}
provides the most sensitive probe to the new physics scale of nTGCs, namely,
the $\,\cut_{G+}^{}$ can be probed up to the range $(1-5)$\,TeV
for the CEPC\,\cite{CEPC}, FCC-ee\,\cite{FCCee} and ILC\,\cite{ILC} colliders with
$\sqrt{s\,} \!=\! (0.25-1)$\,TeV,\, and up to the range $(10-16)$\,TeV
for CLIC\,\cite{CLIC} with $\sqrt{s\,} \!=\! (3-5)$\,TeV,\,
choosing in each case a sample integrated luminosity of
$\,\mL \!=\! 5$\,ab$^{-1}$.\,
We presented systematically new results of the sensitivity reaches for
a set of relevant dimension-8 operators $(\OGP,\,\OGM,\,\OBW,\,\OCP)$
at the $2\sigma$ (exclusion) and $5\sigma$ (discovery) levels
using hadronic $Z$ decays and unpolarized $e^\mp$ beams in Table\,\ref{tab:1},
and for partially-polarized $e^\mp$ beams in Table\,\ref{tab:2},
which can be compared with the results using leptonic $Z$ decays
in Table\,\ref{tab:3}.
We stress that the new sensitivity limits obtained for the dimension-8
pure gauge operator $\OGP$ and for the hadronic $Z$ decay channels in this work
are {\it substantially stronger} than our previous limits for the Higgs-related operator
$\OBW\!$ via the leptonic $Z$ decay channels, as can be clearly seen by
comparing the current Tables\,\ref{tab:1}-\ref{tab:2} versus Table\,\ref{tab:3}.
We have further studied the prospects for simultaneous fits
to pairs of dimension-8 operators, including correlations,
with the results shown in Fig.\,\ref{figchi}.

\vspace*{1mm}

The present findings in Tables\,\ref{tab:1}-\ref{tab:3} and
Figs.\,\ref{fig:8}-\ref{fig:9}
largely go beyond our previous study\,\cite{Ellis:2019zex},
and demonstrate that the reaction $\,e^+ e^- \!\to\! Z \ga\,$
provides an important opportunity to explore the dimension-8 SMEFT contributions
to nTGCs, with sensitivity reaches of the corresponding new physics scales
extending well into the multi-TeV range.
This opportunity provides a new prospect to probing a class of
dimension-8 interactions that is complementary to the other different
dimension-8 terms probed by processes such as the light-by-light scattering in
heavy-ion collisions\,\cite{EMY} and the gluon-gluon\,$\to\!\ga\ga$ in $pp$ collisions\,\cite{EG}.
It would be valuable to extend our present study of the dimension-8 interactions
of nTGCs to the LHC and future high energy $pp$ colliders, and to compare their
prospective sensitivities to those of the $e^+e^-$ colliders in the current study.
We will pursue this research direction in the future work.

\vspace*{8mm}
\noindent
{\bf\large Acknowledgements}
\\[1mm]
We thank Manqi Ruan, Philipp Roloff, Michael Peskin, Tao Han, Tim Barklow and Jie Gao
for useful discussions.
The work of JE was supported in part by United Kingdom STFC Grant ST/P000258/1,
in part by the Estonian Research Council via a Mobilitas Pluss grant, and in part
by the TDLI distinguished visiting fellow programme. The work of HJH and RQX was
supported in part by the National NSF of China (under grants 11675086 and 11835005).
HJH is also supported in part by the CAS Center for Excellence in Particle Physics (CCEPP), by the National Key R\,\&\,D Program of China (No.\,2017YFA0402204),
by the Key Laboratory for Particle Physics, Astrophysics and Cosmology
(Ministry of Education),
and by the Office of Science and Technology, Shanghai Municipal Government
(No.\,16DZ2260200).

\vspace*{8mm}
\appendix

\noindent	
{\bf\large Appendix:}
\vspace*{-3mm}

\section{\large\hspace{-2mm}Helicity Amplitudes for
\boldmath{${Z\gamma}$} Production from
\boldmath{${O}_{G\pm}^{}$} and \boldmath{${O}_{C\pm}^{}$}}
\label{app:A}

In this Appendix, we present our results for the helicity amplitudes of
the reaction $\,e^-e^+\!\to\!Z\ga\,$, including contributions from both
the SM and the dimension-8 operators
$({O}_{G\pm}^{},\,{O}_{C\pm}^{})$.

\vspace*{1mm}

For the final state $Z(\lam)\ga(\lam')$ with helicity combinations
$\lam\lam'\!=\!(--,-+,+-,++)$ and $\lam\lam'\!=\!(0-,0+)$,\,
we find the following SM contributions to the scattering amplitudes:
	\beqs
	\label{eq:Tsm-T+L}
	\begin{eqnarray}
	\hspace*{-2mm}
	\mathcal{T}_{\text{sm}}^{ss'\!,\text{T}}\!\!\left\lgroup\!\!
	\begin{array}{cc}
	-- \!&\! -+ \\
	+- \!&\! ++\\
	\end{array}\!\!\right\rgroup
	\!\!\!\!&=&\!\!\!
    \frac{2e^2}{\,s_W^{}c_W^{}(s\!-\!M_Z^2)\,}\!\!
	\left\lgroup\!\!\!
	\begin{array}{ll}
	\(e_L^{}\!\cot\!\frac{\theta}{2}\!-\!e_R^{}\!\tan\!\frac{\theta}{2}\)\!M_Z^2~
	\!&\!\!
	\(-e_L^{}\!\cot\!\frac{\theta}{2}\!+\!e_R^{}\!\tan\!\frac{\theta}{2}\)\!s
	\\[2mm]
	\(e_L^{}\!\tan\!\frac{\theta}{2}\!-\!e_R^{}\!\cot\!\frac{\theta}{2}\)\!s
	\!&\!\!
	\(-e_L^{}\!\tan\!\frac{\theta}{2}\!+\!e_R^{}\!\cot\!\frac{\theta}{2}\)\!M_Z^2
	\end{array}
	\!\!\!\right\rgroup\!\!,\hspace*{16mm}
	\label{msmT}
	\label{eq:Tsm-T}
	\\[2mm]
	\hspace*{-2mm}
	\mathcal{T}_{\text{sm}}^{ss'\!,\text{L}}(0-,0+)
	\!\!\!&=&\!\!\! \frac{\,2\sqrt{2}(e_L^{}\!\!+\!e_R^{})e^2M_Z^{}\sqrt{s\,}\,}
	{\,s_W^{}c_W^{}(s\!-\!M_Z^2)\,}\(1,\,-1\),
	\label{msm}
	\label{eq:Tsm-L}
	\end{eqnarray}
	\eeqs
where $\,(e_L^{},\,e_R^{}) = (c_L^{}\delta_{\!s,-\frac{1}{2}}^{},\,
c_R^{}\delta_{\!s,\frac{1}{2}}^{})$,\, with the subscript index
$\,s =\mp\fr{1}{2}$\, denoting the initial-state electron helicities.
For the massless initial-state $e^-$ and $e^+$, we have $s=-s'$.\,

\vspace*{1mm}

We then compute the corresponding helicity amplitudes from the new physics
contributions of the dimension-8 operator
$\mO_{G+}^{}\,(\mO_{\!C-}^{})$ as follows:
\beqs
\label{eq:T8}
\begin{eqnarray}
\label{eq:T8-T}
	\mathcal{T}_{(8)}^{ss'\!,\text{T}}
	\!\!\left\lgroup\!\!
	\begin{array}{cc}
	-- \!&\! -+ \\
	+- \!&\! ++\\
	\end{array}\!\!\right\rgroup
	\!\!&=&\!\!
	\frac{\,(e'_L\!\!+\!e'_R) (s\!-\!M_Z^2)s\sin\!\theta\,}{\cut^4}\!\!
	\left\lgroup\!
	\begin{array}{cr}
	1 ~& 0
	\\[1mm]
	0 ~& -1
	\end{array}
	\!\right\rgroup \!\!,
	\label{m8T}
	\\[2mm]
	\mathcal{T}_{(8)}^{ss'\!,\text{L}} (0-,0+)
	\!\!&=&\!\!
    \frac{\,\sqrt{2}M_Z^{}(s\!-\!M_Z^2)\sqrt{s}\,}{\cut^4}\!
	\(\!e'_L\!\sin^2\!\frac{\theta}{2}\!-e'_R\!\cos^2\!\frac{\theta}{2},~
	e'_R\!\sin^2\!\frac{\theta}{2}\!-e'_L\!\cos^2\!\frac{\theta}{2}
	\)\!,
\hspace*{18mm}
\label{eq:T8-L}
\label{m8}
\end{eqnarray}
\eeqs
where $\,(e'_L,\,e'_R) \!= \fr{1}{2}(\delta_{\!s,-\frac{1}{2}}^{},\,0)$.
We see from Eq.\eqref{eq:T8-T} that the transverse amplitudes
$\,\mathcal{T}_{(8)}^{ss'\!,\text{T}}\!\propto\!(\!\sqrt{s\,})^4\,$
in the high energy limit. This can be understood by recalling that the $Z\ga Z^*$
or $Z\ga\ga^*$  vertex of $\mO_{G+}$ has the leading energy dependence
of $(\!\sqrt{s\,})^5$, the $s$-channel $Z^*$\,($\ga^*$) propagator contributes
a leading energy factor $1/s$\,, and the initial state spinor wavefunctions
of $e^-e^+$ contribute an additional leading energy factor $\sqrt{s\,}$.
Hence, the amplitude $\,\mathcal{T}_{(8)}^{ss'\!,\text{T}}\,$
has a leading energy dependence of  $(\!\sqrt{s\,})^4$\,.

\vspace*{1mm}

On the other hand, for the longitudinal gauge boson $Z_L^{}$ in the final state,
we see from Eq.\eqref{eq:T8-L} that the corresponding amplitudes have a lower
energy dependence
$\,\mathcal{T}_{(8)}^{ss'\!,\text{L}}\!\propto\! (\!\sqrt{s\,})^3$.
This is because the longitudinal polarization vector of the final state $Z_L^{}$
has the leading energy behavior
$\,\epsilon_{L}^\mu\!\propto\!q_Z^\mu/M_Z^{}\,$
and the inclusion of $\epsilon_{L}^\mu$
causes a cancellation in the energy factors from $\,s^1\,$ down to $M_Z^2$
in the $Z\ga Z^*$ and $Z\ga\ga^*$  vertices \eqref{eq:Vertex-G+}.
Hence, we find that the leading energy dependence of the longitudinal amplitudes
reduces to  $\,\mathcal{T}_{(8)}^{ss'\!,\text{L}}\!\propto\!(\!\sqrt{s\,})^3\,$.
This is not an accidental cancellation, as we can understand
the leading energy behavior of $\,\mathcal{T}_{(8)}^{ss'\!,\text{L}}$\,
on general grounds by using the equivalence theorem (ET)\,\cite{ET}.
According to the ET, the longitudinal scattering amplitude is connected to
the corresponding Goldstone boson scattering amplitude at high energies.
Thus, in the case of the contribution from $\OGP$, we have
\beqa
\vspace*{-2mm}
\mathcal{T}_{(8)}^{}[Z_L^{},\ga_T^{}] \,=\,
\mathcal{T}_{(8)}^{}[-\ii\pi^0,\ga_T^{}] + B\,,
\label{eq:ET}
\eeqa
where $\pi^0$ is the would-be Goldstone boson absorbed
by $Z_L$ through the Higgs mechanism, and
the residual term $\,B=\mathcal{T}_{(8)}^{}[v^\mu Z_\mu^{},\ga_T^{}]\,$
with $\,v^\mu\equiv\epsilon_L^\mu -q_Z^\mu/M_Z^{}\!=\!O(M_Z^{}/E_Z^{})$
\cite{ET}.
Thus, using the vertex \eqref{eq:Vertex-G+} we readily deduce that the
$\OGP$ contribution to the residual term $B$ has the leading energy behavior
$\,B\!\propto\!(\!\sqrt{s\,})^3$.
Note that the dimension-8 pure gauge operator $\OGP$ defined in Eq.\eqref{eq:OG+}
does not contain any Higgs boson or Goldstone boson field.
This means that $\OGP$ gives vanishing contribution to the Goldstone boson amplitude
at tree-level, $\mathcal{T}_{(8)}^{}[\pi^0,\ga_T^{}]=0$\,.
Hence, the ET \eqref{eq:ET} predicts the following leading energy behavior
of the longitudunal amplitude,
\beqa
\mathcal{T}_{(8)}^{}[Z_L^{}\ga_T^{}] \,=\,
B \propto (\!\sqrt{s\,})^3 \,.
\eeqa
This explains the leading energy dependence of
the scattering amplitudes \eqref{eq:T8-L}
which we obtained by explicit calculations.

\vspace*{1mm}

Next, we derive the new physics contributions of the dimension-8 operators
$\mO_{G-}^{}$, $\mO_{\tilde{B}W}^{}$ and $\mO_{C+}^{}$ as follows:
\beqs
\label{eq:T8-OG-OBW-OC+}
\begin{eqnarray}
\label{eq:T8-TT-OG-OBW-OC+}
\mathcal{T}_{(8)}^{ss'\!,\text{T}}
	\!\!\left\lgroup\!\!
	\begin{array}{cc}
	-- \!&\! -+ \\
	+- \!&\! ++\\
	\end{array}\!\!\right\rgroup
	\!\!\!&=&\!\!\!
	\frac{\,(e'_L\!\!+\!e'_R)\sin\!\theta M_Z^2(s\!-\!M_Z^2)\,}{\cut^4}\!\!
	\left\lgroup\!\!
	\begin{array}{cr}
	1 \!& 0
	\\[1mm]
	0 \!& -1
	\end{array}
	\!\!\right\rgroup \!\!,
	\\[2mm]
\label{eq:T8-LT-OG-OBW-OC+}
	\mathcal{T}_{(8)}^{ss'\!,\text{L}} (0-,0+)
	\!\!&=&\!\!\! \frac{\,\sqrt{2}M_Z^{}(s\!-\!M_Z^2)\sqrt{s}\,}{\cut^4}\!
	\(\!e'_L\!\sin^2\!\frac{\theta}{2}\!-e'_R\!\cos^2\!\frac{\theta}{2},~
	e'_R\!\sin^2\!\frac{\theta}{2}\!-e'_L\!\cos^2\!\frac{\theta}{2}
	\)\!. \hspace*{17mm}
	\end{eqnarray}
	\eeqs	
where the coupling factors
$\,(e'_L,\,e'_R) = s_W^2(\delta_{\!s,-\frac{1}{2}}^{},\,
\delta_{\!s,\frac{1}{2}}^{})$ for $\mO_{G-}$,\, $(e'_L,\,e'_R) = (c_L\delta_{\!s,-\frac{1}{2}}^{},\,
c_R\delta_{\!s,\frac{1}{2}}^{})$ for $\mO_{\tilde BW}$,
and $\,(e'_L,\,e'_R) = \frac{1}{2}(\delta_{\!s,-\frac{1}{2}}^{},\,0)$
for $\mO_{C+}$.

\vspace*{1mm}

We find that the leading energy behaviors of the amplitudes
\eqref{eq:T8-OG-OBW-OC+} agree well with direct power counting estimates,
since there is no extra energy cancellation involved.
For instance, from Eqs.\eqref{eq:vertex-OG-}-\eqref{eq:vertex-OBW},
we see that the contribution of the operator
$\OGM$ or $\OBW$ to the nTGC vertex $Z\ga\ga^*$ or $Z\ga Z^*$
scales as $E^3$ in the high-energy limit, so their contributions to
the amplitude of the $s$-channel process $e^-e^+\!\to\!Z_T^{}\ga_T^{}$
scale as $\,E^2\,$ after taking into account of the energy factors
from the $s$-channel propagator and the spinor wavefunctions of the initial state
$e^-e^+$. This agrees with the leading energy dependence $s^1$
in Eq.\eqref{eq:T8-TT-OG-OBW-OC+}.
The operator $\OCP$ contributes to the $f\bar{f}Z\ga$ contact vertex
\eqref{eq:vertex-OC+} with the leading energy dependence $E^1$.
So, after including the energy factor of the $e^-e^+$ spinor wavefunctions
of the initial states, we find that it contributes to the process
$e^-e^+\!\to\!Z_T^{}\ga_T^{}$ with a leading energy behavior $\propto\!E^2$,
which also coincides with that of Eq.\eqref{eq:T8-TT-OG-OBW-OC+}.
Finally, inspecting the amplitudes \eqref{eq:T8-LT-OG-OBW-OC+} that include
the longitudinally-polarized weak boson $Z_L^{}$ in the final state,
we note that the final state $Z_L^{}$ contributes an additional energy factor $E^1$
to the scattering amplitudes through its longitudinal polarization vector
$\ep_L^\mu\!\propto\! q_Z^\mu/M_Z^{}$\,. Hence, we find that the
longitudinal amplitudes of the reaction $\,e^-e^+\!\!\to\!Z_L^{}\ga_T^{}$\,
should scale like $\,\mathcal{T}_{(8)}^{ss'\!,\text{L}}\!\propto E^3$,\,
in agreement with Eq.\eqref{eq:T8-LT-OG-OBW-OC+}.


\end{document}